\documentclass[10pt,aps,prd,twocolumn,groupedaddress,showpacs]{revtex4-1}
\usepackage{graphicx}
\usepackage{amsmath}
\usepackage{hyperref}
\begin{document}
\title{Gravitational wave background from rotating neutron stars}
\author{Pablo A. \surname{Rosado}}
\email[]{pablo.rosado@aei.mpg.de}
\affiliation{Albert Einstein Institute, Max Planck Institute for Gravitational Physics, 30167 Hanover, Germany}
\date{\today}
\begin{abstract}
The background of gravitational waves produced by the ensemble of rotating neutron stars (which includes pulsars, magnetars, and gravitars) is investigated.
A formula for $\Omega(f)$ (a function that is commonly used to quantify the background, and is directly related to its energy density) is derived,
without making the usual assumption that each radiating system evolves on a short time scale compared to the Hubble time;
the time evolution of the systems since their formation until the present day is properly taken into account.
Moreover, the formula allows one to distinguish the different parts of the background: the unresolvable (which forms a stochastic background or confusion noise, since the waveforms composing it cannot be either individually observed or subtracted out of the data of a detector) and the resolvable.
Several estimations of the background are obtained, for different assumptions on the parameters that characterize neutron stars and their population.
In particular, different initial spin period distributions lead to very different results.
For one of the models, with slow initial spins, the detection of the background by present or planned detectors can be rejected.
However, other models do predict the detection of the background, that would be unresolvable, by the future ground-based gravitational wave detector ET.
A robust upper limit for the background of rotating neutron stars is obtained; it does not exceed the detection threshold of two cross-correlated Advanced LIGO interferometers.
If gravitars exist and constitute more than a few percent of the neutron star population, then they produce an unresolvable background that could be detected by ET.
Under the most reasonable assumptions on the parameters characterizing a neutron star, the background is too faint to be detected.
Previous papers have suggested neutron star models in which large magnetic fields (like the ones that characterize magnetars) induce big deformations in the star, which produce a stronger emission of gravitational radiation.
Considering the most optimistic (in terms of the detection of gravitational waves) of these models, an upper limit for the background produced by magnetars is obtained; it could be detected by ET, but not by BBO or DECIGO.
Simple approximate formulas to characterize both the total and the unresolvable backgrounds are given for the ensemble of rotating neutron stars, and, for completion, also for the ensemble of binary star systems.
\end{abstract}

\maketitle

\newcommand{\om}{\Omega(f)}

\section{\label{sec:intro} Introduction}
The topic of this paper is the gravitational wave background \cite{AllenRomano1999,Maggiore2000} produced by the ensemble of rotating neutron stars in the universe.
These systems are modeled as isolated neutron stars \cite{LattimerPrakash2004} that are formed with an initial spin frequency, and lose energy via electromagnetic dipole emission \cite{Deutsch1955,Pacini1968} and via quadrupolar gravitational radiation \cite{[{Section 9.4.2 (b) of }] HawkingIsrael1987,Prix2009}.
The ensemble of rotating neutron stars contains the populations of pulsars, magnetars, and gravitars.

Pulsars \cite{LorimerKramer2005} are neutron stars that emit electromagnetic radiation in a beam which, if pointing towards Earth, is observed as a ``lighthouse'' of great regularity.
We neglect the contribution of recycled pulsars \cite{Lorimer2008}.

Magnetars \cite{ThompsonDuncan1993,ThompsonDuncan1995,ThompsonEtAl2002,HardingLai2006} are neutron stars with a magnetic field a few orders of magnitude stronger than usual pulsars.
That magnetic field may support large ellipticities \cite{Cutler2002} leading to an enhanced production of gravitational radiation.
We obtain an upper limit for the background produced by the magnetars.

Gravitars \cite{Palomba2005,KnispelAllen2008} are hypothetical neutron stars that have a magnetic field weaker than usual pulsars, and lose rotational energy primarily via gravitational radiation.
There may exist a population of gravitars that cannot yet be detected because they emit very little or no electromagnetic radiation.
A simulation performed in \cite{PopovEtAl2000} shows that the conditions for neutron stars to be gravitars described in \cite{Palomba2005} are possible.
In this paper we investigate the detection prospects for the background produced by such a population.
The ensemble of gravitars provides an upper limit for the background of rotating neutron stars.

This work is a follow-on study to \cite{Rosado2011}, where the background produced by binary systems is studied (including binaries formed by white dwarfs, neutron stars, and black holes).
With both papers, two of the most promising sources of contemporary background are covered.

Other potential sources of contemporary background, not discussed in this paper or in \cite{Rosado2011}, are newborn neutron stars undergoing r-mode instabilities \cite{OwenEtAl1998,ZhuEtAl2011b}, compact objects captured by massive black holes \cite{BarackCutler2004}, inspiralling black hole binaries with intermediate or extreme mass-ratio \cite{AmaroEtAl2007}, supernovae \cite{BuonannoEtAl2005}, and population II and III stars \cite{MarassiEtAl2009,KowalskaEtAl2012}.

Besides the contemporary background, there may exist a primordial one \cite{Allen1996,Maggiore2000,Buonanno2004}, arising from processes in the early history of the universe.

We calculate what part of the total background of rotating neutron stars is unresolvable (commonly named \textit{confusion noise} or \textit{stochastic background}).
The signals composing this part cannot be distinguished from each other or subtracted from the data of a gravitational wave detector (we do not study the problem of the subtraction of resolvable signals, treated, for example, in \cite{CutlerHarms2006,CutlerHolz2009,YagiSeto2011}).
The resolvability of the background is quantified by the \textit{overlap function}, $\mathcal{N}(f,\Delta f,z)$, introduced in \cite{Rosado2011}.
This function gives the expected number of signals, with redshifts smaller than $z$, that are observed within a frequency bin $[f,f+\Delta f]$, where $\Delta f$ is the frequency resolution allowed by the detector and the data analysis method.
When a frequency bin is constantly occupied by one or more overlapping signals, i.e., $\mathcal{N}(f,\Delta f,\infty)\ge 1$, these signals cannot be disentangled, and form an unresolvable background.

The spectral gravitational wave density parameter, or, simply, \textit{spectral function}, $\Omega(f)$, is often used to quantify the background \cite{AllenRomano1999}.
It gives the average energy density of gravitational radiation (per logarithmic frequency interval) divided by the critical density.
The generalized spectral function \cite{Rosado2011}, $\Omega(f,\Delta f,\mathcal{N}_0)$, has the same meaning as $\Omega(f)$, but it quantifies only the part of the background with more than $\mathcal{N}_0$ overlapping signals per frequency bin.
The total background and the unresolvable one are calculated by taking $\mathcal{N}_0=0$ and $\mathcal{N}_0=1$, respectively.
In this paper, the spectral function accounts for the time evolution of the systems, that is not assumed to be short compared to cosmic time scales.

Previous work has studied the gravitational wave background from pulsars \cite{RegimbauFreitas2001} and magnetars \cite{RegimbauFreitas2006,RegimbauMandic2008,MarassiEtAl2011b}.
These articles assume that all neutron stars are formed with the same initial spin frequency.
We show that the results change dramatically if the initial spin frequency follows a probability distribution.
In particular, for one of the distributions considered \cite{FaucherKaspi2006}, the detection of the background by present and planned detectors is rather unrealistic.

For some of the models considered, the detection of the background of rotating neutron stars could be possible by cross-correlating two interferometers of the Einstein Gravitational Wave Telescope (ET), assuming two of the proposed configurations (called ETB and ETD) \cite{PunturoEtAl2010}.
Furthermore, this background is unresolvable.
The current generation of present ground-based detectors \cite{Grote2010,AccadiaSwinkels2010,LIGO2009}, and the advanced version of the Laser Interferometer Gravitational Wave Observatory (aLIGO) \cite{Harry2010}, are not sensitive enough to detect this background.
For future space missions like the Big Bang Observer (BBO) \cite{CutlerHolz2009} and the Decihertz Interferometer Gravitational Wave Observatory (DECIGO) \cite{KawamuraEtAl2011}, the detection is rather unlikely.

The outline of the paper is as follows:

In Section \ref{sec:bg}, the notation and nomenclature of the paper is explained, and the quantification of the gravitational wave background, its resolvability and detectability are briefly reviewed.
A general formula for $\Omega(f,\Delta f,\mathcal{N}_0)$ is derived for a population of systems that emit at different times and locations, without assuming that the evolution of each system is short compared to cosmological time scales.
We also give a formula for $\mathcal{N}_0(f,\Delta f,z)$ which is more general than the one presented in the previous work \cite{Rosado2011}.

In Section \ref{sec:bg2} we expand upon the expressions of $\Omega(f,\Delta f,\mathcal{N}_0)$ and $\mathcal{N}_0(f,\Delta f,z)$, to account for the evolution of the population.
We obtain formulas that depend on the energy and frequency evolution of a system, the initial frequency distribution and the formation rate of the ensemble, and certain cosmological parameters.
Then, assuming that all systems start emitting at the same frequency and evolve in short time scales, we obtain the formula of the spectral function that is commonly used in the literature.

In Section \ref{sec:models} we describe the models assumed for a neutron star and its population.

Section \ref{sec:results} contains the main results of the paper.
We present a robust upper limit for the background of rotating neutron stars, the \textit{gravitar limit}.
We then obtain the background produced by gravitars, and study the likelihood of planned detectors to observe it and to place limits on the abundance of gravitars.
The most realistic expectation of the background of rotating neutron stars is calculated, using a magnetic field and an ellipticity distribution from the literature.
An upper limit on the background produced by magnetars is obtained.
We study the detection prospects of ETB, ETD, BBO and DECIGO, for different assumptions on the initial frequency, magnetic field, and ellipticity of neutron stars.

In Section \ref{sec:discussion} we compare our results with others from the literature.
We also comment on the insensitivity of the spectral function on the choice of star formation rate.

The main results and conclusions are put together in Section \ref{sec:summary}.
First, in Section \ref{sec:summary1}, the technical achievements regarding the calculation of $\Omega(f,\Delta f,\mathcal{N}_0)$ are summarized.
Then, in \ref{sec:summary2}, we compress all results and predictions regarding the detection of the background of rotating neutron stars.
A non-specialized reader interested only in the main conclusions should read the latter section.

In Appendices \ref{sec:rotnssimp} and \ref{sec:binsimp} we give simple approximate formulas for the spectral function of the background of rotating neutron stars, and also for the one of binary systems.
Finally, in Appendix \ref{sec:blandford} we point out a feature in the gravitar limit that is analogous to Blandford's argument \cite{KnispelAllen2008}.

\section{\label{sec:bg} Characterization of the background: an overview}
We follow the notation and terminology explained in Section II of \cite{Rosado2011}.
The index ``$e$'' (for \textit{emitted}) is used for frequencies and energies of the gravitational waves, as well as intervals of time, measured close to the system (for example, a single rotating neutron star) at the time of emission of the radiation.
Observed frequencies, energies and intervals of time (measured here and now) have no index.
Emitted quantities $f_e$, $E_e$, and $\Delta t_e$ (and infinitesimal emitted intervals $df_e$, $dE_e$, and $dt_e$) are affected by the expansion of the universe.
They are related to the observed quantities $f$, $E$, and $\Delta t$ ($df$, $dE$, and $dt$), by
\begin{equation}
\label{eq:redshiftf}
f=[1+z]^{-1}f_e, \quad df=[1+z]^{-1}df_e\,,
\end{equation} 
\begin{equation}
\label{eq:redshifte}
E=[1+z]^{-1}E_e, \quad dE=[1+z]^{-1}dE_e\,,
\end{equation} 
and
\begin{equation}
\label{eq:redshiftt}
\Delta t=[1+z]\Delta t_e, \quad dt=[1+z]dt_e\,,
\end{equation} 
where $z$ is the cosmological redshift.
Any given function $x$ that depends on $f_e$ can be written in terms of observed frequencies.
The notation $x\big|_f$ means that the function $x(f_e)$ must be written in terms of observed frequencies, i.e. $x\big|_f=x(f[1+z])$.

For convenience, a \textit{lookback time} interval is sometimes used, and denoted by an index $L$.
The relation between a lookback time interval $\Delta t^L$ and an ordinary lookforward time interval $\Delta t$ is $\Delta t^L=-\Delta t$.

\subsection{Quantification of the background}
The gravitational wave background is usually characterized by the spectral energy density parameter \cite{AllenRomano1999} (or, simply, \textit{spectral function}),
\begin{equation}
\label{eq:defomega}
\om=\frac{\rho_{\ln}(f)}{\rho_c}=\frac{\varepsilon_{\ln}(f)}{c^2\rho_c}\,,
\end{equation}
where $c$ is the speed of light.
The present critical density of the universe is
\begin{equation}
\rho_c=\frac{3H_0^2}{8 \pi G}\,,
\end{equation} 
where $G$ is the gravitational constant, and $H_0$ is the present Hubble expansion rate, of 74.2\,km\,s$^{-1}$\,Mpc$^{-1}$ \cite{RiessEtAl2009,RiessEtAl2011}.
The function $\varepsilon_{\ln}(f)$ is defined in such a way that $\varepsilon_{\ln}(f)d\ln f$ is the energy per unit volume of gravitational waves between $\ln f$ and $\ln f+d\ln f$.
Thus, $\om$ is related to the total density of gravitational radiation in the universe, that is
\begin{equation}
\label{eq:totalenergydensity}
\rho_{\text{gw}}=\int_0^\infty \rho_{\ln}(f) d\ln f=\rho_c \int_0^\infty \Omega(f) d\ln f \, .
\end{equation}
Here, $\om$ is the spectral function of all sources of gravitational radiation in our past light cone.
For simplicity, we use the same symbol to characterize the background produced only by the systems we are interested in (rotating neutron stars).

The spectral function fully characterizes a Gaussian, stationary, isotropic and unpolarized background \cite{AllenRomano1999}.
As claimed in \cite{Rosado2011}, the spectral function is also the right tool to characterize an unresolvable background.
On the other hand, one loses information when using the spectral function for a resolvable background.

We now derive $\Omega(f)$ for an ensemble of many sources, emitting at different times and locations, that can experience a time evolution.
The radiation we observe today has been produced by many individual systems in the past.
The energy emitted by one system during an infinitesimal interval of time is 
\begin{equation}
dE_e=\frac{dE_e}{dt_e}dt_e=\frac{dE_e}{dt_e}\frac{dt_e}{dt_e^L}dt_e^L=-\frac{dE_e}{dt_e}dt_e^L \, .
\end{equation}
Two waves that reach us now and were emitted at different lookback times $t_e^L$ and $t_e^L+dt_e^L$, have different redshifts $z$ and $z+dz$.
Lookback time intervals can thus be written as redshift intervals,
\begin{equation}
dt_e^L=\frac{dt_e^L}{dz}dz \, .
\end{equation}
The number of systems, per unit comoving volume, contributing to the background with observed frequencies between $\ln f$ and $\ln f+d\ln f$ is
\begin{equation}
dn=\frac{dn}{d\ln f}d\ln f \, .
\end{equation}
The present energy density of gravitational waves, per unit logarithmic frequency interval, produced by the collection of all systems is
\begin{equation}
\label{eq:varep}
\varepsilon_{\ln}(f)=\int_{0}^{t_0} \frac{dE}{dt} \frac{dn}{d\ln f}dt=\int_{0}^{\infty} \frac{dE}{dt} \frac{dn}{d\ln f}\frac{dt^L}{dz}dz \, ,
\end{equation}
where $t_0$ is the current age of the universe.
Using Equations (\ref{eq:redshiftf}), (\ref{eq:redshifte}), and (\ref{eq:redshiftt}), we can write
\begin{equation}
\label{eq:varep2}
\varepsilon_{\ln}(f)=\int_{0}^{\infty} [1+z]^{-1}\frac{dE_e}{dt_e}\bigg|_f \frac{dn}{d\ln f_e}\bigg|_f \frac{dt_e^L}{dz}dz \, .
\end{equation}
Replacing (\ref{eq:varep2}) in (\ref{eq:defomega}), we finally reach the formula for the spectral function of the total background,
\begin{equation}
\label{eq:omegadeflong}
\Omega(f)=\frac{1}{\rho_c c^2}\int_{0}^{\infty} [1+z]^{-1}\frac{dE_e}{dt_e}\bigg|_f \frac{dn}{d\ln f_e}\bigg|_f \frac{dt_e^L}{dz}dz \, .
\end{equation}
The functions $dE_e/dt_e$ and $dn/d\ln f_e$ are obtained in Sections \ref{sec:enerspec} and \ref{sec:evolution2}, respectively, for the ensemble of rotating neutron stars.
The function $dt_e^L/dz$ depends on the choice of the cosmological model; we assume a Lambda-Cold Dark Matter universe, so
\begin{equation}
\label{eq:dtedzdef}
dt_e^L=\frac{1}{[1+z]H_0 \mathcal{E}(z)}dz \, ,
\end{equation}
where
\begin{equation}
\mathcal{E}(z)=\sqrt{\Omega_m[1+z]^3+\Omega_\Lambda} \, .
\end{equation}
Here, $\Omega_m$ and $\Omega_\Lambda$ are the density parameters of matter and dark energy, respectively, whose values \cite{JarosikEtAl2011} are assumed to be $\Omega_m=0.27$ and $\Omega_\Lambda=0.73$.
In Section \ref{sec:resolvability} we explain how to modify the integration limits of Equation (\ref{eq:omegadeflong}) to measure only the unresolvable part of the background.

In the literature, one usually finds the spectral function written as
\begin{equation}
\label{eq:omegalim}
\om =\frac{1}{\rho_cc^2} \int_{0}^{\infty} [1+z]^{-1}\frac{dE_e}{d\ln f_e}\bigg|_f \dot{n}(z) \frac{dt_e^L}{dz}dz
\end{equation} 
(see, for example, Equation (35) of \cite{Rosado2011} or Equation (5) of \cite{Phinney2001}).
In Section \ref{sec:ovfunspecfun}, we show that (\ref{eq:omegalim}) can be derived from (\ref{eq:omegadeflong}) if one assumes that systems are short-lived and start emitting with the same initial frequency.

\subsection{Resolvability of the background}
\label{sec:resolvability}
The overlap function, $\mathcal{N}(f,\Delta f,z)$, allows us to define and quantify the resolvability of the background.
We now define the overlap function more generally than in \cite{Rosado2011}, as
\begin{align}
\label{eq:defovfungen}
\mathcal{N}(f,\Delta f,z)&=\int_0^z \int_{f}^{f+\Delta f} \frac{dn}{d f'}\frac{dV_c}{dz'}df'dz'\nonumber\\
&=\int_0^z \int_f^{f+\Delta f} [1+z']\frac{dn}{df_e}\bigg|_{f'}\frac{dV_c}{dz'}df' dz' \, .
\end{align}
Here, $dV_c$ is the element of comoving volume, given by
\begin{equation}
\label{eq:defcomvol}
dV_c=4\pi \left[\int_0^z \frac{c}{H_0 \mathcal{E}(z')}dz' \right]^2 \frac{c}{H_0 \mathcal{E}(z)}dz \, .
\end{equation} 
The frequency resolution $\Delta f$ can be chosen as the inverse of the observation time (typically of order one year).
The condition of unresolvability is fulfilled from a certain redshift $z$, i.e. $\mathcal{N}(f,\Delta f,z)\ge 1$, when each bin is always filled by one or more signals.
These signals cannot be distinguished, because for that we would need to improve our frequency resolution; we therefore say that they are considered unresolvable \footnote{A more thorough definition of the resolvability, that takes into account the difference in amplitude of the signals, can be the subject of a future work.}.
One can invert $\mathcal{N}(f,\Delta f,z)=\mathcal{N}_0$ with respect to $z$, obtaining a function $\overline{z}=\overline{z}(f,\Delta f,\mathcal{N}_0)$.
Signals with redshifts greater than this produce an overlap greater than $\mathcal{N}_0$.
Using this redshift function as lower limit of the integral in Equation (\ref{eq:omegadeflong}), one considers only the contribution to the background of those signals that produce an overlap greater than $\mathcal{N}_0$.

Therefore, the spectral function of a background with more than $\mathcal{N}_0$ signals per frequency bin $[f,f+\Delta f]$ is
\begin{align}
\label{eq:omegagen}
&\Omega(f,\Delta f,\mathcal{N}_0)\nonumber\\
&=\frac{1}{\rho_c c^2}\int_{\overline{z}(f,\Delta f,\mathcal{N}_0)}^\infty [1+z]^{-1}\frac{dE_e}{dt_e}\bigg|_f \frac{dn}{d\ln f_e}\bigg|_f \frac{dt_e^L}{dz}dz \, .
\end{align}
For simplicity, we assume that the background is unresolvable if the number of overlapping signals in a frequency bin is $\ge$ 1 (other criteria are discussed in Section V.D of \cite{Rosado2011}).
Then, the spectral function of the unresolvable part of the background is given by
\begin{equation}
\label{eq:omunres}
\Omega_{\text{unresolvable}}(f)=\Omega (f,\Delta f,1) \, .
\end{equation} 
On the other hand, the spectral function of the resolvable part is
\begin{equation}
\label{eq:omres}
\Omega_{\text{resolvable}}(f)=\Omega_{\text{total}} (f)-\Omega_{\text{unresolvable}}(f) \, .
\end{equation} 
Here, the spectral function of the total background (which coincides with Equation (\ref{eq:omegadeflong})), is
\begin{equation}
\label{eq:omtotal}
\Omega_{\text{total}}(f)=\Omega (f,\Delta f,0) \, ,
\end{equation} 
where the value of $\Delta f$ becomes irrelevant.

In Section \ref{sec:ovfunspecfun} we prove that Equation (\ref{eq:defovfungen}) leads to the definition of the overlap function given in Equation (41) of \cite{Rosado2011}, if one assumes all systems start emitting with the same initial frequency and the evolution of each system is short compared to cosmic time scales.

The definition of resolvability given in this section is the one that was introduced in \cite{Rosado2011} and that will be used throughout the entire paper.
More thorough definitions can be a topic for future work, for example taking into account the ability of the data analysis method to distinguish individual signals from the instrumental noise, or to even distinguish unresolvable signals with different amplitudes or directions of arrival.
The advantage of our definition is that the resolvability becomes an intrinsic property of the background, i.e., independent of the characteristics of the detector (such as its sensitivity) and of the data analysis method.
The only parameter related to the observation that affects the resolvability is the frequency resolution.
However, the observation time $T_{\text{obs}}$ of any realistic experiment is of the order of a year or a few years; the best frequency resolution achievable (calculated as $\Delta f=T_{\text{obs}}^{-1}$), can thus be considered equal for all possible detectors.

\subsection{Detectability of the background}
\label{sec:detectability}
In practice, the instrumental noise of a detector cannot be modeled with perfect accuracy; if an unresolvable background is present in the data of a detector, it is therefore indistinguishable from instrumental noise (unless one can construct a \textit{null stream}, which is a very advantageous feature of ET \cite{RegimbauEtAl2012}).
The usual technique to cope with this issue is the cross-correlation of the data of two detectors (see, for example, Section 7.8.3 of \cite{Maggiore2008}).

If a background (characterized by a spectral function $\Omega(f)$) is present in the data of two interferometers, one can cross-correlate the outputs of both, that span an interval of time $T_{\text{obs}}$.
Doing this, the resulting \textit{signal-to-noise ratio} (Equation (7.241) of \cite{Maggiore2008}), or SNR, is given by
\begin{equation}
\label{eq:snrdef}
\text{SNR}=\frac{3H_0^2}{4\pi^2}\left[ 2T_{\text{obs}} \int_0^\infty df \frac{\Gamma^2(f) \Omega^2(f)}{f^6 S_{n,1}(f)S_{n,2}(f)}\right]^{1/2} \, .
\end{equation}
Here, $S_{n,1}(f)$ and $S_{n,2}(f)$ are the noise spectral densities of the detectors 1 and 2, respectively, and $\Gamma(f)$ is the non-normalized overlap reduction function (Equation (7.226) of \cite{Maggiore2008}), defined by
\begin{align}
\label{eq:nonnormorf}
\Gamma(f)=&\frac{1}{8\pi^2}\int_0^{2\pi}d\phi \int_0^\pi \sin(\theta)d\theta \cos \left( \frac{2\pi f \vec{u} (\theta,\phi)\cdot \vec{\Delta x}}{c} \right)\nonumber\\
&\times \int_0^{2\pi} d\psi \sum_{p=+,\times} F_1^p(\theta,\phi,\psi)F_2^p(\theta,\phi,\psi) \, .
\end{align}
In this definition, $\vec{\Delta x}=\vec{x_2}-\vec{x_1}$, where $\vec{x_d}$ is the position of the detector $d$, and $\vec{u}(\theta,\phi)$ is a unit vector pointing to the direction defined by the angles $\theta$ and $\phi$.
The function $F_d^p(\theta,\phi,\psi)$ (for the detector $d=$1 or 2, and for the polarization $p=+$ or $\times$) is the antenna pattern function, evaluated at the direction $(\theta,\phi)$, for a wave with a polarization angle $\psi$.
The antenna pattern functions can be found in Section II.B of \cite{RegimbauEtAl2012} for ET, and in Section 2.1 of \cite{Schutz2011} for aLIGO.
Notice that Equation (\ref{eq:snrdef}) is equivalent to Equation (3.75) of \cite{AllenRomano1999}; however, the normalized overlap reduction function $\gamma(f)$ defined in \cite{Maggiore2008} and in \cite{AllenRomano1999} are only equivalent for the case of an L-shaped detector.
A detailed study on the overlap reduction function can be found in \cite{FinnEtAl2009}.

Following \cite{Maggiore2008}, the non-normalized overlap reduction function can be written as
\begin{equation}
\Gamma(f)=F_{1,2}\gamma(f) \, .
\end{equation}
For two colocated and coaligned detectors, $\gamma(f)=1$ for all frequencies.
For the correlation between two interferometric V-shaped detectors like ET, one obtains $F_{1,2}=3/10$, whereas for L-shaped detectors like aLIGO, $F_{1,2}=2/5$.
In Section \ref{sec:results} the SNR is calculated for two interferometers of ET, and for two aLIGO interferometers (one at Livingston and one at Hanford), using the full overlap reduction function (Equation (\ref{eq:nonnormorf})) and assuming an observation time of one year.
For simplicity, the SNR for two interferometers of BBO or DECIGO will be calculated by using $F_{1,2}=3/10$ and $\gamma(f)=1$, and an observation time of one year.

\begin{figure}
\includegraphics{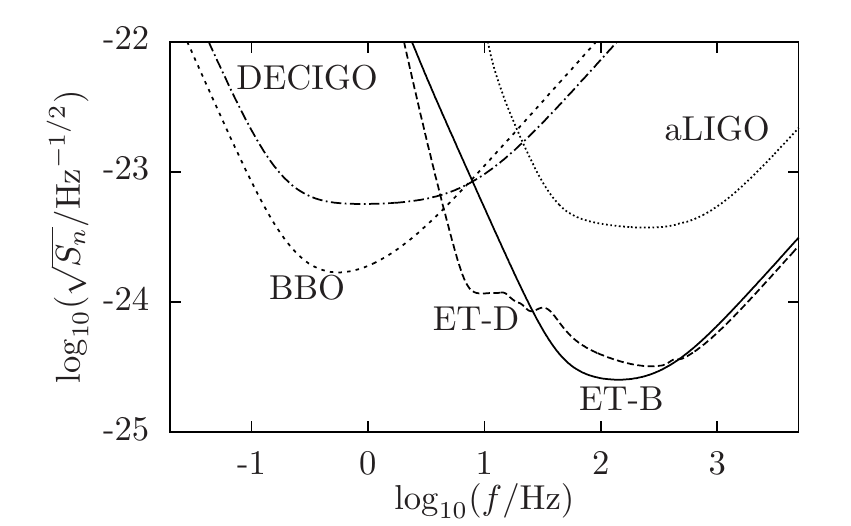}
\caption{Spectral strain sensitivity of aLIGO, two possible configurations of ET (named ETB and ETD), DECIGO and BBO. The sources of the curves are given in the text.}
\label{fig:strain}
\end{figure}

Figure \ref{fig:strain} shows the spectral strain sensitivity $\sqrt{S_n(f)}$ of aLIGO \cite{[{Taken from }] LIGO2010e}, two possible configurations of ET \footnote{The spectral strain sensitivities of ETB and ETD were kindly provided by Tania Regimbau in a private communication.}, DECIGO \cite{[{The spectral strain sensitivities of DECIGO and BBO are obtained by using the fitting formulas given in }]NishizawaEtAl2012} and BBO \cite{[{The spectral strain sensitivities of DECIGO and BBO are obtained by using the fitting formulas given in }]NishizawaEtAl2012}.
We consider two pairs of detectors:
the two aLIGO detectors, at Hanford and Livingston, and two V-shaped ET detectors sharing one arm of the triangle.

A background is said to be \textit{detectable} if it produces SNR larger than a certain threshold value.
Be aware that a background can be resolvable without being detectable; it would consist of signals that are separated in a frequency-time plot but would be buried in noise (for example, instrumental noise, or confusion noise produced by another background).

\section{\label{sec:bg2} Characterization of the background: a detailed derivation of $\Omega(f,\Delta f,\mathcal{N}_0)$ for an evolving population of systems}

\subsection{Formation rate of systems}
\label{sec:signalrate}
The comoving density rate of systems formed $\dot{n}(z)$ (or, simply, \textit{rate}), is defined such that $\dot{n}(z)dz$ is the number of systems formed per unit emitted interval of time, $dt_e$, per unit comoving volume, $dV_c$, between redshifts $z$ and $z+dz$.

Sometimes it is convenient to write the rate as a function of time, instead of redshift.
We define a function, $\mathcal{T}(z)$, that gives the interval of time elapsed between the formation of the first systems (at redshift $z_{\text{max}}$), and the formation of the systems at redshift $z$.
This function can be derived, for our cosmological model, using the formulas given in Section II.13 of \cite{Peebles1993},
\begin{equation}
\label{eq:timez}
\mathcal{T}(z)=\frac{2}{3H_0 \sqrt{\Omega_\Lambda}}\left[\text{asinh}\left( \sqrt{\frac{\Omega_\Lambda}{\Omega_m}}[1+z]^{-3/2}\right) -\xi \right].
\end{equation}
Here, we have introduced the constant
\begin{equation}
\xi =\text{asinh}\left( \sqrt{\frac{\Omega_\Lambda}{\Omega_m}}[1+z_{\text{max}}]^{-3/2}\right),
\end{equation}
that imposes a time offset between the Big Bang and the formation of the first systems \footnote{By imposing $\xi=0$, the function $\mathcal{T}(z)$ gives the age of the universe at the instant when the waves of redshift $z$ were emitted, as in Equation (13.20) of \cite{Peebles1993}.}.
One can invert $\mathcal{T}(z)=\Delta t$ with respect to the redshift and obtain another useful formula,
\begin{equation}
\label{eq:redt}
\mathcal{Z}(\Delta t)=\left[ \sqrt{\frac{\Omega_m}{\Omega_\Lambda}} \text{sinh}\left( \frac{3H_0 \sqrt{\Omega_\Lambda}\Delta t}{2}+\xi \right) \right]^{-2/3}-1 \, .
\end{equation}
This gives the redshift observed in a signal that was emitted an interval of time $\Delta t$ after the formation of the first systems.
Using Equation (\ref{eq:redt}), one can write the rate as a function of time, $\dot{n}(\mathcal{Z}(t))$.

\subsection{Time evolution of the ensemble}
\label{sec:evolution2}
We now explain how to calculate the term $[dn/d\ln f_e]\big|_f$ in Equation (\ref{eq:omegadeflong}), that is the number of systems per unit comoving volume per unit logarithmic frequency interval emitting around an observed frequency $f$.

Suppose a gravitational wave of redshift $z$ was emitted by a system an interval of time $\mathcal{T}(z)$ after the formation of the first systems (recall the definition of the function $\mathcal{T}(z)$, in Section \ref{sec:signalrate}).
At the instant of emission, the system had already evolved during a certain interval of time $t_e$ (smaller than $\mathcal{T}(z)$).
The system was thus formed an interval of time $\mathcal{T}(z)-t_e$ after the formation of the first systems.
At the instant of formation, the system was emitting waves that have now a redshift $\mathcal{Z}(\mathcal{T}(z)-t_e)$.
The formation rate at that instant was $\dot{n}(\mathcal{Z}(\mathcal{T}(z)-t_e))$.
Then, during an infinitesimal interval of time $dt_e$, the number of systems formed per unit comoving volume that emit waves of redshift $z$ is $\dot{n}(\mathcal{Z}(\mathcal{T}(z)-t_e))dt_e$.

Assume a probability density function $p_e(f_e,t_e)$, such that $p_e(f_e,t_e)df_e$ is the probability of a system to emit between $f_e$ and $f_e+df_e$ after a time evolution $t_e$.
Then, the number of systems formed during $dt_e$ per unit comoving volume that emit waves of redshift $z$ in the frequency interval $[f_e,f_e+df_e]$ is $p_e(f_e,t_e)\dot{n}(\mathcal{Z}(\mathcal{T}(z)-t_e))dt_edf_e$.
The total number of systems per unit comoving volume per unit frequency interval emitting with redshift $z$ and frequency $f_e$ is
\begin{equation}
\label{eq:dndfedef}
\frac{dn}{df_e}=\int_{0}^{\mathcal{T}(z)} p_e(f_e, t_e)\dot{n}(\mathcal{Z}(\mathcal{T}(z)-t_e))dt_e \, .
\end{equation}
We now show how to calculate the probability density function $p_e(f_e,t_e)$.
For this purpose, we follow a similar approach as \cite{KnispelAllen2008,Knispel2011}, although we do not use any distribution of systems in the galaxy, but rather assume that systems are homogeneously distributed in the universe.

Additionally, suppose we know the initial probability density function $p_{\text{ini}}(f_{\text{ini}})$, such that $p_{\text{ini}}(f_{\text{ini}})df_{\text{ini}}$ is the probability of a system to emit between $f_{\text{ini}}$ and $f_{\text{ini}}+df_{\text{ini}}$ at the instant of formation.
Now we make the following assumption: a formed system never stops emitting gravitational waves.
This means that all systems that were initially emitting in the frequency range $[f_{\text{ini}},f_{\text{ini}}+df_{\text{ini}}]$ are now emitting in $[f_e,f_e+df_e]$.
Then, in order to conserve the number of systems,
\begin{equation}
\label{eq:pefete}
p_e(f_e,t_e)df_e=p_{\text{ini}}(f_{\text{ini}})df_{\text{ini}}
\end{equation} 
must be fulfilled.

The radiation we observe now from a system, at frequency $f$, was emitted in the past at frequency $f_e$; that system was formed an interval of time $t_e$ before emitting at $f_e$.
The frequency at which the system was emitting at its formation is given by $f_{\text{ini}}=f_{\text{ini}}(f_e,t_e)$.
Using this function we can rewrite Equation (\ref{eq:pefete}),
\begin{equation}
\label{eq:pefetedef}
p_e(f_e,t_e)=p_{\text{ini}}(f_{\text{ini}}(f_e,t_e))\frac{\partial f_{\text{ini}}}{\partial f_e}(f_e,t_e) \, .
\end{equation} 
Equations (\ref{eq:dndfedef}) and (\ref{eq:pefetedef}) can be combined into
\begin{align}
\label{eq:dndfe2}
\frac{dn}{df_e}=&\int_{0}^{\mathcal{T}(z)} dt_e p_{\text{ini}}(f_{\text{ini}}(f_e,t_e))\nonumber\\
&\times\frac{\partial f_{\text{ini}}}{\partial f_e}(f_e,t_e)\dot{n}(\mathcal{Z}(\mathcal{T}(z)-t_e)) \, .
\end{align}
Finally, we can rewrite Equation (\ref{eq:dndfe2}) in terms of observed frequencies per logarithmic frequency interval to obtain
\begin{align}
\label{eq:dndlnfef}
\frac{dn}{d\ln f_e}\bigg|_f=&\left[f_e\frac{dn}{d f_e}\right]\bigg|_f \nonumber\\
=&f[1+z] \int_{0}^{\mathcal{T}(z)} dt_e p_{\text{ini}}(f_{\text{ini}}(f[1+z],t_e))\nonumber\\
&\times \frac{\partial f_{\text{ini}}}{\partial f_e}(f[1+z],t_e)\dot{n}(\mathcal{Z}(\mathcal{T}(z)-t_e)) \, .
\end{align}
In Section \ref{sec:ovfunspecfun}, we derive general formulas for the spectral function and for the overlap function, that depend on the shape of the function $p_{\text{ini}}(f_{\text{ini}})$;
then, we highlight a special case where all systems emit at the same frequency at the instant of formation.

\subsection{Spectral function and overlap function}
\label{sec:ovfunspecfun}

\subsubsection{General case}
The spectral function of a background, with more than $\mathcal{N}_0$ signals per frequency bin $\Delta f$, produced by an ensemble that follows an initial frequency distribution $p_{\text{ini}}(f_{\text{ini}})$, is given by
\begin{align}
\label{eq:omgendist}
\Omega(f,\Delta f,\mathcal{N}_0)=&\frac{f}{\rho_c c^2}\int_{\overline{z}(f,\Delta f,\mathcal{N}_0)}^{z_{\text{max}}}dz \frac{dE_e}{dt_e}\bigg|_f \frac{dt_e^L}{dz}\nonumber\\ &\times \int_{0}^{\mathcal{T}(z)} dt_e p_{\text{ini}}(f_{\text{ini}}(f[1+z], t_e))\nonumber\\
&\times \frac{\partial f_{\text{ini}}}{\partial f_e}(f[1+z],t_e)\dot{n}(\mathcal{Z}(\mathcal{T}(z)-t_e)) \, .
\end{align}
This is obtained by replacing Equation (\ref{eq:dndlnfef}) in (\ref{eq:omegagen}); the upper limit of the integral has been replaced by $z_{\text{max}}$, since we assume that no systems were formed at larger redshifts.

The overlap function of a background produced by an ensemble that follows an initial frequency distribution $p_{\text{ini}}(f_{\text{ini}})$ is given by
\begin{align}
\label{eq:ovfungeneral}
\mathcal{N}(f,\Delta f,z)=&\int_0^z dz'\int_f^{f+\Delta f} df'[1+z']\frac{dV_c}{dz'}\nonumber\\
&\times \int_{0}^{\mathcal{T}(z')}dt_ep_{\text{ini}}(f_{\text{ini}}(f'[1+z'],t_e))\nonumber\\
&\times \frac{\partial f_{\text{ini}}}{\partial f_e}(f'[1+z'],t_e)\dot{n}(\mathcal{Z}(\mathcal{T}(z')-t_e)) \, .
\end{align}
This is obtained by replacing $[dn/df_e]|_f$ (from Equation (\ref{eq:dndlnfef})) in (\ref{eq:defovfungen}).
The frequency resolution is typically much smaller than the range of frequencies of interest for rotating neutron stars.
Equation (\ref{eq:ovfungeneral}) can thus be simplified with the approximation
\begin{align}
\label{eq:ovfungenapp}
\mathcal{N}(f,\Delta f,z)\approx &\int_0^z dz' \Delta f [1+z']\frac{dV_c}{dz'}\nonumber\\
&\times \int_{0}^{\mathcal{T}(z')}dt_e p_{\text{ini}}(f_{\text{ini}}(f[1+z'],t_e))\nonumber\\
&\times \frac{\partial f_{\text{ini}}}{\partial f_e}(f[1+z'],t_e)\dot{n}(\mathcal{Z}(\mathcal{T}(z')-t_e)) \, ,
\end{align}
which is accurate as long as $\Delta f\ll f$.

\subsubsection{For a fixed initial frequency}
\label{sec:fixedfreq}
From all possible initial frequency distributions $p_{\text{ini}}(f_{\text{ini}})$, we now study a particular case with the form
\begin{equation}
\label{eq:pinidelta}
p_{\text{ini}}(f_{\text{ini}})=\delta(f_{\text{ini}}-f_{\text{fix}}) \, .
\end{equation}
Using this distribution, one assumes that all systems start emitting at a fixed initial frequency $f_{\text{fix}}$.
Suppose we have a function $f_{\text{ini}}(f_e,t_e)$ (that was introduced in Section \ref{sec:evolution2}, and will be derived in Section \ref{sec:evolution} for the case of a rotating neutron star).
Using one of the properties of the Dirac delta function, one can rewrite Equation (\ref{eq:pinidelta}) as
\begin{align}
\label{eq:pinidelta2}
p_{\text{ini}}(f_{\text{ini}}(f_e,t_e))=\frac{\partial t_e}{\partial f_{\text{ini}}}(f_e,\tau_e(f_{\text{fix}},f_e))\delta(t_e-\tau_e(f_{\text{fix}},f_e)).
\end{align}
Here, $\tau_e(f_{\text{fix}},f_e)$ is the interval of time that a system spends emitting between $f_{\text{fix}}$ and $f_e$ such that
\begin{equation}
f_{\text{ini}}(f_e,\tau_e(f_{\text{fix}},f_e))=f_{\text{fix}}
\end{equation}
is fulfilled.
The transformation of the Dirac delta function performed is valid as long as $\partial f_{\text{ini}}/\partial t_e$ is non-zero for all values of $f_e$ and $t_e$.
This condition holds for rotating neutron stars \footnote{One can prove it by partially differentiating Equation (\ref{eq:finidef}) with respect to $t_e$.}.
By replacing Equation (\ref{eq:pinidelta2}) in (\ref{eq:dndfe2}), one obtains
\begin{align}
\frac{dn}{df_e}=&  \frac{\partial t_e}{\partial f_{\text{ini}}} (f_e,\tau_e(f_{\text{fix}},f_e))\frac{\partial f_{\text{ini}}}{\partial f_e}(f_e,\tau_e(f_{\text{fix}},f_e))\nonumber\\
&\times \dot{n}(\mathcal{Z}(\mathcal{T}(z)-\tau_e(f_{\text{fix}},f_e))) \Theta(z,f_e) \, ,
\end{align}
where we have introduced the function
\begin{equation}
\label{eq:thetadef}
\Theta (z,f_e)=\theta(\mathcal{T}(z)-\tau_e(f_{\text{fix}},f_e))\theta(\tau_e(f_{\text{fix}},f_e)-0).
\end{equation}
Here, $\theta(x-y)$ is the Heaviside step function; it is equal to one for $x>y$, and equal to zero for $x<y$.
For rotating neutron stars, $\partial t_e/\partial f_e$ is not a function of $t_e$.
Therefore, without loss of generality, we write $\partial t_e/\partial f_e(f_e)$.
Then,
\begin{equation}
\label{eq:dndfe}
\frac{dn}{df_e}=\frac{\partial t_e}{\partial f_e}(f_e)\dot{n}(\mathcal{Z}(\mathcal{T}(z)-\tau_e(f_{\text{fix}},f_e))) \Theta(z,f_e) \, .
\end{equation}
We now substitute this result in the formulas of the overlap function and the spectral function.

Inserting Equation (\ref{eq:dndfe}) in (\ref{eq:defovfungen}), we obtain
\begin{align}
\label{eq:ovfunapp2}
&\mathcal{N}(f,\Delta f,z)=\int_0^z dz'\int_{f}^{f+\Delta f} df'[1+z']\frac{dt_e}{d f_e}(f'[1+z'])\nonumber\\
&\times \frac{dV_c}{dz'}\dot{n}(\mathcal{Z}(\mathcal{T}(z')-\tau_e(f_{\text{fix}},f'[1+z']))) \Theta(z',f'[1+z']) \, ,
\end{align}
which is the overlap function of a population of systems that start emitting at the same initial frequency $f_{\text{fix}}$.
Similarly, inserting Equation (\ref{eq:dndfe}) in (\ref{eq:omegagen}), we obtain the spectral function
\begin{align}
\label{eq:omegaffix}
\Omega(f,\Delta f,\mathcal{N}_0)=\frac{1}{\rho_c c^2} \int_{\overline{z}(f,\Delta f,\mathcal{N}_0)}^{z_{\text{max}}} [1+z]^{-1}  \frac{dE_e}{d \ln f_e}\bigg|_f \nonumber\\
\times \dot{n}(\mathcal{Z}(\mathcal{T}(z)-\tau_e(f_{\text{fix}},f[1+z]))) \Theta(z,f[1+z]) \frac{dt_e^L}{dz}dz \, .
\end{align}
Performing the same approximation as in Equation (\ref{eq:ovfungenapp}), the overlap function can be simplified as
\begin{align}
\label{eq:ovfunappfix}
&\mathcal{N}(f,\Delta f,z)\approx \Delta f\int_0^z dz'[1+z']\frac{dV_c}{dz'}\frac{dt_e}{d f_e}(f[1+z'])\nonumber\\
&\times \dot{n}(\mathcal{Z}(\mathcal{T}(z')-\tau_e(f_{\text{fix}},f[1+z']))) \Theta(z',f[1+z']) \, ,
\end{align}
which is accurate for $\Delta f\ll f$.

\subsubsection{For short-lived systems with fixed initial frequency}
We now prove that if systems are assumed to evolve rapidly compared to cosmic time scales, then one obtains the definition of $\mathcal{N}(f,\Delta f,z)$ and $\Omega(f,\Delta f,\mathcal{N}_0)$ given in Equations (41) and (44) of \cite{Rosado2011}, respectively.
Under this assumption, $\tau_e(f_{\text{fix}},f_e)$, which is the time a system has evolved since formation, is much smaller than $\mathcal{T}(z)$, and therefore,
\begin{equation}
\label{eq:shortsignals}
\dot{n}(\mathcal{Z}(\mathcal{T}(z)-\tau_e(f_{\text{fix}},f_e)))\approx \dot{n}(\mathcal{Z}(\mathcal{T}(z)))=\dot{n}(z) \, .
\end{equation}
Using this, we can rewrite Equation (\ref{eq:ovfunapp2}) to obtain
\begin{align}
\label{eq:ovfunapp3}
\mathcal{N}(f,\Delta f,z)\approx &\int_0^z dz' \dot{n}(z')\frac{dV_c}{dz'} \nonumber\\
&\times \int_{f[1+z']}^{[f+\Delta f][1+z']} \frac{dt_e}{d f_e} (f_e) \Theta(z',f_e)df_e\nonumber\\
=&\int_{z_{\text{low}}(f)}^z \tau_e(f,\Delta f,z') \dot{n}(z')\frac{dV_c}{dz'} dz' \, .
\end{align}
Here, the function $\tau_e(f,\Delta f,z)$ gives the interval of time that a system, whose radiation is now observed with redshift $z$, spends emitting between observed frequencies $f$ and $f+\Delta f$.
Alternatively, inserting Equation (\ref{eq:shortsignals}) in (\ref{eq:omegaffix}), the spectral function becomes
\begin{align}
\label{eq:omclass}
\Omega(f)&\approx \frac{1}{\rho_c c^2} \int_{\overline{z}(f,\Delta f,\mathcal{N}_0)}^{z_{\text{upp}}(f)} [1+z]^{-1}  \frac{dE_e}{d\ln f_e}\bigg|_f \dot{n}(z) \frac{dt_e^L}{dz}dz \, .
\end{align}
In Equations (\ref{eq:ovfunapp3}) and (\ref{eq:omclass}), the functions $z_{\text{low}}(f)$ and $z_{\text{upp}}(f)$ ensure that the integration is performed only where $\Theta(z,f[1+z])$ is non-zero.

Let us examine the spectral function of the total background, i.e., the one obtained by imposing no restriction ($\mathcal{N}_0=0$) in the number of signals per frequency bin.
By inverting $\mathcal{N}(f,\Delta f,z)=0$ (Equation (\ref{eq:ovfunapp3})) with respect to the redshift, one obtains $\overline{z}(f,\Delta f,0)=z_{\text{low}}(f)$.
Replacing this in Equation (\ref{eq:omclass}), the canonical formula for the spectral function (used for example in \cite{Phinney2001,Rosado2011,[{}][{. See corrected version in arXiv: astro-ph/1101.2762v3.}] Regimbau2011}) is recovered,
\begin{align}
\label{eq:omclass2}
\Omega(f)&\approx \frac{1}{\rho_c c^2} \int_{z_{\text{low}}(f)}^{z_{\text{upp}}(f)} [1+z]^{-1}  \frac{dE_e}{d\ln f_e}\bigg|_f \dot{n}(z) \frac{dt_e^L}{dz}dz \, .
\end{align}
By carefully studying the limits of this integral, one realizes that the redshift functions $z_{\text{low}}(f)$ and $z_{\text{upp}}(f)$ are not exactly the same as the ones defined in Equations (37) and (38) of \cite{Rosado2011}, or in Equations (10) and (9) of \cite{[{}][{. See corrected version in arXiv: astro-ph/1101.2762v3.}] Regimbau2011}.
The difference, however, occurs only at the low-frequency part of the spectrum, at which the time scales needed for the systems to evolve are comparable to cosmic ones \footnote{In this regard, one should read Section \ref{sec:minmaxfreq}; the plots in this section, for instance the ones in Figure (\ref{fig:zfplotmd}), can be qualitatively compared with Figures (2) and (4) of \cite{Rosado2011}.}.

\section{\label{sec:models} Model for the ensemble of rotating neutron stars}
\subsection{Neutron star model}
\label{sec:nsmodel}
A neutron star is modeled as a rigid rotating ellipsoid of mass $m$.
Its semiaxes with respect to the coordinate axes $x$, $y$, and $z$ have lengths $a$, $b$, and $c$, respectively.
The rotation occurs around the $z$-axis at an angular velocity $\omega$, which slowly decreases in time.
Assuming a uniform density, the moment of inertia about the $z$-axis is
\begin{equation}
\label{eq:defmomin}
I_3=I=\frac{m}{5}[a^2+b^2] \, .
\end{equation} 
The \textit{ellipticity} is defined by
\begin{equation}
\label{eq:ellipticitydef}
\epsilon=[I_1-I_2]/I_3 \, ,
\end{equation}
where $I_1$ and $I_2$ are the moments of inertia about the $x$ and $y$ axes, respectively.
The average ellipticity of the ensemble of neutron stars is very uncertain; reasonable values for $\epsilon$ can range from $10^{-8}$ to $10^{-4}$.
For such values, $\epsilon \approx [b-a]/a$, and, replacing it in (\ref{eq:defmomin}), we can very accurately approximate
\begin{equation}
\label{eq:defmomin2}
I\approx \frac{2\,m\,a^2}{5} \, .
\end{equation} 
We assume $m=1.4\,M_\odot$ (where $M_\odot$ is the solar mass) and $a=12\,$km \cite{Lattimer2010,Truemper2011}, obtaining a moment of inertia of $I\approx 1.6\times 10^{38}\,$kg\,m$^2$ (see Sections 3.1.1 and 3.1.2 of \cite{LorimerKramer2005} and references therein for a discussion about these values).
Other mechanisms of gravitational wave emission, like dynamical bar-mode \cite{Brown2000} or r-mode \cite{OwenEtAl1998} instabilities, are not taken into account in this work.

A neutron star behaves like a rotating dipole magnet.
The value of the magnetic field at the magnetic pole is $B$, where it forms an angle $\alpha$ (assumed, for simplicity, of $\alpha=\pi/2$) with the direction of the rotation.
We do not consider any magnetic field decay \cite{[{See Section 3.5 of }][{, and references therein.}]Lorimer2011}; $B$ is the same during the entire life of the star.

With this toy model, one can infer the average value of $B$ by looking at the rotating period and its time derivative (the so-called $P-\dot{P}$ diagram) of a pulsar catalogue \cite{[{ATNF catalogue: }] ManchesterEtAl2005}.
Without taking into account recycled pulsars, a reasonable average value for pulsars is $B=10^{8}\,$T.
For magnetars, larger values (of $B\approx 10^{10}\,$T) can be reached.

The maximum gravitational wave frequency at which a rotating neutron star can emit is estimated by
\begin{equation}
\label{eq:fmaxescape}
f_{\text{max}}^{\text{escape}}=\frac{1}{\pi}\sqrt{\frac{Gm}{a^3}}\approx 3.3\,\text{kHz} \, .
\end{equation} 
Above this frequency, the material at the equator would have enough velocity to escape the gravitational potential, since the latter becomes lower than the centrifugal potential \cite{Lattimer2010}.
A gravitational wave frequency of 3.3\,kHz corresponds to a rotation period of 0.6\,ms, which is roughly the half of the fastest rotation period known in a pulsar \cite{BackerEtAl1982}.
A more realistic estimate \cite{[{See Section 7.4 of }]ShapiroTeukolsky1983} of the maximum frequency is
\begin{equation}
\label{eq:fmaxroche}
f_{\text{max}}^{\text{Roche}}=\left[\frac{2}{3} \right]^{3/2} f_{\text{max}}^{\text{escape}} \approx 1.8\,\text{kHz} \, .
\end{equation} 
This frequency takes into account the deformation of the equatorial radius because of the rotation.
From now on, we make the choice
\begin{equation}
\label{eq:fmaxchoice}
f_{\text{max}}=f_{\text{max}}^{\text{escape}} \, ,
\end{equation}
which leads to the most optimistic results, regarding the detection of the background.
The conclusions of this paper would be unaffected, however, by choosing the alternative maximum frequency in Equation (\ref{eq:fmaxroche}).

\subsection{Formation rate of neutron stars}
The amount of mass converted into stars per unit emitted interval of time per unit comoving volume between redshifts $z$ and $z+dz$ is given by $\dot{\rho}(z)dz$, where $\dot{\rho}(z)$ is the \textit{star formation rate}.
The models for the star formation rate \cite{MadauEtAl1998,PorcianiMadau2001,2dFGRS2001,StrolgerEtAl2004,HopkinsBeacom2006,NagamineEtAl2006,FardalEtAl2007,WilkinsEtAl2008} usually present a similar shape: $\dot{\rho}(z)$ increases from its local value (at $z=0$) until $z\approx 1$ or 2, and then decays, reaching negligible values for redshifts larger than 5 or 6.
For this reason, the range of redshifts considered in the calculations is $[0,z_{\text{max}}]$, with
\begin{equation}
z_{\text{max}}=5 \, .
\end{equation}
All calculations shown in the plots of Section \ref{sec:results} are obtained by assuming the star formation rate given in Section 5.4 of \cite{2dFGRS2001},
\begin{equation}
\label{eq:sfr}
\dot{\rho}(z)=h\frac{a+bz}{1+\left[\frac{z}{c}\right]^d}\,M_\odot \text{yr}^{-1}\text{Mpc}^{-3} \, ,
\end{equation}
with the parameters found in Table I of \cite{HopkinsBeacom2006}, namely $(a,b,c,d)=(0.0170,0.13,3.3,5.3)$, and $h=H_0/[100\,$km\,s$^{-1}$Mpc$^{-1}]=0.742$.
In Section \ref{sec:discussionrate} we comment on the fact that the specific choice of star formation rate does not affect the spectral function significantly.
Furthermore, the results would not be affected by the use of a constant rate.

Only a fraction $\lambda$ of all stars formed become neutron stars, so the rate is
\begin{equation}
\label{eq:signalrate}
\dot{n}(z)=\lambda \dot{\rho}(z) \, .
\end{equation} 
The fraction of stars formed with masses between $m$ and $m+dm$ is $\Phi(m)dm$, where $\Phi(m)$ is the \textit{initial mass function}.
We assume a Salpeter initial mass function \cite{Salpeter1955},
\begin{equation}
\Phi(m)=Am^{-2.35} \, ,
\end{equation} 
where the value of the normalization constant $A$ turns out to be unnecessary, as we now see.
The fraction of stellar mass converted into neutron stars is
\begin{equation}
\lambda=\frac{\int_{8\,M_\odot}^{20\,M_\odot}\Phi(m)dm}{\int_{0.1\,M_\odot}^{100\,M_\odot}m\Phi(m)dm}=5\times 10^{-3} \,M_\odot^{-1} \, .
\end{equation} 
The denominator is the average mass of a star (considering all stars with masses between $0.1\,M_\odot$ and $100\,M_\odot$), and the numerator is the fraction of stars that can be progenitors of neutron stars (namely stars with masses between $8\,M_\odot$ and $20\,M_\odot$).
The value of $\lambda$ tells us that, for each $10^3\,M_\odot$ of gas converted into stellar mass, 5 neutron stars are produced.
We assume that this fraction is the same at all redshifts.

\subsection{Energy evolution}
\label{sec:enerspec}
The rotational energy of a system is given by
\begin{equation}
\label{eq:eesys}
E_{e,\text{rot}}=\frac{1}{2} I \omega_e^2 \, .
\end{equation} 
For convenience, we write the angular velocity $\omega_e$ in terms of the frequency of the emitted gravitational waves, $f_e$, which fulfills
\begin{equation}
\label{eq:angfreq}
\omega_e=\pi f_e \, .
\end{equation} 
Differentiating (\ref{eq:eesys}) with respect to the time, and using (\ref{eq:angfreq}), one obtains
\begin{equation}
\label{eq:deesysdterot}
\frac{dE_{e,\text{rot}}}{dt_e}=\pi^2 I f_e \frac{df_e}{dt_e} \, .
\end{equation} 
In the following we show that $df_e/dt_e$ is negative, thus, $dE_{e,\text{rot}}/dt_e$ is also negative; the system loses rotational energy with the time.
This energy loss is due to the emission of both electromagnetic and gravitational radiation.

Suppose that the system lost energy only via the magnetic dipole emission.
Such a system is studied in \cite{Deutsch1955}.
Rewriting Equation (15) of that paper with our notation, 
\begin{equation}
\label{eq:deesysdtemd}
\frac{dE_{e,\text{md}}}{dt_e}=-\frac{2\pi^5 a^6 B^2 \sin^2(\alpha)}{3c^3 \mu_0}f_e^4=-\frac{\pi^2 I}{2\delta_\text{md}}f_e^4 \, ,
\end{equation} 
where the index `md' stands for \textit{magnetic dipole} and
\begin{equation}
\delta_\text{md}=\frac{3\mu_0 c^3 I}{4 \pi^3 B^2 a^6 \sin^2(\alpha)} \, .
\end{equation} 
Here, $\mu_0$ is the magnetic permeability of the vacuum.
We have used that $B=\mu_0 H$, where $H$ is called $R_1$ in \cite{Deutsch1955}.
The magnetic field is often expressed in Gaussian units \cite{RegimbauMandic2008,MarassiEtAl2011b}.
For clarity, recall that  $B_{\text{Gauss}}=\sqrt{\frac{4\pi}{\mu_0}}B$.
Equation (\ref{eq:deesysdtemd}) gives the rate at which the system loses energy by emitting electromagnetic waves.

Suppose now that the system emitted only gravitational radiation.
This system is studied, for example, in Section 4.2.1 of \cite{Maggiore2008}.
Rewriting Equation (4.227) of \cite{Maggiore2008} with our notation,
\begin{equation}
\label{eq:deesysdtegr}
\frac{dE_{e,\text{gr}}}{dt_e}=-\frac{32\pi^6G\epsilon^2 I^2}{5c^5}f_e^6=-\frac{\pi^2 I}{4 \delta_\text{gr}}f_e^6 \, ,
\end{equation} 
where the index `gr' stands for \textit{gravitational radiation}, and
\begin{equation}
\delta_\text{gr}=\frac{5c^5}{128 \pi^4 G\epsilon^2I}.
\end{equation} 
Equation (\ref{eq:deesysdtegr}) gives the rate at which the system loses energy by emitting gravitational waves.

The system we study loses energy via both magnetic dipole emission and gravitational radiation.
Thus, the total loss of energy (which is a loss in rotational energy) fulfills
\begin{equation}
\label{eq:deesysdte}
\frac{dE_{e,\text{rot}}}{dt_e}=\frac{dE_{e,\text{md}}}{dt_e}+\frac{dE_{e,\text{gr}}}{dt_e} \, .
\end{equation} 
Rewriting Equation (\ref{eq:deesysdte}) in terms of Equations (\ref{eq:deesysdterot}), (\ref{eq:deesysdtemd}), and (\ref{eq:deesysdtegr}), we obtain
\begin{equation}
\label{eq:dfedte}
\frac{df_e}{dt_e}=-\frac{1}{2\delta_\text{md}}f_e^3-\frac{1}{4 \delta_\text{gr}}f_e^5 \, ,
\end{equation} 
where we see that the frequency of the rotation (written in terms of gravitational wave frequencies) decreases with time.
The energy lost by the rotating star is the energy gained by the gravitational waves, so
\begin{equation}
\label{deedte}
\frac{dE_e}{dt_e}=- \frac{dE_{e,\text{gr}}}{dt_e}=\frac{\pi^2 I}{4 \delta_\text{gr}}f_e^6
\end{equation}
is the amount of gravitational wave energy produced by a system per unit time.
The amount of gravitational wave energy produced by one system contained in an infinitesimal logarithmic interval of frequency $d\ln f_e$ is
\begin{equation}
\label{eq:deedfeabs}
\frac{dE_e}{d\ln f_e}=f_e \frac{dE_e}{df_e} =f_e \bigg| \frac{dE_e}{dt_e} \left[\frac{df_e}{dt_e}\right]^{-1}\bigg| \, ,
\end{equation}
where we have used the chain rule.
Using Equations (\ref{eq:dfedte}) and (\ref{deedte}),
\begin{equation}
\label{eq:deedfe}
\frac{dE_e}{d\ln f_e}=\frac{\pi^2 I f_e^4}{f_C^2+f_e^2} \, ,
\end{equation}
where we have introduced the \textit{cut frequency},
\begin{equation}
\label{eq:fcdef}
f_C=\sqrt{\frac{2\delta_\text{gr}}{\delta_\text{md}}} \, .
\end{equation} 
At this frequency, both terms on the right side of Equation (\ref{eq:dfedte}) become equal; this is the frequency at which both mechanisms of energy loss ``cut'' each other.
The absolute value in Equation (\ref{eq:deedfeabs}) is used because $dE_e$ must be a positive quantity; it represents the amount of gravitational wave energy within a logarithmic frequency interval, regardless of whether the energy of the system increases or decreases with the frequency.

The three main expressions of this section are in Equations (\ref{eq:dfedte}), (\ref{deedte}), and (\ref{eq:deedfe}).
They can be rewritten, using (\ref{eq:redshiftf}), in terms of observed frequencies.
The resulting formulas are:
\begin{equation}
\frac{df_e}{dt_e}\bigg|_f =-\frac{1}{2\delta_\text{md}}f^3[1+z]^3-\frac{1}{4\delta_\text{gr}}f^5[1+z]^5 \, ,
\end{equation}
\begin{equation}
\label{eq:deedtef}
\frac{dE_e}{dt_e}\bigg|_f=\frac{\pi^2 I}{4\delta_\text{gr}}f^6[1+z]^6 \, ,
\end{equation}
and
\begin{equation}
\label{eq:enerspecobs}
\frac{dE_e}{d\ln f_e}\bigg|_f=\frac{\pi^2 I f^4[1+z]^4}{f_C^2+f^2[1+z]^2} \, ,
\end{equation}
respectively.

We can distinguish two frequency intervals: one where the magnetic dipole emission dominates (let us call it the \textit{md-range}) and one where the gravitational radiation dominates (the \textit{gr-range}).
The frequency at which both mechanisms are equally dominant is $f_C$.
For simplicity, some of the next calculations are performed in the two frequency intervals separately.
The energy spectrum can be approximated by
\begin{equation}
\label{eq:deemddfe}
\frac{dE_e}{d\ln f_e}\bigg|_f\approx \frac{\pi^2 I}{f_C^2} f^4[1+z]^4
\end{equation}
in the md-range, and by
\begin{equation}
\label{eq:deegrdfe}
\frac{dE_e}{d\ln f_e}\bigg|_f\approx \pi^2 I f^2[1+z]^2
\end{equation}
in the gr-range.

The \textit{braking index} $n_b$ is defined by \cite{LorimerKramer2005}
\begin{equation}
\frac{df_e}{dt_e}= -K (f_e)^{n_b} \, ,
\end{equation} 
where $K$ is a constant.
Equation (\ref{eq:dfedte}) shows that the braking index is equal to 3 in the md-range and equal to 5 in the gr-range for all neutron stars.
Observational measurements of the braking index, however, obtain very different values.
For example, in Table 4 of \cite{FaucherKaspi2006} $n_b$ is smaller than 3 for some known pulsars.
Alternatively, the braking index measured in other pulsars can be orders of magnitude larger than 3, or even negative \cite{JohnstonGalloway1999}.
The results of this paper would differ considerably if one used models with different braking indices.
The consideration of such other models is out of the purposes of this work.

\subsection{Frequency evolution}
\label{sec:evolution}
We now calculate the lapse of time $\tau_e(f_{e,1},f_{e,2})$ that a system spends emitting within a certain frequency interval $[f_{e,1},f_{e,2}]$.
This is achieved by integrating Equation (\ref{eq:dfedte}), which leads to the analytical formula
\begin{align}
\label{eq:taue}
&\tau_e(f_{e,1},f_{e,2})=\nonumber\\
&\delta_\text{md}\left[f_{e,2}^{-2}-f_{e,1}^{-2}+f_C^{-2}\ln \left(\frac{f_{e,2}^2 [f_C^2+f_{e,1}^2]}{f_{e,1}^2[f_C^2+f_{e,2}^2]} \right) \right].
\end{align} 
A system that starts emitting at an initial frequency $f_{\text{ini}}$, needs an interval of time $\tau_e(f_{\text{ini}},f_e)$ to reach the frequency $f_e$.

It is useful to obtain a function $f_{\text{ini}}=f_{\text{ini}}(f_e,\Delta t_e)$, that gives the frequency at which a system, that now emits at $f_e$, was emitting an interval of time $\Delta t_e$ before.
One cannot invert Equation (\ref{eq:taue}) with respect to $f_{\text{ini}}$ analytically.
For this reason, it is more convenient to approximate $\tau_e(f_{\text{ini}},f_e)$ by
\begin{align}
\label{eq:taueapp}
&\tau_e(f_{\text{ini}},f_e)=\nonumber\\
&\left\{ \begin{array}{lc}
\delta_\text{gr} [f_e^{-4}-f_{\text{ini}}^{-4}] & f_C\le f_e <f_{\text{ini}}\\
\delta_\text{gr} [f_C^{-4}-f_{\text{ini}}^{-4}]+\delta_\text{md} [f_e^{-2}-f_C^{-2}] & f_e<f_C< f_{\text{ini}}\\
\delta_\text{md}[f_e^{-2}-f_{\text{ini}}^{-2}] & f_e<f_{\text{ini}}\le f_C
\end{array}
\right. .
\end{align} 
With this approximation, one can analytically invert $\tau(f_{\text{ini}},f_e)=\Delta t_e$ with respect to $f_{\text{ini}}$, obtaining
\begin{widetext}
\begin{equation}
\label{eq:finidef}
f_{\text{ini}}(f_e,\Delta t_e)=
\left\{ \begin{array}{lccc}
\left[f_e^{-4}-\frac{\Delta t_e}{\delta_\text{gr}} \right]^{-1/4} & f_C\le f_e & \& & f_e<\left[\frac{\Delta t_e}{\delta_\text{gr}} \right]^{-1/4}\\
\left[f_C^{-2} [ 2f_e^{-2}-f_C^{-2} ] -\frac{\Delta t_e}{\delta_\text{gr}} \right]^{-1/4} \qquad & \left[f_C^{-2}+\frac{\Delta t_e}{\delta_\text{md}} \right]^{-1/2}<f_e<f_C & \& &  f_e<\left[\frac{1}{2}f_C^{-2}+\frac{\Delta t_e}{\delta_\text{md}} \right]^{-1/2} \\
\left[f_e^{-2}-\frac{\Delta t_e}{\delta_\text{md}} \right]^{-1/2} & f_e\le \left[f_C^{-2}+\frac{\Delta t_e}{\delta_\text{md}} \right]^{-1/2} & &
\end{array}
\right. .
\end{equation} 
\end{widetext}
In this equation, the conditions $f_e<\left[\frac{1}{2}f_C^{-2}+\frac{\Delta t_e}{\delta_\text{md}} \right]^{-1/2}$ and $f_e<\left[\frac{\Delta t_e}{\delta_\text{gr}} \right]^{-1/4}$ are introduced to avoid unphysical values for $f_{\text{ini}}$.

\subsection{Initial frequency distribution}
\label{sec:initialspin}
We consider three simple initial frequency distributions $p_{\text{ini}}(f_{\text{ini}})$ in the calculations.

The first one was already introduced in Equation (\ref{eq:pinidelta}).
Let us call it \textit{Distribution 0}.

\textit{Distribution 1} is obtained from the log-normal initial period distribution given in \cite{ArzoumanianEtAl2002},
\begin{equation}
\label{eq:dist1}
p_P(P_{\text{ini}})=\frac{1}{\sqrt{2\pi}\sigma P_{\text{ini}}} \exp\left(-\frac{[\ln(P_{\text{ini}}/\text{s})-\mu]^2}{2\sigma^2}\right),
\end{equation} 
where $\mu=\ln(0.005)$, $\sigma=0.3/\log_{10}(e)\approx 0.69$.
This distribution hence assumes that the average initial spin period is of 5\,ms.
The initial period, $P_{\text{ini}}$, is related to the initial spin frequency $f_{\text{ini}}$ (in terms of gravitational wave frequencies) by
\begin{equation}
P_{\text{ini}}=\frac{2}{f_{\text{ini}}}.
\end{equation} 
Therefore,
\begin{equation}
\label{eq:pinifini}
p_{\text{ini}}(f_{\text{ini}})=\frac{2}{f_{\text{ini}}^2}p_P\left(\frac{2}{f_{\text{ini}}}\right)
\end{equation} 
is the corresponding probability density function of the initial frequency.

Lastly, \textit{Distribution 2} is obtained from the normal initial period distribution given in \cite{FaucherKaspi2006},
\begin{equation}
\label{eq:dist2}
p_P(P_0)=\frac{1}{\sqrt{2\pi}\sigma} \exp \left(-\frac{[P_0-\mu]^2}{2\sigma^2}\right),
\end{equation} 
with $\mu=$300\,ms and $\sigma=$150\,ms.
Similar distributions to this one are used in \cite{RegimbauFreitas2000} and in \cite{PopovEtAl2010}.
To obtain the corresponding probability density function of the initial frequency, one can again use Equation (\ref{eq:pinifini}).

Initial frequency distributions like Distributions 0 and 1 are more favorable for the detection of the background than Distribution 2.
Some studies do predict large initial frequencies for the population of magnetars \cite{ThompsonDuncan1993}; on the other hand, an initial period of 5\,ms (like the average of Distribution 1) or shorter may be considered too small to properly describe the ensemble of known pulsars \cite{RegimbauFreitas2000,PernaEtAl2008}.
Another possible distribution, used in \cite{GonthierEtAl2011}, could be a Gaussian distribution like that of Equation (\ref{eq:dist2}) with $\mu=$50\,ms and $\sigma=$50\,ms.
This distribution leads to intermediate results between those of Distributions 1 and 2.
Regarding gravitars, our current knowledge about their population statistics is so poor that any of the previous distributions is equally plausible.

\subsection{Magnetic field and ellipticity distributions}
Some of the calculations in Section \ref{sec:results} are performed using a magnetic field distribution and an ellipticity distribution.
The formulas for the overlap function (Equation (\ref{eq:ovfungenapp})) and the spectral function (Equation (\ref{eq:omgendist})) can be modified to take into account these distributions.
The overlap function becomes
\begin{align}
&\hat{\mathcal{N}}(f,\Delta f,z)=\nonumber\\
&\int_{\epsilon_{\text{min}}}^{\epsilon_{\text{max}}} d\epsilon p_{\epsilon}(\epsilon)\int_{B_{\text{min}}}^{B_{\text{max}}}dBp_B(B)\mathcal{N}(f,\Delta f,z) \, .
\end{align}
Inverting $\hat{\mathcal{N}}(f,\Delta f,z)=\mathcal{N}_0$ with respect to the redshift, one obtains a function $\hat{\overline{z}}(f,\Delta f,\mathcal{N}_0)$.
This function can be used as a lower limit of the redshift integral in (\ref{eq:omgendist}), to obtain $\Omega'(f,\Delta f,\mathcal{N}_0)$.
The spectral function is obtained by solving
\begin{align}
&\hat{\Omega}(f,\Delta f,\mathcal{N}_0)\nonumber\\
&=\int_{\epsilon_{\text{min}}}^{\epsilon_{\text{max}}}d\epsilon p_\epsilon (\epsilon) \int_{B_{\text{min}}}^{B_{\text{max}}}dB p_B(B) \Omega'(f,\Delta f,\mathcal{N}_0) \, .
\end{align}
In these two formulas, the magnetic field and ellipticity distributions are assumed to be independent.
We point out that these distributions could in fact be correlated; as an example, in Section \ref{sec:magnetars} we mention that a high magnetic field can increase the ellipticity.
We now specify the magnetic field and ellipticity distributions used.

The probability density function of the magnetic field is taken from \cite{ArzoumanianEtAl2002} (also used in \cite{Palomba2005,KnispelAllen2008}),
\begin{equation}
\label{eq:magdis}
p_B(B)=\frac{1}{\sqrt{2\pi} \sigma B} \exp\left(-\frac{[\ln(B/\text{T})-\mu]^2}{2 \sigma^2}\right) \, , 
\end{equation} 
with $\mu=\ln(10^{8.35})$ and $\sigma=0.4/\log_{10}(e)\approx 0.9$.
This means that the average magnetic field is of $10^{8.35}\,$T.
The distribution is normalized to unity between a minimum value of $B_{\text{min}}=10^{7.2}\,$T and a maximum value of $B_{\text{max}}=10^{9.8}\,$T.

The probability density function of the ellipticity is taken from \cite{Palomba2005},
\begin{equation}
\label{eq:ellipdis}
p_\epsilon(\epsilon)=A\frac{\exp\left( -\frac{\epsilon}{\tau}\right)}{\tau\left[1-\exp\left(-\frac{\epsilon_{\text{max}}}{\tau}\right)\right]} \, ,
\end{equation} 
where $\tau$ is the solution of
\begin{equation}
\overline{\epsilon}=\tau-\frac{\epsilon_{\text{max}}}{\exp\left(\frac{\epsilon_{\text{max}}}{\tau}\right)-1} \, .
\end{equation} 
The values for $\overline{\epsilon}$ and $\epsilon_{\text{max}}$ are $10^{-7}$ and $2.5\times10^{-6}$, respectively.
The normalization constant $A$ is obtained by imposing
\begin{equation}
\int_{\epsilon_{\text{min}}}^{\epsilon_{\text{max}}} p_\epsilon (\epsilon)d\epsilon=1 \, ,
\end{equation}
where the minimum ellipticity is $\epsilon_{\text{min}}\approx 0$.

\subsection{\label{sec:minmaxfreq} Minimum and maximum frequencies}
In Section \ref{sec:fixedfreq}, a formula for the spectral function is obtained (in Equation (\ref{eq:omegaffix})), assuming a fixed initial frequency; $\Omega(f,\Delta f,\mathcal{N}_0)$ contains the function $\Theta(z,f[1+z])$, defined in Equation (\ref{eq:thetadef}), that determines the redshifts and observed frequencies of the systems that can contribute to the background.
Introducing (\ref{eq:redshiftf}) in (\ref{eq:thetadef}), one gets
\begin{align}
\label{eq:thetadef2}
&\Theta(z,f[1+z]) \nonumber\\
&=\theta(\mathcal{T}(z)-\tau(f_{\text{fix}},f[1+z]))\theta(\tau(f_{\text{fix}},f[1+z])-0) \, .
\end{align}
We now study the limits that this function sets on the possible observed frequencies and redshifts of the gravitational waves, for the ensemble of rotating neutron stars.
For that, the fixed initial frequency can be replaced by the maximum frequency (in Equation (\ref{eq:fmaxchoice})), i.e. $f_{\text{fix}}=f_{\text{max}}$.

The first Heaviside step function in (\ref{eq:thetadef2}) becomes zero for a certain observed frequency $f=f_{\text{low}}(z)$.
This function gives the minimum observed frequency that a gravitational wave with redshift $z$ can have.
Using Equations (\ref{eq:timez}) and (\ref{eq:taueapp}), the condition $\mathcal{T}(z)=\tau(f_{\text{max}},f_{\text{low}}(z)[1+z])$ leads to
\begin{equation}
\label{eq:flowdef}
f_{\text{low}}(z)=
\left\{ \begin{array}{lc}
f_1(z) & f_C\le f_1(z) \\
f_2(z) & f_1(z)<f_C<f_{\text{max}} \\
f_3(z) & f_{\text{max}}\le f_C
\end{array}
\right. ,
\end{equation}
where
\begin{equation}
\label{eq:f1func}
f_1(z)=\left[\frac{\mathcal{T}(z)}{\delta_{\text{gr}}}+f_{\text{max}}^{-4} \right]^{-1/4}[1+z]^{-1} \, ,
\end{equation}
\begin{equation}
\label{eq:f2func}
f_2(z)=\left[\frac{\mathcal{T}(z)-\delta_{\text{gr}}[f_C^{-4}-f_{\text{max}}^{-4}]}{\delta_{\text{md}}}+f_C^{-2} \right]^{-1/2}[1+z]^{-1} \, ,
\end{equation}
and
\begin{equation}
\label{eq:f3func}
f_3(z)=\left[\frac{\mathcal{T}(z)}{\delta_{\text{md}}}+f_{\text{max}}^{-2} \right]^{-1/2}[1+z]^{-1} \, .
\end{equation}
One should notice that $f_{\text{low}}(z)$ is an observed frequency, unlike $f_C$ and $f_{\text{max}}$, that are emitted frequencies; the $e$-index in the two latter quantities has been omitted to ease the notation.

The second Heaviside step function in Equation (\ref{eq:thetadef2}) becomes zero when evaluated at the observed frequency $f=f_{\text{upp}}(z)$.
This function gives the maximum observed frequency that a gravitational wave with redshift $z$ can have.
Using Equation (\ref{eq:taueapp}), the condition $\tau(f_{\text{max}},f_{\text{upp}}(z)[1+z])=0$ leads to
\begin{equation}
\label{eq:fuppdef}
f_{\text{upp}}(z)=f_{\text{max}}[1+z]^{-1} \, .
\end{equation}
Again, notice that $f_{\text{upp}}$ is an observed frequency, whereas $f_{\text{max}}$ is an emitted frequency.

With the previous results, we can calculate the maximum and minimum observed frequencies possible.
The maximum observed frequency, as Equation (\ref{eq:fuppdef}) clearly shows, is achieved at redshift 0, and is precisely $f_{\text{max}}$.
On the other hand, to find the minimum observed frequency, one has to minimize Equation (\ref{eq:flowdef}).
The redshift at which $f_{\text{low}}(z)$ is minimum is the solution of
\begin{equation}
\label{eq:zsol1}
\left[\mathcal{T}(z)+\frac{\delta_{\text{md}}}{f_{\text{max}}^2} \right]^{-1}\frac{d\mathcal{T}}{dz}(z)+2[1+z]^{-1}=0 \, ,
\end{equation}
if $f_C\le f_1(z)$, of
\begin{equation}
\label{eq:zsol2}
\left[\mathcal{T}(z)+\frac{\delta_{\text{md}}^2}{4\delta_{\text{gr}}}+\frac{\delta_{\text{gr}}}{f_{\text{max}}^4} \right]^{-1}\frac{d\mathcal{T}}{dz}(z)+2[1+z]^{-1}=0 \, ,
\end{equation}
if $f_1(z)<f_C<f_{\text{max}}$, and of
\begin{equation}
\label{eq:zsol3}
\left[\mathcal{T}(z)+\frac{\delta_{\text{gr}}}{f_{\text{max}}^4} \right]^{-1}\frac{d\mathcal{T}}{dz}(z)+4[1+z]^{-1}=0 \, ,
\end{equation}
if $f_{\text{max}}\le f_C$.
As a good approximation, one can assume that $\mathcal{T}(z)\gg \delta_{\text{md}}f_{\text{max}}^{-2}$ and $\mathcal{T}(z)\gg \delta_{\text{gr}}f_{\text{max}}^{-4}$. 
Doing this, Equations (\ref{eq:zsol1}) and (\ref{eq:zsol3}) depend only on cosmological parameters, and their numerical solutions are
\begin{equation}
\label{eq:zgrdef}
z_{\text{gr}}\approx 3.39 \, ,
\end{equation}
and
\begin{equation}
\label{eq:zmddef}
z_{\text{md}}\approx 1.54 \, ,
\end{equation}
respectively.
The solution of Equation (\ref{eq:zsol2}) will depend on the values of the astrophysical parameters ($B$, $\epsilon$, et cetera), but must lie between $z_{\text{md}}$ and $z_{\text{gr}}$.
As an example, for a rotating neutron star with $B=10^8\,$T and $\epsilon=10^{-7}$, the cut frequency is $f_C>f_{\text{max}}$, so the minimum observed frequency is given by $f_3(z_{\text{md}})\approx 86\,$mHz.

\begin{figure}
\includegraphics{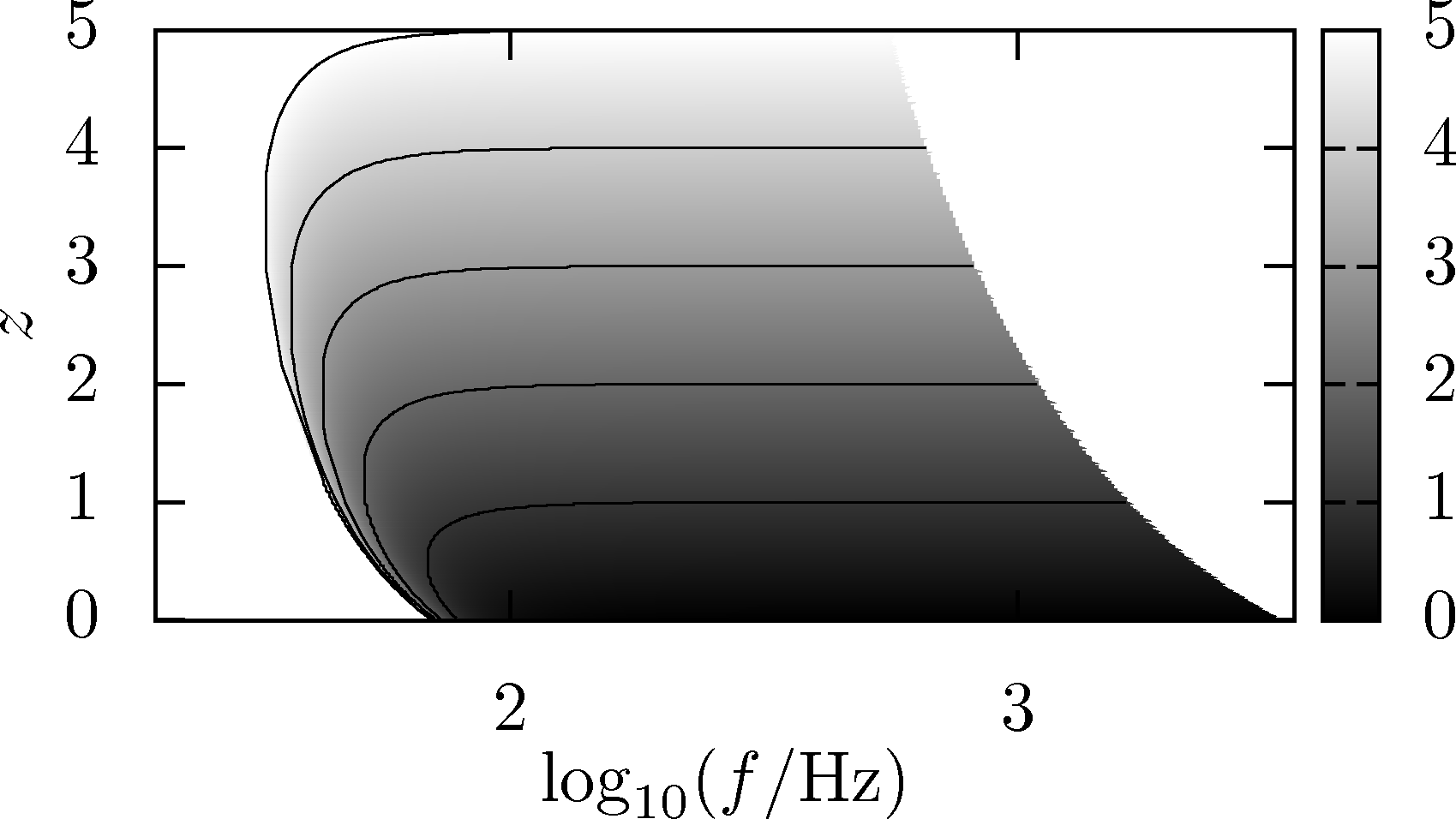}
\includegraphics{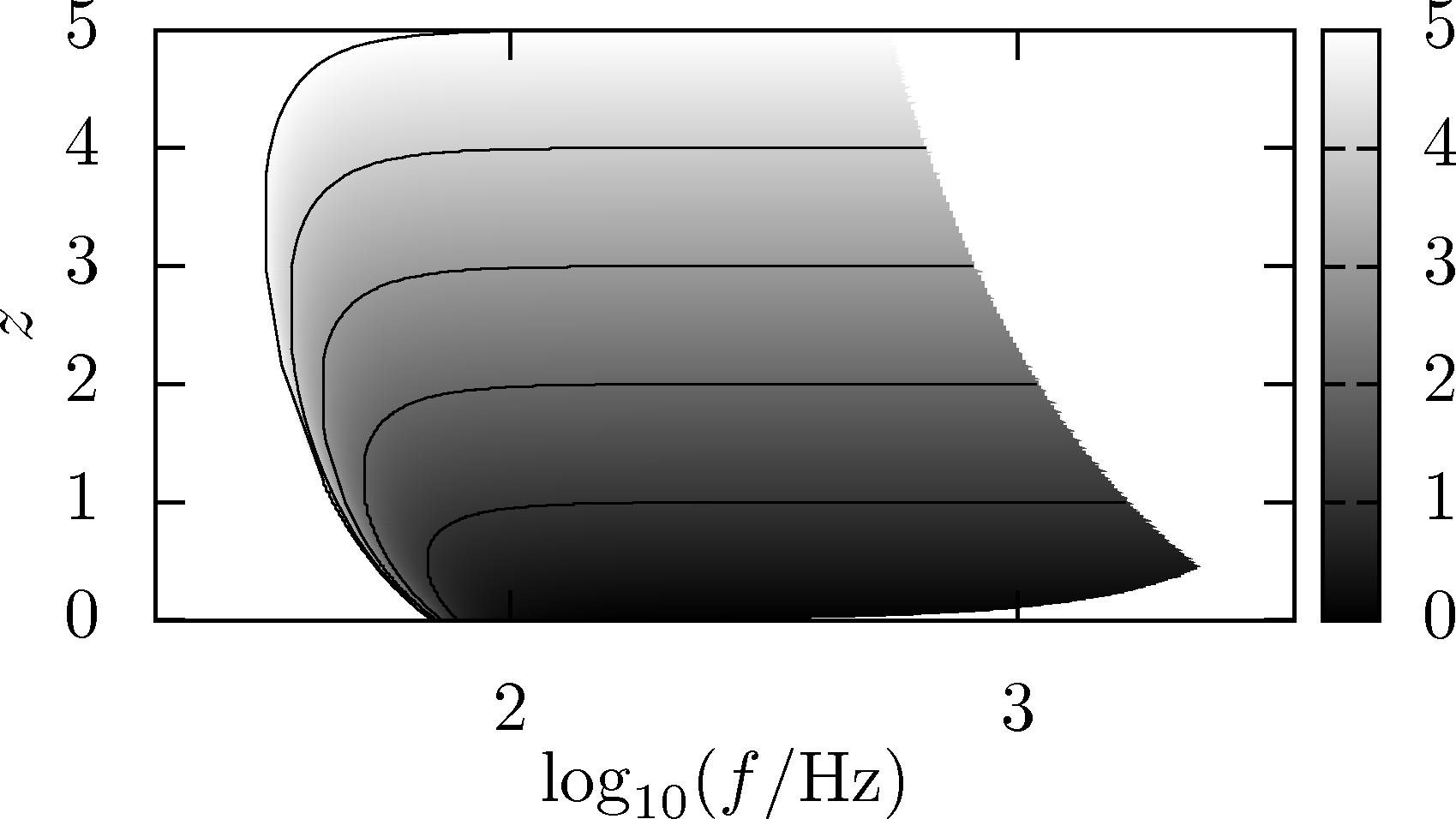}
\caption{Redshift versus observed frequency of the gravitational waves produced by the ensemble of gravitars, assuming an ellipticity of $\epsilon=10^{-7}$.
The vertical axis gives the redshift of the gravitational waves observed today (redshift at present).
Those waves were emitted by gravitars that started radiating at the redshifts given by the gray scale (redshift of formation).
The solid lines follow points of equal redshifts of formation (corresponding, from bottom to top, to redshifts 1, 2, 3, 4, and 5).
The upper plot accounts for all signals in the universe, whereas the lower plot accounts only for unresolvable signals.}
\label{fig:zfplotgr}
\end{figure}
\begin{figure}
\includegraphics{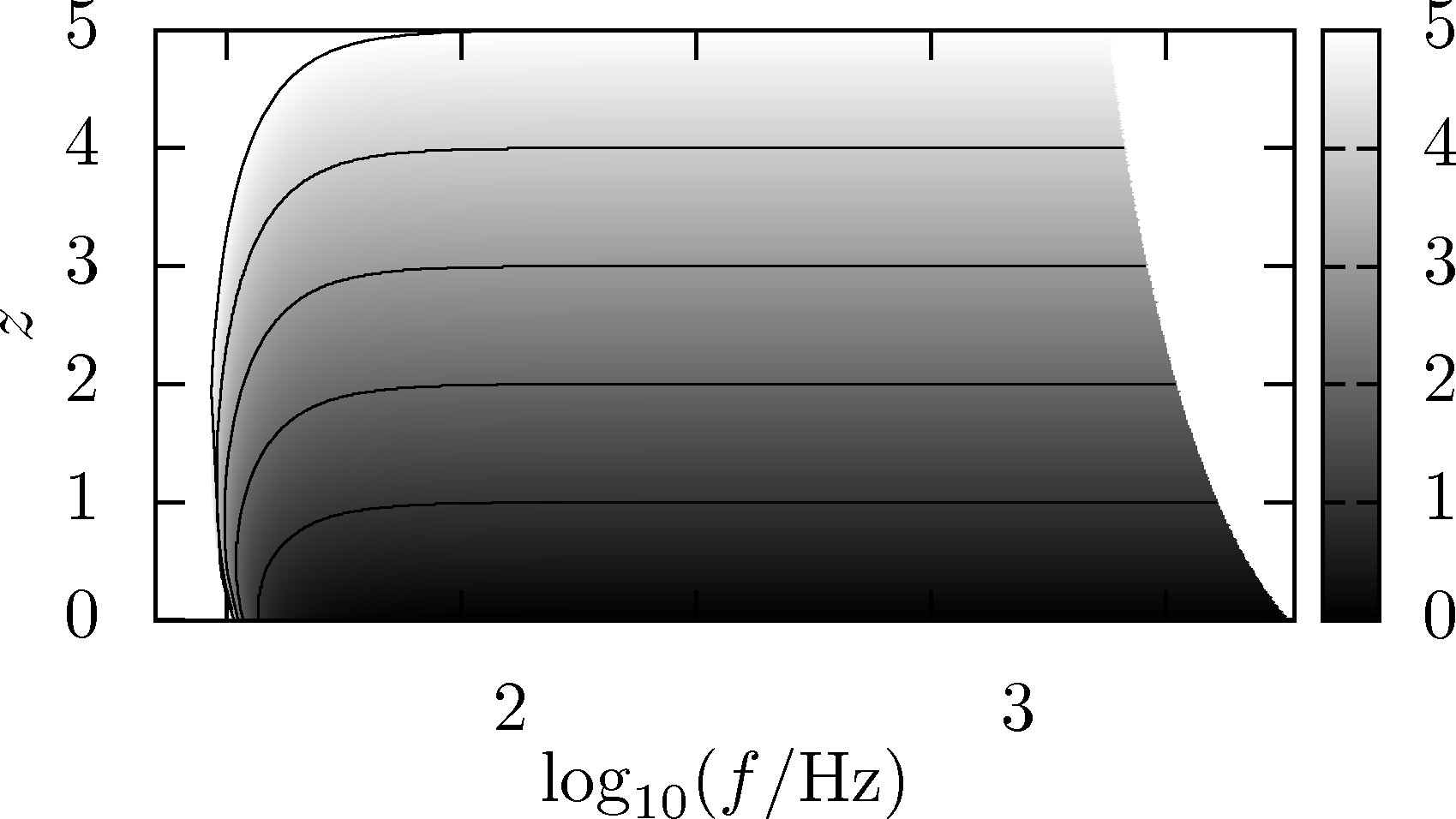}
\includegraphics{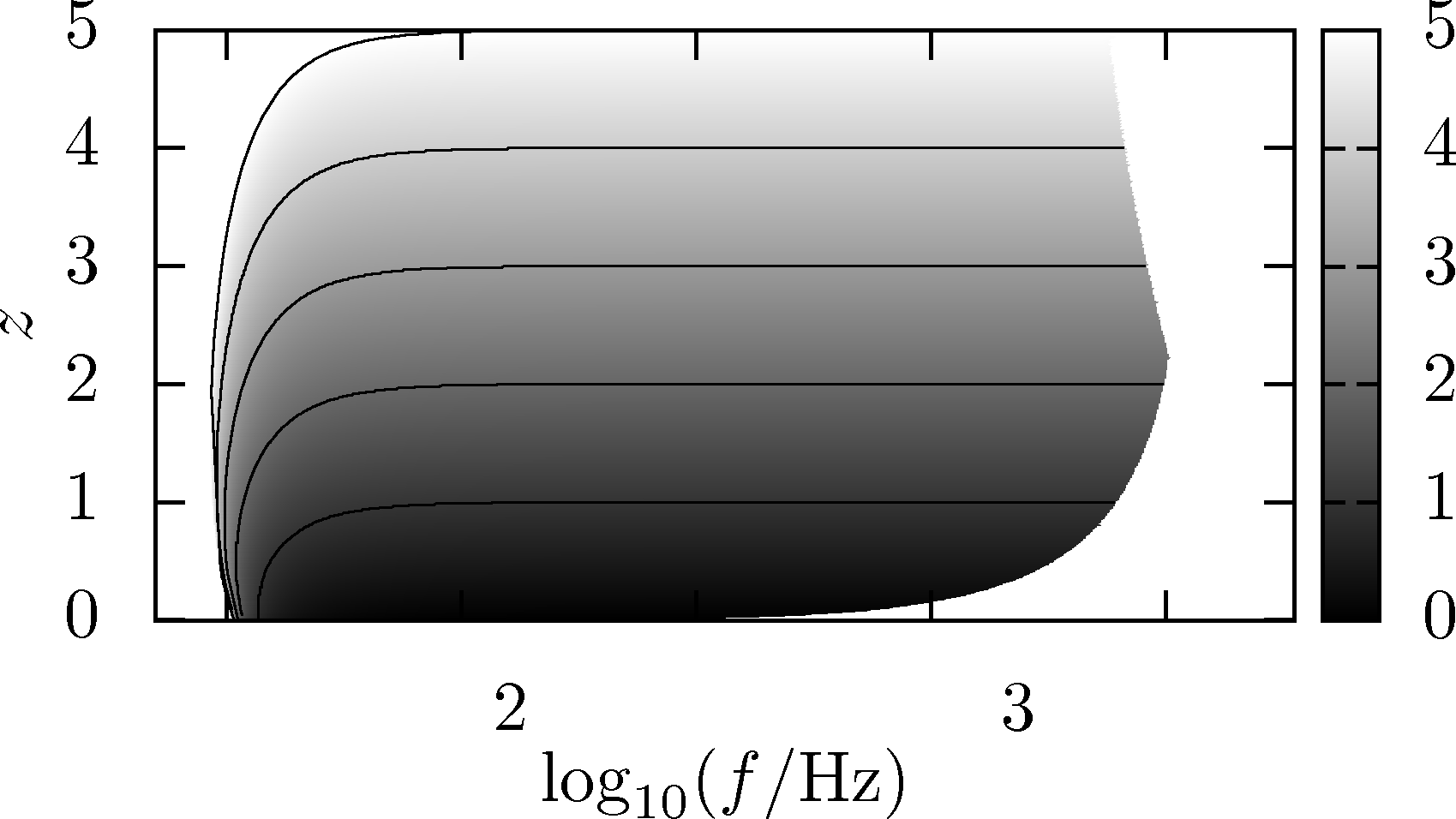}
\caption{Same as Figure \ref{fig:zfplotgr}, but for the ensemble of rotating neutron stars with magnetic field $B=10^8\,$T and ellipticity $\epsilon=10^{-7}$.}
\label{fig:zfplotmd}
\end{figure}

The upper plot in Figure \ref{fig:zfplotgr} shows the redshifts and observed frequencies that the gravitational waves from the ensemble of gravitars with ellipticity $\epsilon=10^{-7}$ can have.
The lower plot is the same, but only for the unresolvable waves (those that produce an overlap larger than $\mathcal{N}_0=1$ in a frequency bin $\Delta f=1\,\text{yr}^{-1}$).
A point in the graph tells the present redshift observed in a gravitational wave emitted by a gravitar, and the gray scale tells the redshift of the waves emitted by that same gravitar at the instant of its formation.
Figure \ref{fig:zfplotmd} is analogous to \ref{fig:zfplotgr}, but for the ensemble of rotating neutron stars with $B=10^8\,$T and $\epsilon=10^{-7}$.

\section{\label{sec:results}Results}

\subsection{Gravitars}
\label{sec:gravitars}
Let us first give a definition of gravitar: it is a rotating neutron star that emits gravitational waves at a frequency $f_e>f_C$, i.e., the dominating mechanism for the loss of rotational energy is the emission of gravitational waves (see the definition of the cut frequency in Section \ref{sec:enerspec}).

The gravitar limit is an upper limit on the gravitational wave background produced by rotating neutron stars.
For simplicity, and in order to obtain a robust upper limit, we obtain the gravitar limit under the following assumptions: all neutron stars are gravitars;
all gravitars start emitting with the same initial frequency; this frequency is infinite; and the spectrum can be extended to arbitrarily low frequencies, as if signals had an infinite amount of time to evolve.
Under these unrealistic assumptions, the energy spectrum can be approximated by Equation (\ref{eq:deegrdfe}), and the spectral function of the gravitar limit (using Equation (\ref{eq:omclass})) becomes
\begin{equation}
\label{eq:omegagr}
\Omega_{\text{GL}}(f)=\frac{\pi^2I}{\rho_c c^2H_0} f^2 \int_{0}^{z_{\text{max}}} \dot{n}(z) \mathcal{E}^{-1}(z)dz \, . 
\end{equation}
Using the star formation rate of Equation (\ref{eq:sfr}), the background yielded by such an ensemble would produce a SNR (Equation (\ref{eq:snrdef})) of $\sim$1.3 for aLIGO, after one year of observation.
We use the gravitar limit as a reference in the following plots.

We now justify that the background of rotating neutron stars cannot be larger than the gravitar limit.
The spectral function in Equation (\ref{eq:omegagr}) depends only on the rate $\dot{n}(z)$ and on the average moment of inertia $I$.
The latter is well constrained by present neutron star equations of state.
The abundance and even the existence of gravitars is unknown, but certainly not all neutron stars are gravitars, so the rate of gravitars must certainly be smaller than $\dot{n}(z)$.
Equation (\ref{eq:omegagr}) is obtained by assuming that $f_C=0$.
If the cut frequency were not zero, at frequencies lower than $f_C$ the spectral function would be proportional to $f^4$, reaching its maximum around the cut frequency.
There is hence no choice of the parameters $B$, $\epsilon$, and $\alpha$, and there is no frequency at which the spectral function can be larger than (\ref{eq:omegagr}), as long as the rate and the momentum of inertia (as well as the cosmological parameters) remain unchanged.

\begin{figure}
\includegraphics{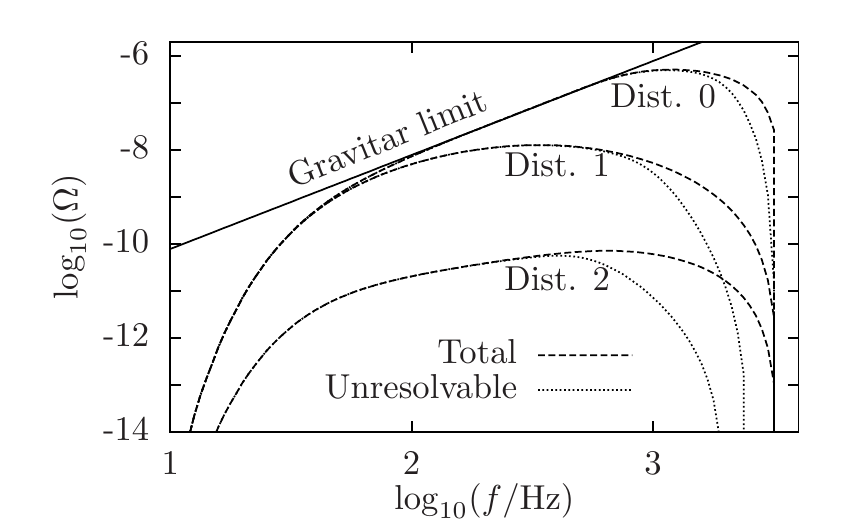}
\caption{Spectral function, versus observed frequency, of the total and unresolvable backgrounds produced under the assumption that all rotating neutron stars are gravitars, i.e. neutron stars which spin down primarily by emitting gravitational waves. The gravitar limit is also shown, as a robust upper limit of the background from rotating neutron stars. The background is calculated by using three different initial frequency distributions (see Section \ref{sec:initialspin}). The unresolvable background is calculated with $\mathcal{N}_0=1$ and $\Delta f=1\,\text{yr}^{-1}$.}
\label{fig:gravitars}
\end{figure}

If all rotating neutron stars were gravitars, the background they would produce would be different than the gravitar limit.
First, their initial frequency is finite, and second, they had a finite amount of time to evolve, so they cannot emit at arbitrarily low frequencies.
In Figure \ref{fig:gravitars}, besides the gravitar limit, we show the background that would be produced if all rotating neutron stars were gravitars, assuming the three initial frequency distributions considered in Section \ref{sec:initialspin}.
The curves for Distribution 0 are obtained by using Equation (\ref{eq:omegaffix}), whereas those for Distributions 1 and 2 are obtained by evaluating Equation (\ref{eq:omgendist}).
In all cases, the star formation rate is the one in Equation (\ref{eq:sfr}), the magnetic field is approximately zero, and the ellipticity follows the distribution given in Equation (\ref{eq:ellipdis}).
The obtained background turns out to be almost entirely unresolvable.

The SNR produced by the total background, assuming Distribution 0, is of 0.64, 6.6$\times 10^2$, and 3.5$\times 10^2$, for aLIGO, ETB, and ETD, respectively (assuming one year of observation).
With Distribution 1, these numbers are 0.56, 4.1$\times 10^2$, and 1.8$\times 10^2$.
Finally, with Distribution 2, the values of SNR are 4.0$\times 10^{-3}$, 2.0, and 0.81.
We can thus claim that aLIGO is not sensitive enough to either detect the background of rotating neutron stars, or to set upper limits on the fraction of neutron stars that are gravitars.
The SNR for BBO and DECIGO is in all cases much smaller than 1.
Obviously, not all neutron stars are gravitars; if only a certain fraction of the population of neutron stars were gravitars, the values of the spectral function in Figure \ref{fig:gravitars}, as well as the values of SNR, would be multiplied by that fraction.
Hence, if only $1\%$ of neutron stars were gravitars, they would produce a background that could be detected by ETB with SNR 6.6, assuming Distribution 0, and 4.1, assuming Distribution 1.
On the other hand, with Distribution 2, even if all neutron stars were gravitars the detection statistics of all detectors are below the detection threshold.

\subsection{\label{sec:realistic} A more realistic expectation}

\begin{figure}
\includegraphics{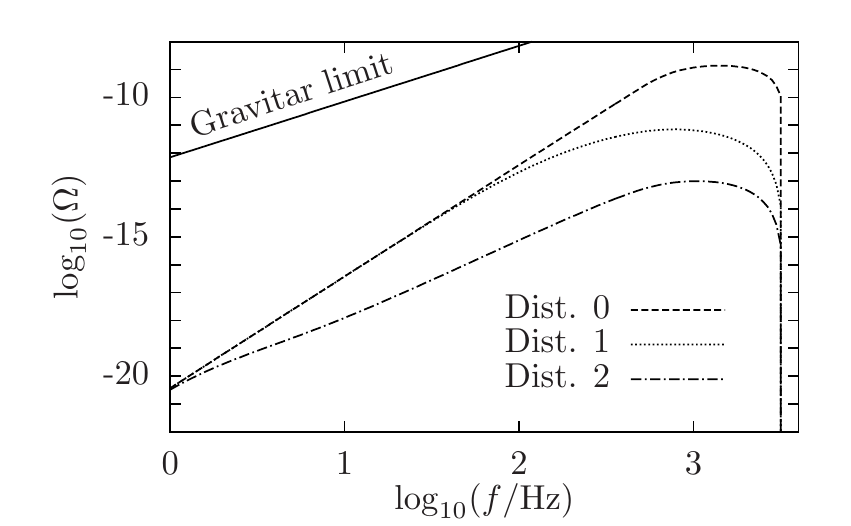}
\caption{Spectral function, versus observed frequency, of the total background produced by rotating neutron stars, assuming the magnetic field distribution of Equation (\ref{eq:magdis}) (which is taken from \cite{ArzoumanianEtAl2002}) and the ellipticity distribution of Equation (\ref{eq:ellipdis}) (from \cite{Palomba2005}). Three initial frequency distributions are used (see Section \ref{sec:initialspin}). None of the present or planned gravitational wave detectors is sensitive enough to observe such a background.}
\label{fig:rotns}
\end{figure}

In Figure \ref{fig:rotns}, we show the spectral function of the total background of rotating neutron stars, calculated by assuming the magnetic field distribution of Equation (\ref{eq:magdis}), and the ellipticity distribution of (\ref{eq:ellipdis}).
The star formation rate used is the one of Equation (\ref{eq:sfr}).
The obtained SNR for such a background is much smaller than 1, for all detectors considered.

\subsection{Upper limit for magnetars}
\label{sec:magnetars}
There are two facts that make the detection of the magnetars background difficult:
First, since they have a larger magnetic field, the electromagnetic emission dominates over the gravitational wave emission.
Second, the fraction of magnetars is believed to be of order $10\%$ of the total population of neutron stars \cite{RegimbauFreitas2006,PopovEtAl2010}.
On the other hand, large magnetic fields can deform a neutron star \cite{Cutler2002}, increasing its ellipticity.
If the ellipticity is large enough, the contribution of gravitational waves can be important.
Furthermore, the larger the magnetic field, the faster systems evolve towards lower frequencies, entering the band of highly sensitive detectors like BBO and DECIGO.
We now investigate if the background of magnetars has a good chance to be detected.

\begin{figure}
\includegraphics{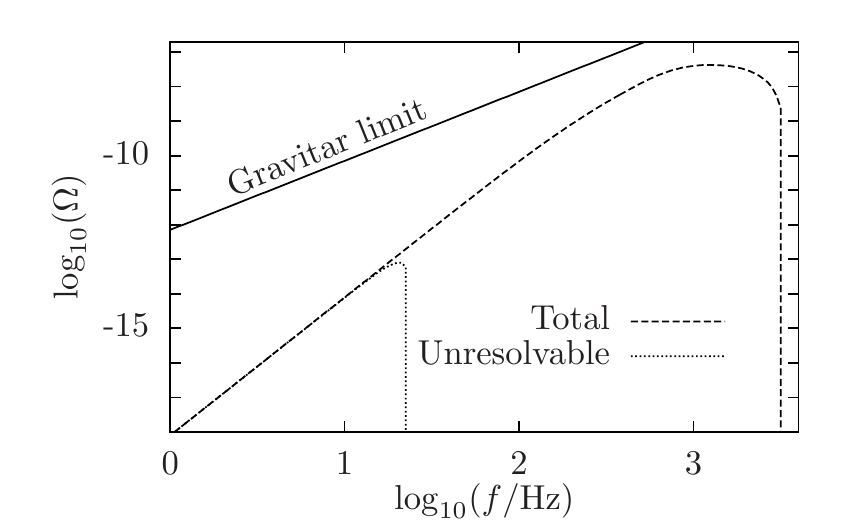}
\caption{Spectral function, versus observed frequency, of the total and unresolvable backgrounds produced by magnetars.
The rate of magnetars is assumed to be $10\%$ of the one of neutron stars, the average magnetic field is $B=10^{10}\,$T, and the average ellipticity, $\epsilon=-6.4\times 10^{-4}$. This corresponds to the TD model described in \cite{MarassiEtAl2011b}, which is the most optimistic model (regarding the detection) considered in that paper.
Other models predict levels of background several orders of magnitude lower.
The total background plotted here can thus be considered an upper limit for the background of magnetars.}
\label{fig:magnetars}
\end{figure}

In \cite{MarassiEtAl2011b}, different models for the population of magnetars are compared.
One of the models, with a dominating toroidal magnetic field, produces a gravitational wave background that can be detected by ET.
This model predicts a poloidal magnetic field of $B=10^{10}\,$T and an ellipticity $\epsilon=-6.4\times 10^{-4}$.
In Figure \ref{fig:magnetars} we show the background produced with this model, assuming that all magnetars start emitting gravitational waves of the same frequency, $f_{\text{max}}$ \footnote{Some studies predict that magnetars are formed with fast initial spins \cite{ThompsonDuncan1993}. Since we are interested in obtaining an upper limit, we assume that all magnetars start emitting at $f_{\text{max}}$.}.
Other models in \cite{MarassiEtAl2011b} predict levels of background orders of magnitude lower than the one in Figure \ref{fig:magnetars}.
The total background shown in this figure can thus be considered an optimistic upper limit for the background of magnetars.

The SNR with which the total background of Figure \ref{fig:magnetars} would be detected by ETB and ETD is of 14 and 11, respectively (for one year of observation time).
For aLIGO, as well as for BBO and DECIGO, the values of SNR are negligible.
On the other hand, the SNR of the unresolvable background is lower than $10^{-2}$ for all detectors.
This means that the magnetars background is resolvable in the band of ET.

\subsection{\label{sec:prospects} Detection prospects}

In Section \ref{sec:gravitars}, we have claimed that the background produced by gravitars could be detected even if they constituted only a $1\%$ of the neutron star population.
Nevertheless, the existence of gravitars is questionable.
We now show that similar detection claims can be achieved for certain (plausible) choices of $B$ and $\epsilon$.

\begin{figure}
\includegraphics{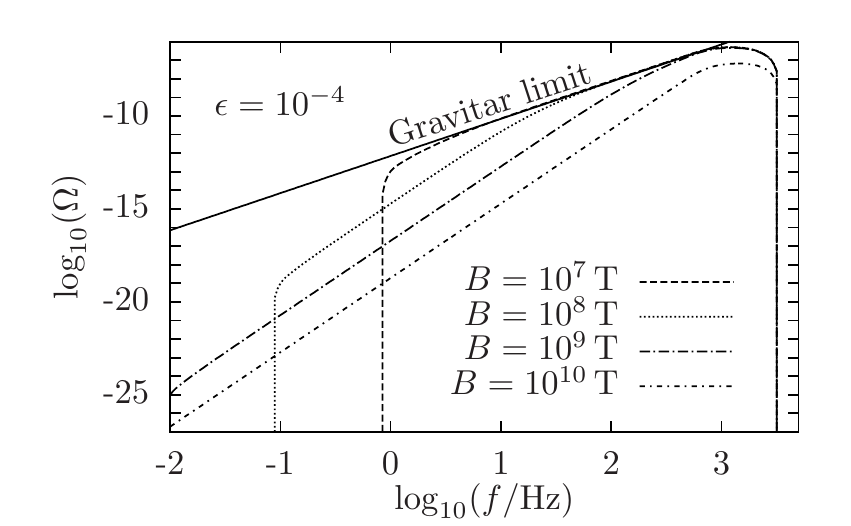}
\includegraphics{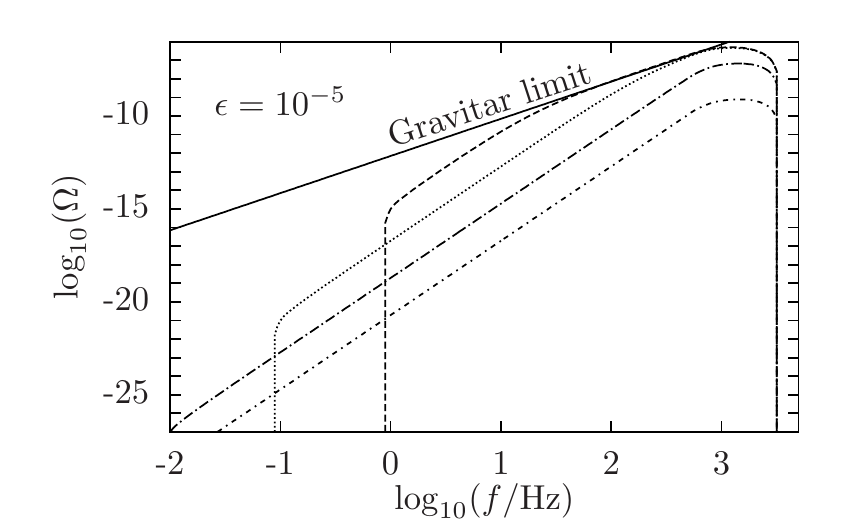}
\includegraphics{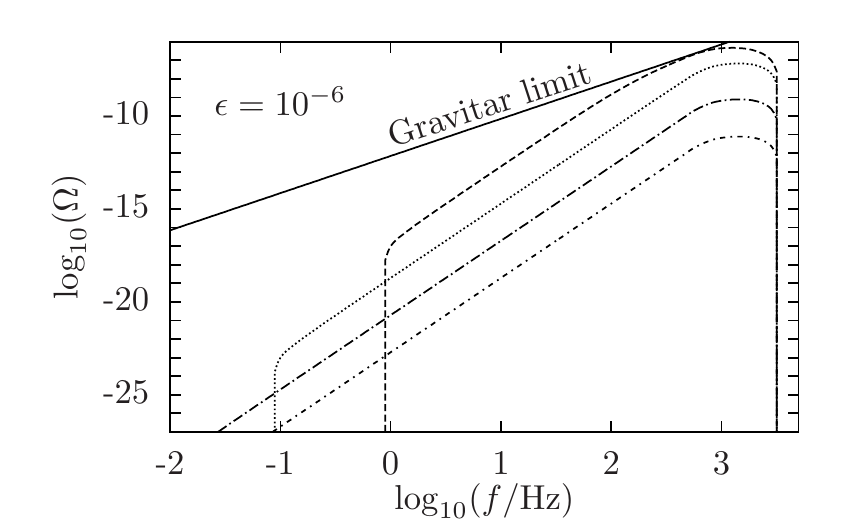}
\includegraphics{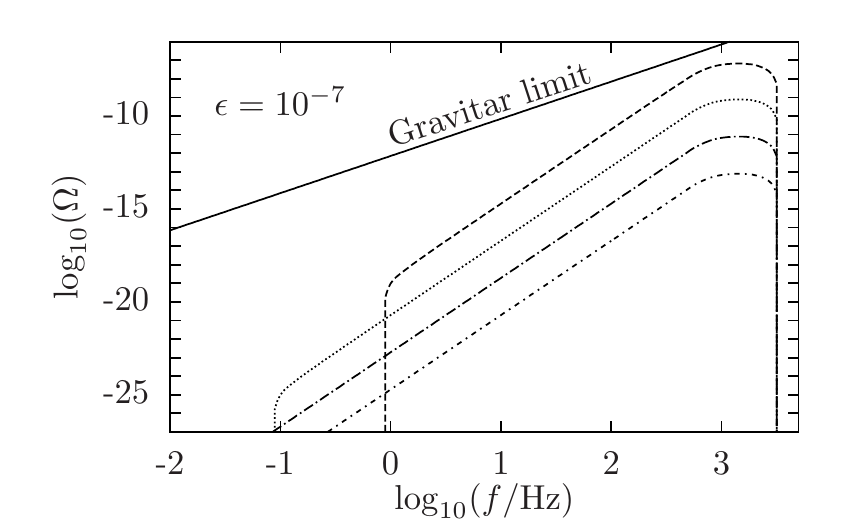}
\caption{Spectral function, versus observed frequency, of the total background produced if all rotating neutron stars had the same magnetic field and ellipticity.
Different line types correspond to different magnetic fields (as the legend of the upper plot describes).
Each plot corresponds to a certain ellipticity.}
\label{fig:rotnst}
\end{figure}

\begin{figure}
\includegraphics{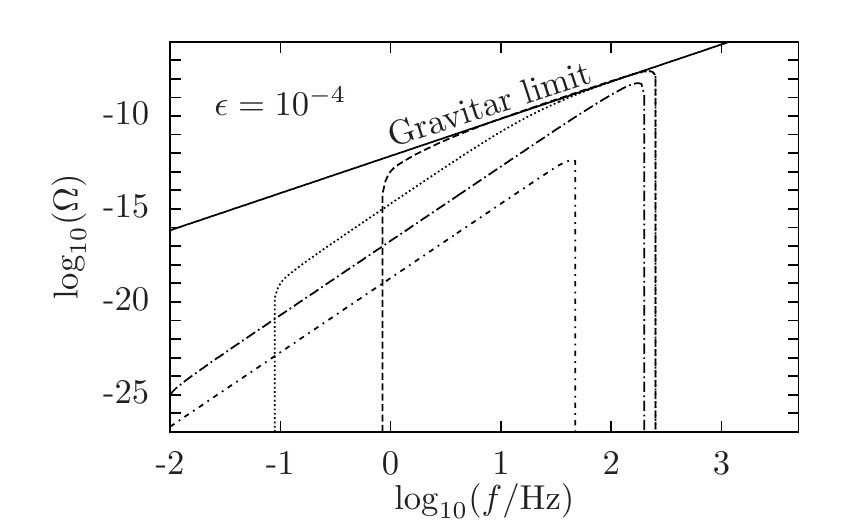}
\includegraphics{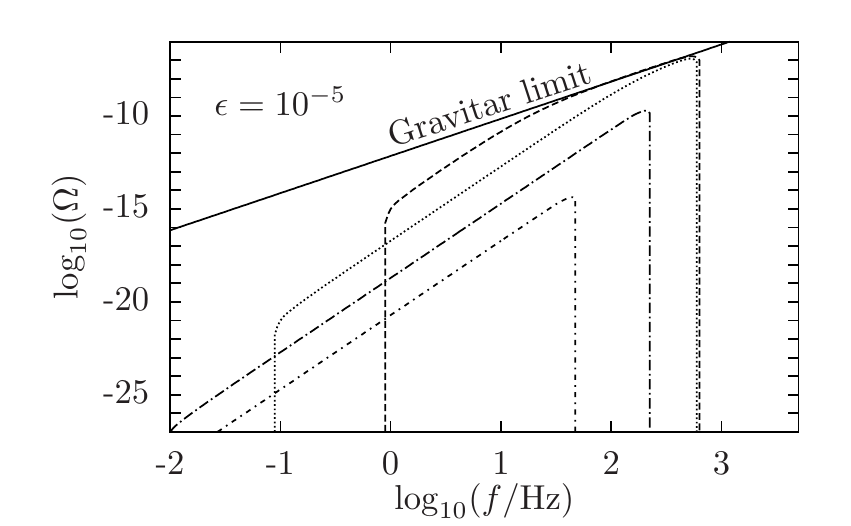}
\includegraphics{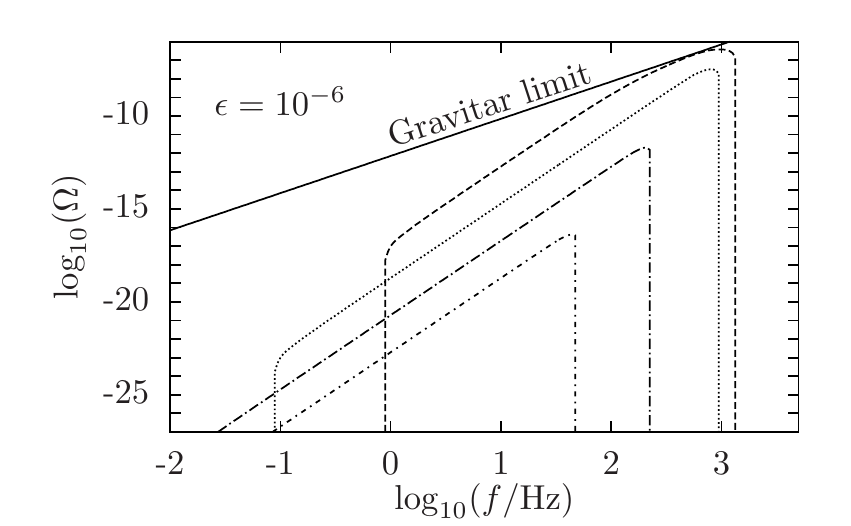}
\includegraphics{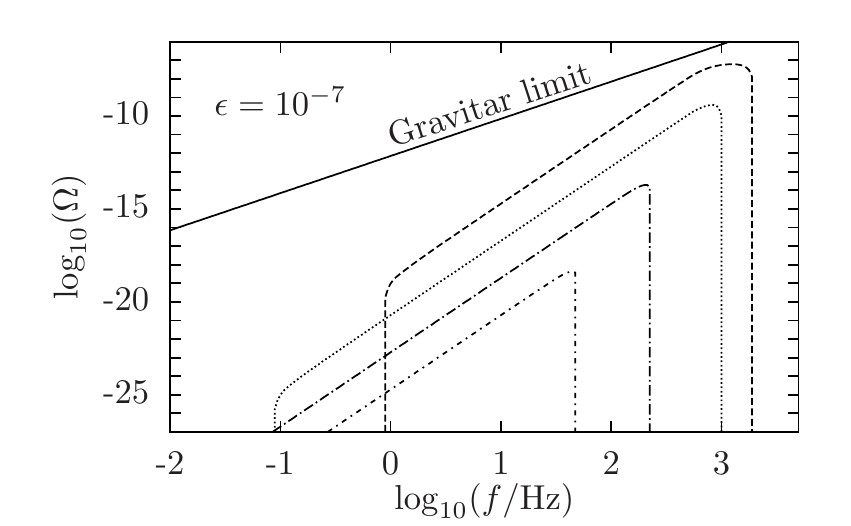}
\caption{Same as Figure \ref{fig:rotnst}, but for the unresolvable part of the background.}
\label{fig:rotnsu}
\end{figure}

In Figures \ref{fig:rotnst} and \ref{fig:rotnsu}, the total and unresolvable backgrounds are plotted, respectively, assuming that all rotating neutron stars have the same magnetic field and ellipticity.
These plots show that the spectral function is larger for smaller magnetic fields and larger ellipticities, as it was expected.
Furthermore, with larger magnetic fields, lower frequencies are achieved, and a bigger part of the background becomes resolvable.

\begin{figure}
\includegraphics{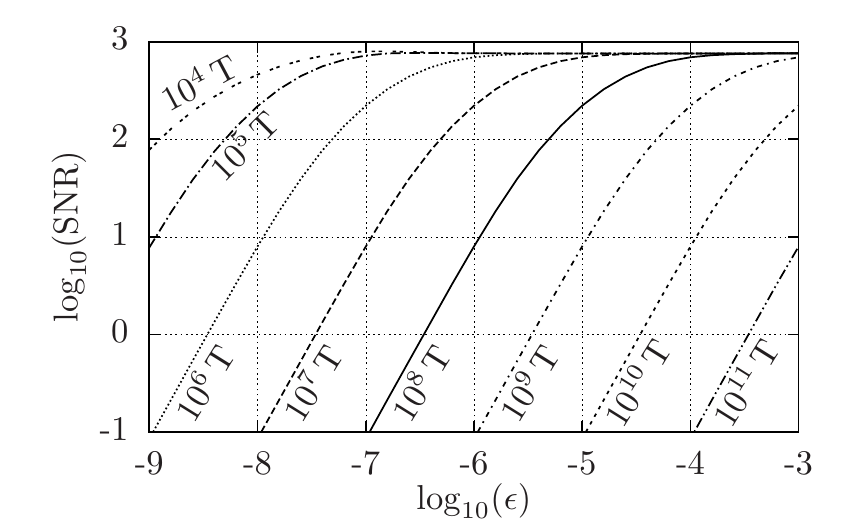}
\includegraphics{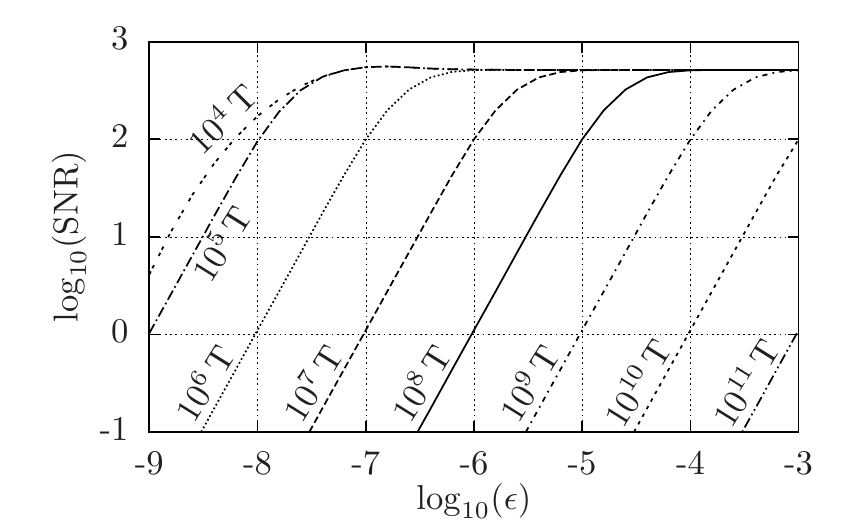}
\includegraphics{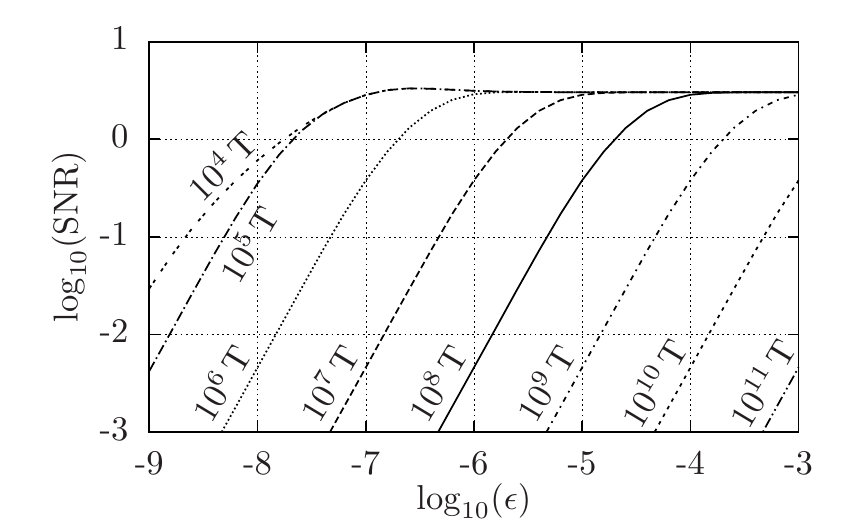}
\caption{Signal-to-noise ratio produced by the total background of rotating neutron stars, assuming that all of them have the same magnetic field and ellipticity.
Each curve corresponds to a certain magnetic field (specified on top of each curve), and each point on a curve corresponds to a certain ellipticity (specified on the horizontal axis).
These values of SNR are obtained by cross-correlating 1 year of data of two interferometers of ETB.
Upper, middle, and lower plots are obtained using the initial frequency distribution called Distribution 0, 1, and 2, respectively, in Section \ref{sec:initialspin}.}
\label{fig:snretb}
\end{figure}

\begin{figure}
\includegraphics{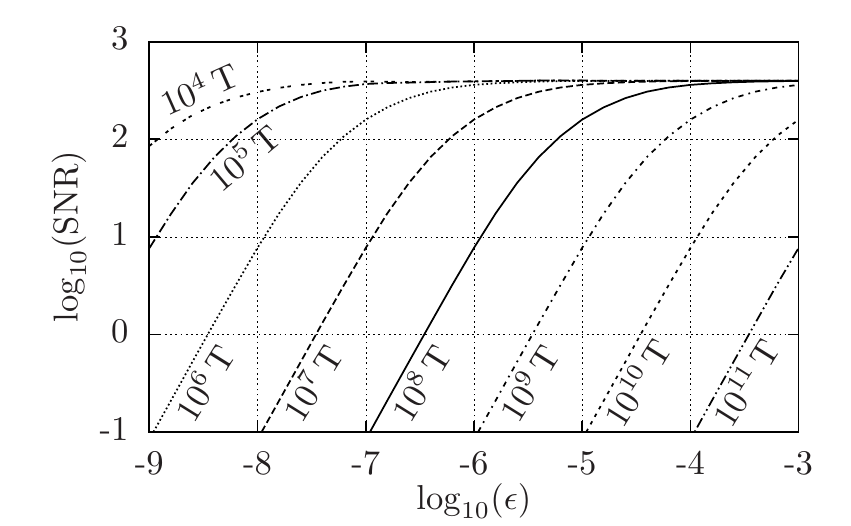}
\includegraphics{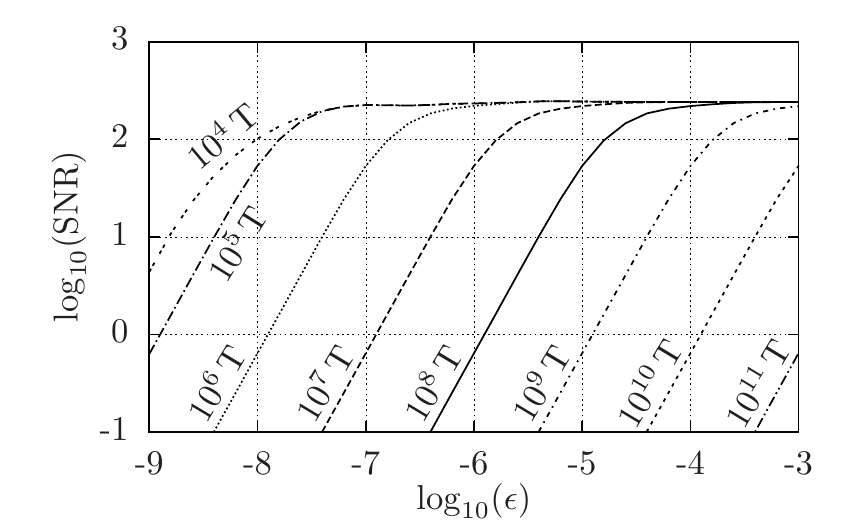}
\includegraphics{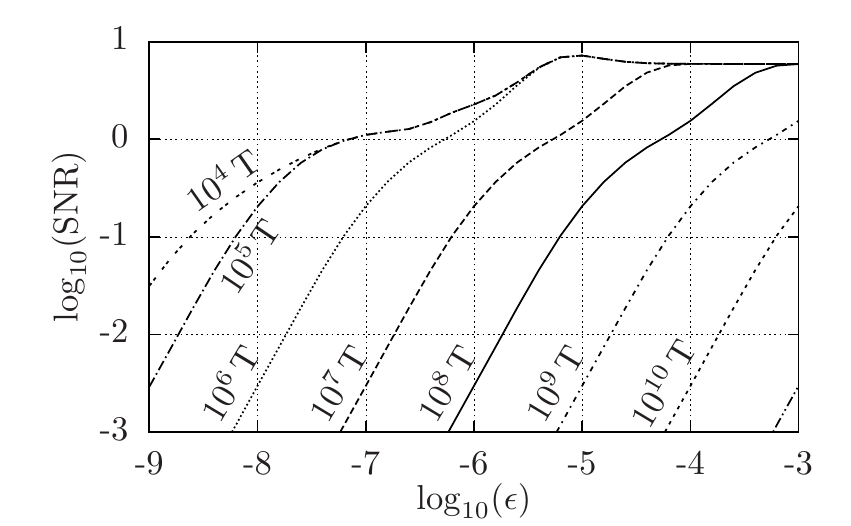}
\caption{Same as Figure \ref{fig:snretb}, calculated in this case for ETD.}
\label{fig:snretd}
\end{figure}

\begin{figure}
\includegraphics{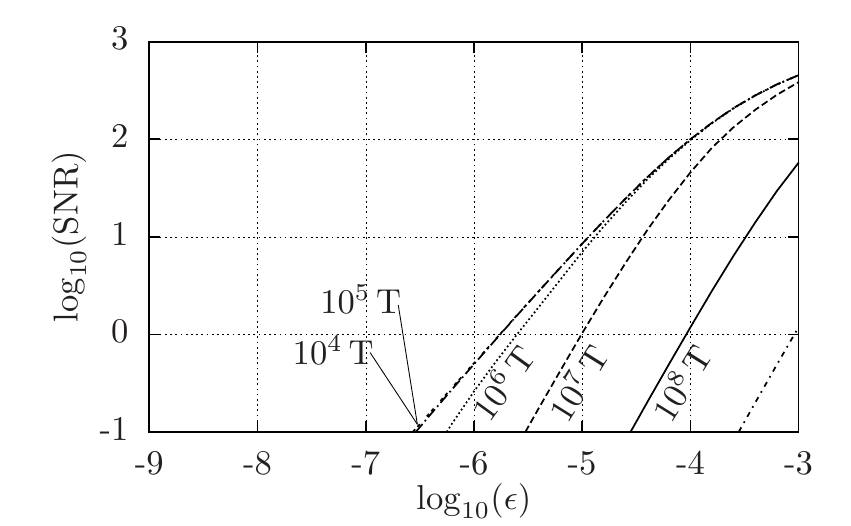}
\includegraphics{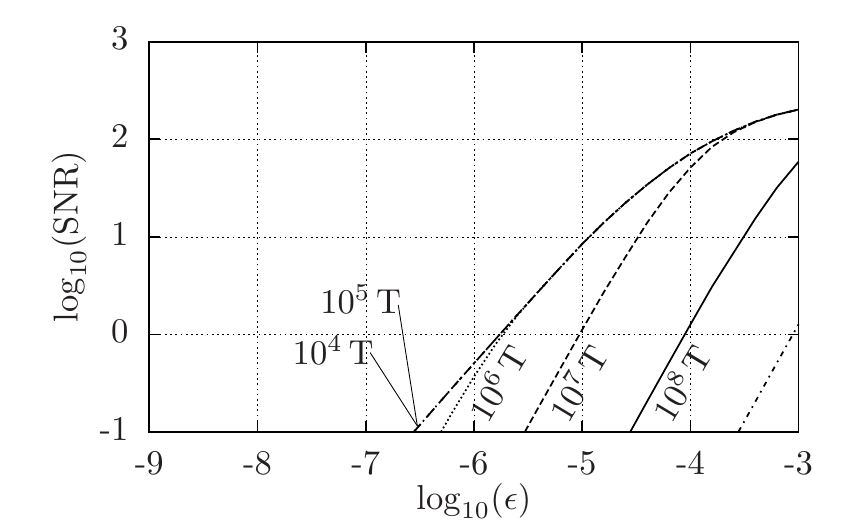}
\includegraphics{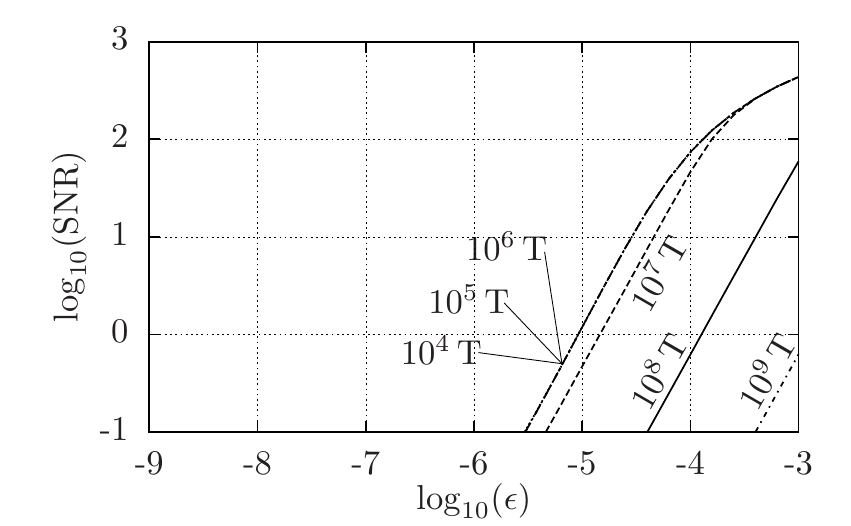}
\caption{Same as Figure \ref{fig:snretb}, but calculated for two interferometers of BBO.
The observation time assumed is also of 1 year.}
\label{fig:snrbbo}
\end{figure}

\begin{figure}
\includegraphics{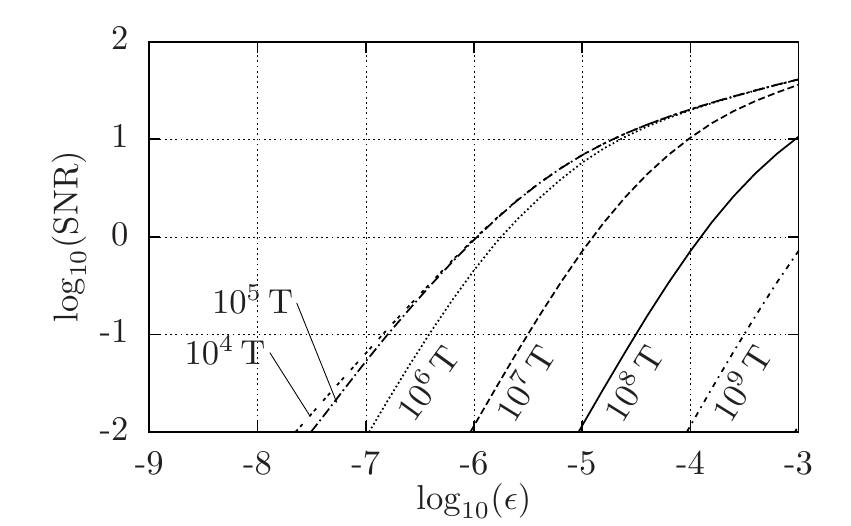}
\includegraphics{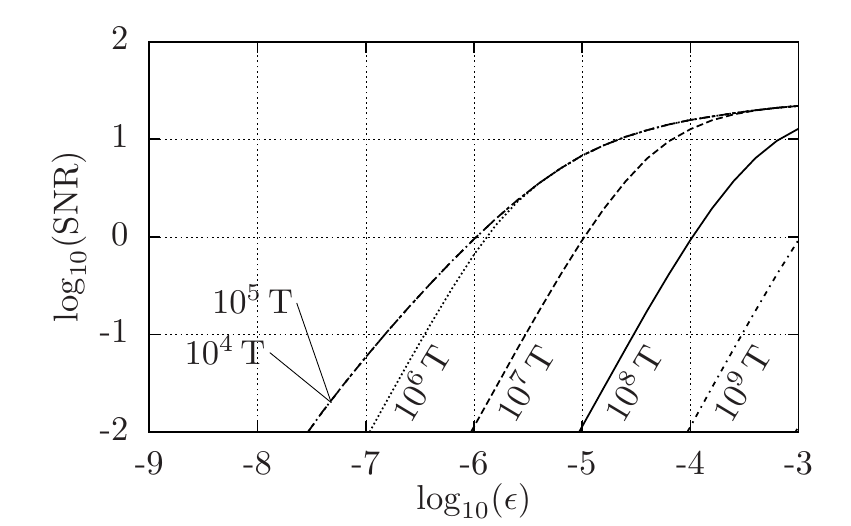}
\includegraphics{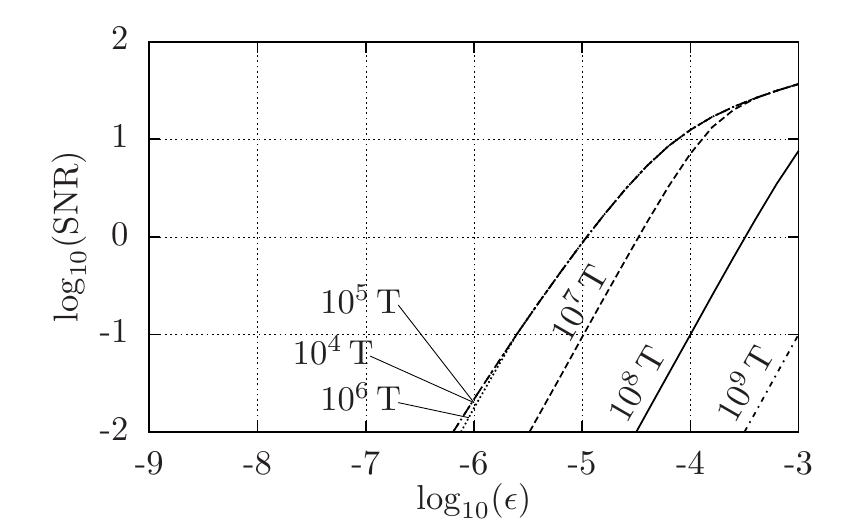}
\caption{Same as Figure \ref{fig:snretb}, but for two interferometers of DECIGO.
The observation time assumed is also of 1 year.}
\label{fig:snrdec}
\end{figure}

In Figures \ref{fig:snretb}, \ref{fig:snretd}, \ref{fig:snrbbo}, and \ref{fig:snrdec}, we plot the SNR obtained by cross-correlating two interferometers of ETB, ETD, BBO or DECIGO, respectively, assuming one year of observation.
To obtain the corresponding SNR for an interval of observation time $T_{\text{obs}}$, one can just multiply those values by $\sqrt{T_{\text{obs}}/[1\,\text{yr}]}$.
Each point on each curve corresponds to one realization of the background, in which all neutron stars have the same magnetic field and the same ellipticity.
We know, of course, that not all neutron stars are equal.
However, these plots are interesting for the following reason:
the SNR (as well as the spectral function) is proportional to the rate.
So, all neutron stars may not have the same certain values of $B$ and $\epsilon$, but if only a given fraction does, the SNR produced would be the one of the plots, multiplied by that fraction.
This allows us to draw a few interesting conclusions.

If Distribution 2 accurately describes the initial distribution of frequencies, the detection of the background of rotating neutron stars seems unlikely;
instead, if Distributions 0 or 1 are accurate, the chances of detection (by ET) are high.
For example, we see in Figure \ref{fig:snretb} (assuming Distribution 1) that, if all neutron stars have an ellipticity of $\epsilon=10^{-6}$ and a magnetic field of $B=10^7\,$T, the obtained SNR is $1.0\times 10^2$ for ETB; then, if at least a few percent of neutron stars have ellipticities larger than $10^{-6}$ and magnetic fields smaller than $10^7\,$T, ETB would detect the produced background with SNR of at least a few.
Suppose now that neutron stars cannot have ellipticities larger than $10^{-7}$.
Even in this case, SNR of a few would be obtained for ETB if only a few percent of the rotating neutron stars have magnetic fields lower than $10^6\,$T.

As Figures \ref{fig:snrbbo} and \ref{fig:snrdec} show, the SNR calculated for BBO and DECIGO reach relevant values for magnetic fields smaller than $\sim 10^{7}$\,T, and, in Figure \ref{fig:rotnst}, we see that the spectral function, for such a magnetic field, has support only at frequencies larger than $\approx$1\,Hz.
The main contribution to the SNR of BBO and DECIGO thus comes from frequencies between 1\,Hz and 10\,Hz.
In \cite{NishizawaEtAl2010}, the overlap reduction function is calculated for different configurations of the spacecraft constellations of BBO and DECIGO.
Almost all configurations produce an overlap reduction function close to zero between 1\,Hz and 10\,Hz.
The assumption made in Section \ref{sec:detectability} of an overlap reduction function equal to one is therefore very crude.
The SNR obtained with a more realistic overlap reduction function would reasonably be much lower.

\begin{figure}
\includegraphics{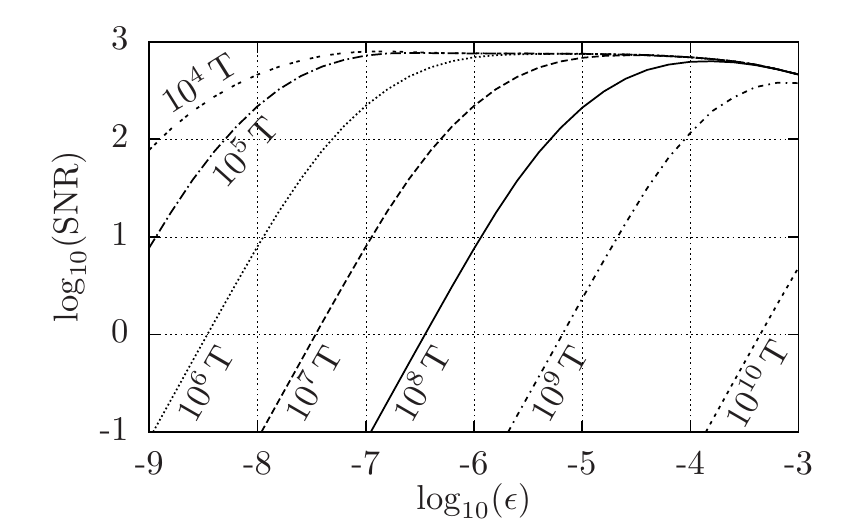}
\includegraphics{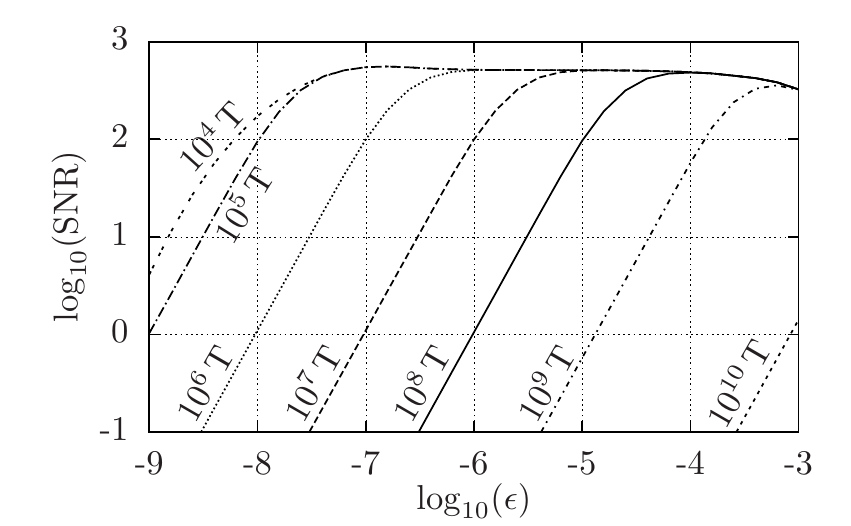}
\includegraphics{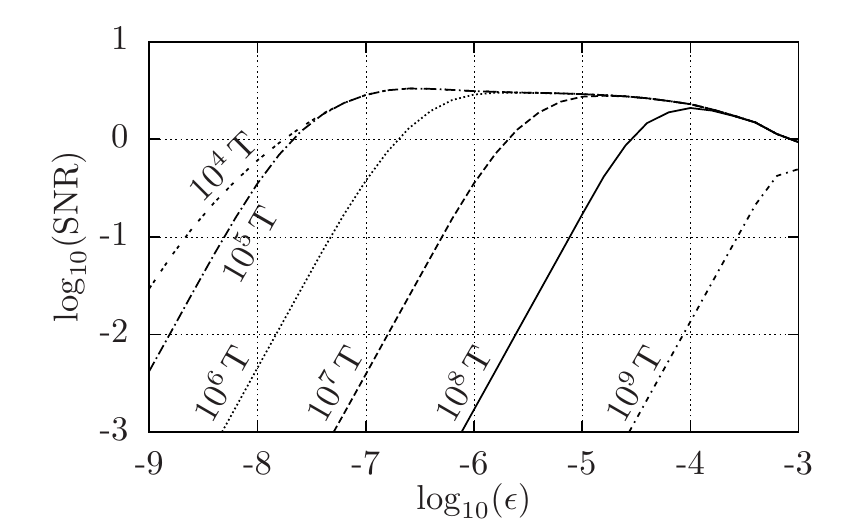}
\caption{Same as Figure \ref{fig:snretb}, but for the unresolvable part of the background (using $\mathcal{N}_0=1$ and $\Delta f=1\,\text{yr}^{-1}$).
These values of SNR are obtained by cross-correlating 1 year of data of two interferometers of ETB.}
\label{fig:snrunresetb}
\end{figure}

\begin{figure}
\includegraphics{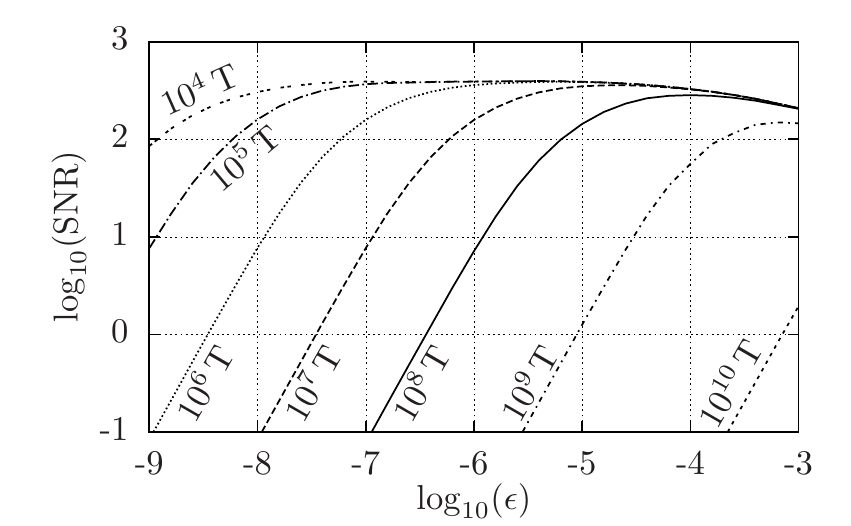}
\includegraphics{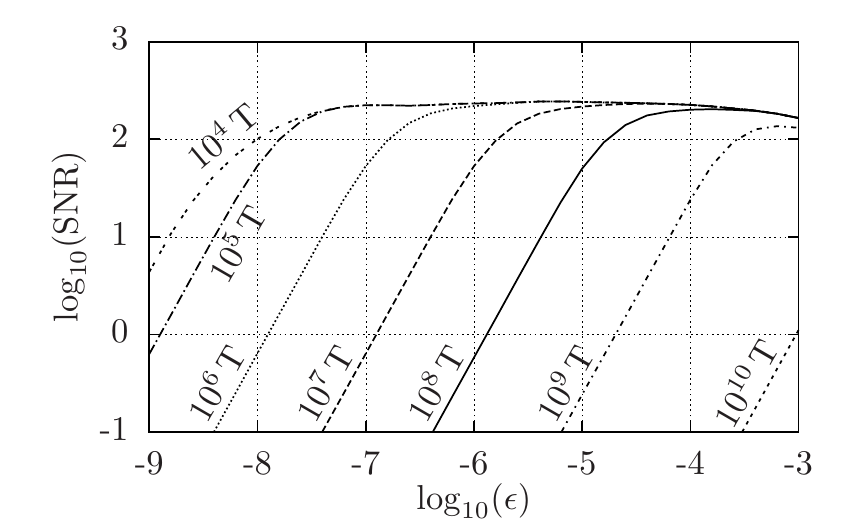}
\includegraphics{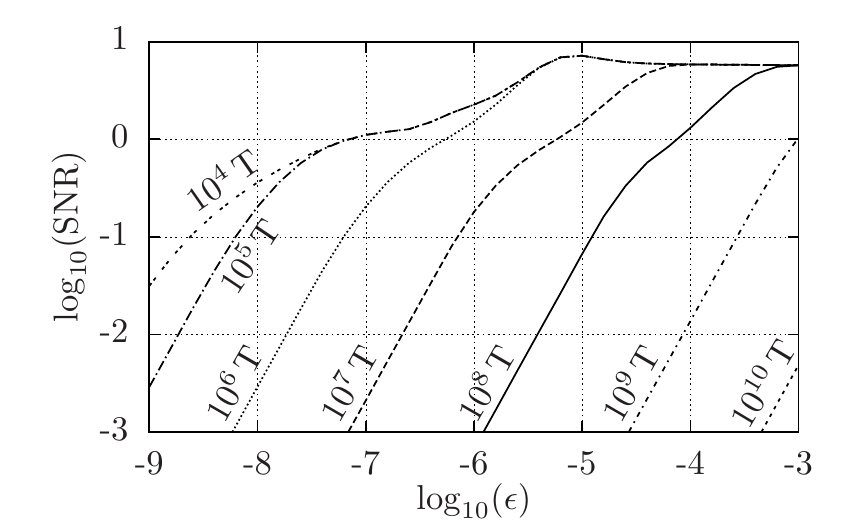}
\caption{Same as Figure \ref{fig:snrunresetb}, but for ETD.}
\label{fig:snrunresetd}
\end{figure}

In Figures \ref{fig:snrunresetb} and \ref{fig:snrunresetd} we show the same SNR calculations as in Figures \ref{fig:snretb} and \ref{fig:snretd}, but for the unresolvable part of the background.
By comparison, one can conclude that, for ellipticities smaller than $\epsilon \sim 10^{-4}$ and magnetic fields smaller than $B\sim 10^{8}\,$T, the background is almost entirely unresolvable.

\section{Discussion}
\label{sec:discussion}

\subsection{\label{sec:comparison} Comparison with previous work}
The background produced by magnetars has recently been calculated in \cite{MarassiEtAl2011b}, assuming different models.
Our upper limit is obtained by assuming the magnetic field and ellipticity of one of the models given in that paper: the one that predicts the largest spectral function.
All other models in that work, as well as the models used in previous papers \cite{RegimbauFreitas2006,RegimbauMandic2008} produce smaller levels of background.

In \cite{Kotera2011}, the gravitational wave background is calculated, assuming a population of magnetars that could fit the ultrahigh energy cosmic ray spectrum.
The most optimistic of the expectations for the spectral function in that paper is, in a certain range of frequencies, a factor of $\approx 8$ larger than the upper limit calculated by us in Section \ref{sec:magnetars}.
That expectation of the background is said to possibly reach the sensitivities of BBO and DECIGO, but not that of ET; however, no calculation of SNR is performed in that paper.
We now assert that, performing the SNR calculations, the claim is the opposite.
The SNR that our magnetar upper limit would produce on BBO and DECIGO (assuming an overlap reduction function equal to one, which is already too optimistic), after one year of observation time, is of the order of $10^{-3}$; these values are too low to claim a possible detection, even if multiplied by that factor of $\approx 8$.
However, the upper limit of \cite{Kotera2011} would produce an SNR of $\approx 5$ and $\approx 2$ on ETB and ETD, respectively.
Therefore, one can conclude that the upper limit of magnetars (either with the estimate of \cite{Kotera2011} or with ours) is out of the reach of BBO and DECIGO, but could be detected by ET.

All the papers mentioned in this section use the so-called \textit{duty cycle} to account for the statistical properties of the background.
As commented in Section V.F.3 of \cite{Rosado2011}, the overlap function (which is a generalization of the duty cycle) is the right tool to quantify the resolvability of the background.
Moreover, the duty cycle can only be used for short (burst-like) signals, not for long signals, like the ones produced by rotating neutron stars.
In Figures \ref{fig:magnetars}, \ref{fig:snrunresetb}, and \ref{fig:snrunresetd}, one sees that, even if having a duty cycle much larger than 1 (as reported in \cite{MarassiEtAl2011b}), the background produced by magnetars is resolvable in the band of ET.

\subsection{On the formation rate}
\label{sec:discussionrate}
The spectral function turns out to be rather insensitive to the shape of the rate.
In this section we compare the spectral function obtained using the different star formation rates $\dot{\rho}(z)$ of \cite{MadauEtAl1998,PorcianiMadau2001,2dFGRS2001,StrolgerEtAl2004,HopkinsBeacom2006,NagamineEtAl2006,FardalEtAl2007,WilkinsEtAl2008}, and a star formation rate that has the same value over all cosmological epochs.

Let us first obtain a reasonable value for the constant rate $\dot{n}(z)=R$.
Given one star formation rate $\dot{\rho}_i(z)$, one can calculate its average value over all redshifts,
\begin{equation}
\langle \dot{\rho}_i \rangle=\frac{\int_{0}^{z_{\text{max}}}\dot{\rho}_i(z) dz}{\int_0^{z_{\text{max}}}dz} \, .
\end{equation}
Considering the $N=11$ star formation rates from \cite{MadauEtAl1998,PorcianiMadau2001,2dFGRS2001,StrolgerEtAl2004,HopkinsBeacom2006,NagamineEtAl2006,FardalEtAl2007,WilkinsEtAl2008}, the mean value of $\langle \dot{\rho}_i \rangle$ is
\begin{equation}
\overline{\langle \dot{\rho} \rangle}=\frac{1}{N}\sum_{i=1}^N \langle \dot{\rho}_i\rangle=0.10\,M_\odot\,\text{yr}^{-1}\,\text{Mpc}^{-3} \, .
\end{equation}
Finally, replacing this star formation rate in Equation (\ref{eq:signalrate}), one obtains the value of the constant rate $R$,
\begin{equation}
\label{eq:rate}
R=\lambda \overline{\langle \dot{\rho} \rangle} =5\times 10^{-4}\,\text{yr}^{-1}\,\text{Mpc}^{-3} \, .
\end{equation}
This is the approximate number of neutron stars formed per unit emitted interval of time per unit comoving volume, at any cosmological epoch.

\begin{figure}
\includegraphics{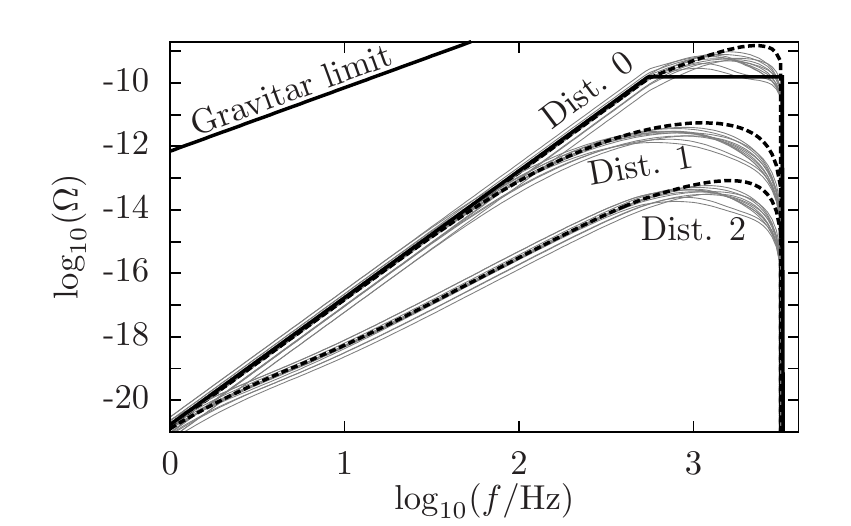}
\caption{Spectral function, versus observed frequency, of the total background produced by the ensemble of rotating neutron stars, using the three initial frequency distributions (Distribution 0, 1, and 2, of Section \ref{sec:initialspin}). All systems are assumed to have $B=10^8\,$T, and $\epsilon=10^{-7}$. The light-gray curves are obtained with different star formation rates \cite{MadauEtAl1998,PorcianiMadau2001,2dFGRS2001,StrolgerEtAl2004,HopkinsBeacom2006,NagamineEtAl2006,FardalEtAl2007,WilkinsEtAl2008}, whereas the black dashed lines use the constant star formation rate of Equation (\ref{eq:rate}). The black solid line is obtained with the approximate formula of Equation (\ref{eq:omegasimpt}).}
\label{fig:comptotal}
\end{figure}

In Figure \ref{fig:comptotal}, we show the spectral function, with the three initial frequency distributions of Section \ref{sec:initialspin}, obtained with constant and non-constant rates.
One sees that the spectral function using different rates differs just by a constant factor at almost all frequencies; only at frequencies close to the maximum one, the spectral function presents different shapes depending on the specific rate assumed.
Given the lack of information on the population statistics (which is evident when comparing the results of different initial frequency distributions), the choice of one or another rate is irrelevant.
A similar conclusion is drawn for binary systems in Section V.A of \cite{Rosado2011}.

\section{\label{sec:summary}Summary and conclusions}

\subsection{\label{sec:summary1} Regarding the calculation of the spectral function}
We have derived a formula for the spectral function $\Omega(f,\Delta f,\mathcal{N}_0)$, i.e., the density per logarithmic frequency interval (in units of critical density), around the observed frequency $f$, of a background made of signals that produce an overlap of $\mathcal{N}_0$ signals per frequency bin $\Delta f$.
This formula (in Equation (\ref{eq:omegagen})) can be used for a population that evolves over long time scales.
We also present an expression for the overlap function $\mathcal{N}(f,\Delta f,z)$, that gives the number of signals with observed frequency $f$ per frequency bin $\Delta f$, with redshifts smaller than $z$.
This overlap function (in Equation (\ref{eq:defovfungen})) is more general than the one introduced in \cite{Rosado2011}.

A more detailed expression of the spectral function is given in Equation (\ref{eq:omgendist}), where the dependence on the initial frequency distribution of the ensemble is explicitly shown.
Similarly, a more explicit formula for the overlap function is presented in Equation (\ref{eq:ovfungeneral}) (or, in a more convenient form, in Equation (\ref{eq:ovfungenapp})).
Assuming that all systems start emitting at the same frequency, the overlap function and the spectral function become the ones of Equations (\ref{eq:ovfunapp2}) (or, more conveniently, (\ref{eq:ovfunappfix})), and (\ref{eq:omegaffix}), respectively.
Equation (\ref{eq:ovfunapp3}) gives the overlap function calculated with the additional assumption that the evolution of the systems is short compared to cosmological time scales.
This formula coincides with the definition given in \cite{Rosado2011}.
The spectral function is then obtained under the same assumptions (Equation (\ref{eq:omclass})).
If one applies the latter formula to calculate the total background (with $\mathcal{N}_0=0$), one obtains Equation (\ref{eq:omclass2}), which is the expression that usually appears in the literature.

In Section \ref{sec:models} we have described a simple but useful model for the energy and frequency evolution of rotating neutron stars.
Figures \ref{fig:zfplotgr} and \ref{fig:zfplotmd} show the collection of possible redshifts and observed frequencies of gravitars and of normal pulsars, respectively, assuming in both cases that all systems have the same ellipticity, $\epsilon=10^{-7}$.
In these plots one can see the frequency range in which the usual assumption of short-lived systems (which has not been adopted for our results) is inaccurate.

In Appendices \ref{sec:rotnssimp} and \ref{sec:binsimp}, we propose simple approximate formulas for the spectral function of rotating neutron stars and, for completion, for binary systems too.
Given the uncertainties in some astrophysical parameters, like the rate and the initial frequency distribution, these approximate formulas can be used as a reasonably good estimation for the levels of contemporary backgrounds.

\subsection{\label{sec:summary2} Regarding the detection of the background of rotating neutron stars}

The three initial spin period distributions considered in the calculations (see Section \ref{sec:initialspin}), lead to very disparate results.
The first one (called Distribution 0), assumes that all systems are formed with the same initial frequency.
The other distributions (called Distributions 1 and 2) are taken from the literature \cite{ArzoumanianEtAl2002,FaucherKaspi2006}.
If Distribution 2 is accurate, the detection of the background by present and planned detectors can be discarded.

In Section \ref{sec:gravitars} we have justified that there is a robust upper limit (the gravitar limit, in Equation (\ref{eq:omegagr}), or, more simply in (\ref{eq:gravlim})) on the level of background produced by rotating neutron stars.
In Appendix \ref{sec:blandford}, we point out an interesting characteristic of the background obtained under the unrealistic assumptions of the gravitar limit: its characteristic amplitude $h_c$ is independent of the ellipticity of the systems and of their spin frequency.
A similar feature was predicted by Blandford, for the expected gravitational wave amplitude of the nearest system of a uniform galactic population of gravitars.
This feature disappears with more realistic models, as it occurs with Blandford's argument.

In Figure \ref{fig:gravitars}, the spectral functions of the total and the unresolvable background are calculated, under the unrealistic assumption that all neutron stars are gravitars.
One sees that the background is almost entirely unresolvable.
Such a background, assuming Distributions 0 and 1, can be detected with ETB and ETD, but not with aLIGO, by using the typical cross-correlation method.
If at least 1$\%$ of neutron stars behave like gravitars, they will produce an unresolvable background that can be detected by ET.

Using a magnetic field and an ellipticity distribution (from \cite{ArzoumanianEtAl2002,Palomba2005}), a reasonable level of background of rotating neutron stars is obtained (see Figure \ref{fig:rotns}); it is below the detection capabilities of any existing or planned instrument.

We have obtained an upper limit on the total background of magnetars (Figure \ref{fig:magnetars}), using one of the models presented in \cite{MarassiEtAl2011b}.
This background can only be detected by ET.
However, other models predict levels of background several orders of magnitude lower.
Hence, we claim that magnetars are not the most promising rotating neutron stars, regarding the detection of the background.

With Figures \ref{fig:rotnst} and \ref{fig:rotnsu} one can get an idea of how the amplitude and the resolvability of the background depend on the values of the magnetic field and ellipticity (assuming that these values are equal for all neutron stars).

Figures \ref{fig:snretb}, \ref{fig:snretd}, \ref{fig:snrbbo}, and \ref{fig:snrdec} summarize the prospects of ETB, ETD, BBO, and DECIGO, respectively, to detect the total background of rotating neutron stars.
The values of SNR in these plots are calculated for the cross-correlation of the data of two interferometers during 1 year (to obtain the values after 3 or 5 years, it is enough to multiply the vertical axis by $\sqrt{3}$ or $\sqrt{5}$).
These graphs are useful because the SNR is proportional to the fraction of stars that are neutron stars.
Suppose that some given values of magnetic field $B$ and ellipticity $\epsilon$ are associated with SNR equal to $S$ in the plots; then, if only a fraction $x$ of all neutron stars have $B$ and $\epsilon$, they will produce a background observed with SNR equal to $x\times S$.
For example, one can conclude from Figure \ref{fig:snretb} that, if at least a few percent of neutron stars have $\epsilon \ge 10^{-6}$ and $B \le 10^7\,$T, the background would be observed by ETB with SNR of a few.
Another conclusion from Figure \ref{fig:snretb} is that, if the maximum ellipticity of neutron stars is of $\epsilon=10^{-7}$, ETB will still observe a background of SNR of a few, if just a few percent of neutron stars have $B\le 10^6\,T$.

We point out that the SNR values of BBO and DECIGO are obtained by assuming an overlap reduction function (see Section \ref{sec:detectability}) identically equal to 1.
This is quite inaccurate between 1\,Hz and 10\,Hz (see \cite{NishizawaEtAl2010}, where the overlap reduction function is calculated for different configurations of the spacecrafts), which is the frequency interval where the background contributes the most to the SNR of BBO and DECIGO.
The detection prospects of BBO and DECIGO should therefore be smaller than what Figures \ref{fig:snrbbo} and \ref{fig:snrdec} suggest.

Figures \ref{fig:snrunresetb} and \ref{fig:snrunresetd} are analogous to \ref{fig:snretb} and \ref{fig:snretd}, respectively, but for the unresolvable part of the background.
They are included to show that the total and the unresolvable backgrounds are identical for all configurations with $\epsilon \le 10^{-4}$ and $B\le 10^8\,$T; on the other hand, the background produced by magnetars (with larger magnetic fields) is mostly resolvable.

This work, together with \cite{Rosado2011}, covers two of the most promising sources of contemporary gravitational wave background.
If the most reasonable estimate of the background (in Section \ref{sec:realistic}) is accurate, or if rotating neutron stars form with initial spin frequencies well described by Distribution 2 (in Equation (\ref{eq:dist2})), then we can conclude that ground-based detectors operate in a frequency window that is free of contemporary unresolvable background from binaries and rotating neutron stars.
However, if at least a few percent of neutron stars behave as gravitars, or if any of the configurations proposed in Section \ref{sec:prospects} that produce high SNR values is in good agreement with the real neutron star population, an unresolvable background of rotating neutron stars can obscure the searches of ET for other sources of background.

\begin{acknowledgments}
I thank Bruce Allen for his guidance and recommendations during the development and writing of the paper, Benjamin Knispel for his help and fruitful comments, Evan Goetz for his corrections, and Tania Regimbau for her suggestions and careful reading of the paper.
This work was supported by the IMPRS on Gravitational Wave Astronomy.
\end{acknowledgments}

\appendix

\section{\label{sec:rotnssimp} Simple formulas for the background of rotating neutron stars}
In this section we present some approximate formulas for the spectral function of the total and the unresolvable background of rotating neutron stars.
They are useful to obtain a simple estimate of the background; nevertheless, these formulas were not used to obtain the results of Section \ref{sec:results}.

We first need to define the constants
\begin{equation}
C_{\text{gr}}=\int_{0}^{z_{\text{max}}}\mathcal{E}^{-1}(z)dz\approx 1.9 \, ,
\end{equation} 
\begin{equation}
C_{\text{md}}=\int_{0}^{z_{\text{max}}}[1+z]^2\mathcal{E}^{-1}(z)dz=\frac{2\mathcal{E}(z_{\text{max}})}{3\Omega_m}\approx 19.0 \, ,
\end{equation} 
\begin{equation}
\overline{C}_{\text{md}}=\int_{0}^{z_{\text{max}}} [1+z]^{-2} \left[\int_0^{z} \mathcal{E}^{-1}(z')dz'\right]^2\mathcal{E}^{-1}(z)dz\approx 0.23 \, ,
\end{equation} 
and
\begin{equation}
\overline{C}_{\text{gr}}=\int_{0}^{z_{\text{max}}} [1+z]^{-4}\left[\int_0^{z} \mathcal{E}^{-1}(z')dz'\right]^2\mathcal{E}^{-1}(z)dz\approx 0.042 \, ,
\end{equation} 
which depend only on cosmological parameters.

A formula for the spectral function of the gravitar limit can be very easily obtained.
Using a constant rate $\dot{n}(z)=R$, Equation (\ref{eq:omegagr}) becomes
\begin{equation}
\label{eq:gravlim}
\Omega_{\text{GL}}(f)=\frac{8\pi^3GIRC_{\text{gr}}}{3H_0^3 c^2} f^2 \, .
\end{equation}
A value for $R$ can be found in Section \ref{sec:discussionrate}.

In the following, we assume a fixed initial frequency $f_{\text{max}}$, and obtain the spectral function by solving Equation (\ref{eq:omclass}) for rotating neutron stars.
Moreover, we assume a constant rate $\dot{n}(z)=R$, and perform the approximation introduced in Section \ref{sec:enerspec} of distinguishing md- and gr-ranges.
The spectral function of the background of rotating neutron stars, under these assumptions, has an analytical form.
To account for the gr- and the md- regimes, we define the function
\begin{equation}
\label{eq:omegasimp}
\overline{\Omega}(f)=\frac{\pi^2 IR}{\rho_c H_0c^2}\left\{ \begin{array}{lc}
C_{\text{gr}} f^2 & \sqrt{\frac{C_{\text{gr}}}{C_{\text{md}}}}f_C<f \\
C_{\text{md}} f^4 & f\le \sqrt{\frac{C_{\text{gr}}}{C_{\text{md}}}}f_C \\
\end{array}
\right. ,
\end{equation} 
where the cut frequency $f_C$ is the one in Equation (\ref{eq:fcdef}).

The total background can then be estimated using
\begin{equation}
\label{eq:omegasimpt}
\Omega_{\text{total}}(f)=\left\{ \begin{array}{lc}
0 & f<f_{\text{min}} \\
\overline{\Omega}(f) & f_{\text{min}}\le f \le \frac{f_{\text{max}}}{1+z_{\text{max}}} \\
\overline{\Omega}(\frac{f_{\text{max}}}{1+z_{\text{max}}}) & \frac{f_{\text{max}}}{1+z_{\text{max}}} < f\le f_{\text{max}} \\
0 & f_{\text{max}}<f
\end{array}
\right. .
\end{equation}
The maximum frequency is the one defined in Equation (\ref{eq:fmaxchoice}), and the minimum frequency is
\begin{equation}
\label{eq:fminapp}
f_{\text{min}}=
\left\{ \begin{array}{lc}
f_1(z_{\text{gr}}) & f_C\le f_1(z_{\text{gr}}) \\
f_2(\frac{z_{\text{gr}}+z_{\text{md}}}{2}) & f_1(z_{\text{gr}})<f_C<f_{\text{max}} \\
f_3(z_{\text{md}}) & f_{\text{max}}\le f_C
\end{array}
\right. ,
\end{equation}
where $f_1(z)$, $f_2(z)$, $f_3(z)$, $z_{\text{gr}}$, and $z_{\text{md}}$ are given in Equations (\ref{eq:f1func}), (\ref{eq:f2func}), (\ref{eq:f3func}), (\ref{eq:zgrdef}), and (\ref{eq:zmddef}), respectively.
Equation (\ref{eq:fminapp}) is an approximation to the minimum value of $f_{\text{low}}(z)$, given by Equation (\ref{eq:flowdef}); $f_{\text{min}}$ is, therefore, defined as an observed frequency (unlike $f_{\text{max}}$, that is an emitted frequency).

The spectral function of the unresolvable background can be approximated by
\begin{equation}
\label{eq:omegasimpu}
\Omega_{\text{unresolvable}}(f)=\left\{ \begin{array}{lc}
0 & f<f_{\text{p,min}} \\
\Omega_{\text{total}}(f) & f_{\text{p,min}}\le f \le f_{\text{p,max}} \\
0 & f_{\text{p,max}}<f
\end{array}
\right. .
\end{equation} 
In this equation we have introduced the \textit{limiting frequencies} (see Section III.E.4 of \cite{Rosado2011}); $f_{\text{p,min}}$ ($f_{\text{p,max}}$) is the minimum (maximum) frequency above (below) which the unresolvable background is present.
The limiting frequencies can be obtained from
\begin{equation}
\label{eq:fpmin}
f_{\text{p,min}}=f_{\text{min}} \, ,
\end{equation} 
and
\begin{equation}
\label{eq:fpmax}
f_{\text{p,max}}=\left\{ \begin{array}{lc}
 \min \left(\chi,f_{\text{max}}\right) & \chi \le f_C \\
\min \left(\chi^{3/5}\left[\sqrt{\frac{\overline{C}_{\text{gr}}}{\overline{C}_{\text{md}}}}f_C\right]^{2/5},f_{\text{max}}\right) & f_C<\chi
\end{array}
\right. ,
\end{equation}
where
\begin{equation}
\chi=\left[\frac{8\pi c^3 R \Delta f \delta_{\text{md}}\overline{C}_{\text{md}}}{H_0^3\mathcal{N}_0}\right]^{1/3}.
\end{equation} 
The upper case in (\ref{eq:fpmax}) occurs when the unresolvable background is restricted solely to the md-range, and the lower case occurs when the unresolvable background is either restricted to the gr-range or partially in both ranges.
Instead of $f_{\text{p,min}}$ and $f_{\text{p,max}}$, one can use $f_{\text{d,min}}$ and $f_{\text{d,max}}$ in Equation (\ref{eq:omegasimpu}); $f_{\text{d,min}}$ ($f_{\text{d,max}}$) is the minimum (maximum) frequency above (below) which the unresolvable background dominates over the resolvable.
The unresolvable background is said to dominate over the resolvable when the spectral function of the former is larger than the spectral function of the latter.
One can prove that $f_{\text{d,min}}\approx f_{\text{p,min}}$, whereas $f_{\text{d,max}}$ is constrained to the interval
\begin{equation}
\label{eq:fdmax}
Ff_{\text{p,max}}\le f_{\text{d,max}}<f_{\text{p,max}} \, .
\end{equation} 
The factor $F$ depends only on cosmological parameters, and is $F\approx 0.9$ (both in the md and in the gr ranges).
With Equation (\ref{eq:fdmax}) we can conclude that, as soon as an unresolvable background appears at a certain frequency $f_{\text{p,max}}$, that background dominates over the resolvable background below $\approx 0.9 f_{\text{p,max}}$.

\section{\label{sec:binsimp} Simple formulas for the background of binary systems}
To have a complete estimate of some of the most promising sources of contemporary backgrounds, we also give some simple approximate formulas regarding the background of stellar binary systems (those systems composed of neutron stars, white dwarfs, or stellar-mass black holes).
These formulas are based on the calculations of \cite{Rosado2011}.

The spectral function of the total background can be calculated again using Equation (\ref{eq:omegasimpt}), but with different definitions of $\overline{\Omega}(f)$, $f_{\text{max}}$, and $f_{\text{min}}$.
For binary systems, we have
\begin{equation}
\label{eq:omegabinsimp}
\overline{\Omega}(f)=\frac{R_{\text{bin}} [G\pi]^{2/3} m_1m_2 C_{\text{bin}}}{3\rho_c c^2 H_0[m_1+m_2]^{1/3}} f^{2/3} \, .
\end{equation} 
Here, $m_1$ and $m_2$ are the masses of the two components of the binary, $R_{\text{bin}}$ is the binary rate (summarized in Table I of \cite{Rosado2011}), and
\begin{equation}
C_{\text{bin}}=\int_0^{z_{\text{max}}}[1+z]^{-4/3}\mathcal{E}^{-1}(z)dz\approx 0.75 \, .
\end{equation}
The maximum frequency can be the frequency of the last stable orbit, which is given by
\begin{equation}
f_{\text{max}}=\frac{c^3}{6\sqrt{6}\pi G[m_1+m_2]} \, .
\end{equation}
For systems containing a white dwarf, a better choice of the maximum frequency is
\begin{equation}
f_{\text{max}}^{\text{WD}}=\sqrt{\frac{G[m_1+m_2]}{\pi^2 [r_1+r_2]^3}} \, ,
\end{equation}
where $r_1$ and $r_2$ are the radii of the components; this frequency corresponds to a separation of the two components equal the sum of their radii.
On the other hand, the minimum frequency is approximately given by
\begin{equation}
\label{eq:fminbin}
f_{\text{min}}=\left[ \frac{256\mathcal{T}(z_{\text{max}})\pi^{8/3}G^{5/3}m_1m_2}{5c^5[m_1+m_2]^{1/3}} \right]^{-3/8}[1+z_{\text{max}}]^{-1} \, ,
\end{equation}
where the function $\mathcal{T}(z)$ is the same function that has been used in the previous sections, defined in Equation (\ref{eq:timez}).
The minimum frequency for binary systems has been defined as an observed frequency (unlike in \cite{Rosado2011}) in analogy to the minimum frequency defined for rotating neutron stars, in Equation (\ref{eq:fminapp}).

To obtain the spectral function of the unresolvable background, one can once more use Equation (\ref{eq:omegasimpu}), with the definition of $\overline{\Omega}(f)$ given in Equation (\ref{eq:omegabinsimp}), and with the limiting frequencies given below.
First, $f_{\text{p,min}}\approx f_{\text{min}}$, which is given in Equation (\ref{eq:fminbin}).
Second, the limiting frequency $f_{\text{p,max}}$ is approximately
\begin{align}
&f_{\text{p,max}}\nonumber\\
&\approx \min \left(\left[\frac{5\Delta f \overline{C}_{\text{bin}}c^8R_{\text{bin}}[m_1+m_2]^{1/3}}{24\pi^{5/3}G^{5/3}m_1m_2H_0^3\mathcal{N}_0} \right]^{3/11},f_{\text{max}}\right),
\end{align} 
where
\begin{align}
&\overline{C}_{\text{bin}}\nonumber\\
&=\int_0^{z_{\text{max}}}[1+z]^{-8/3}\left[ \int_0^z\mathcal{E}^{-1}(z')dz'\right]^2\mathcal{E}^{-1}(z)dz\approx 0.12 \, .
\end{align} 
In Equation (\ref{eq:omegasimpu}), instead of $f_{\text{p,min}}$ and $f_{\text{p,max}}$, one can use $f_{\text{d,min}}$ (which is approximately $f_{\text{p,min}}$) and $f_{\text{d,max}}$; the latter also fulfills Equation (\ref{eq:fdmax}), but, in the case of binaries, the value of the $F$-factor is approximately $0.6$.
$$$$

\section{On the connexion between the gravitar limit and Blandford's argument}
\label{sec:blandford}
Let us consider again the gravitar limit, described by the spectral function in Equation (\ref{eq:gravlim}).
If a stochastic background was characterized by such a spectral function, the characteristic amplitude of the gravitational wave spectrum \cite{Phinney2001} would be
\begin{equation}
h_c = \sqrt{fS_h(f)} \, ,
\end{equation}
where
\begin{equation}
S_h(f)=\frac{3H_0^2}{4\pi^2}f^{-3}\Omega_{\text{GL}}(f) \, .
\end{equation}
Using Equation (\ref{eq:gravlim}), we obtain a characteristic amplitude of the form
\begin{equation}
\label{eq:hcgl}
h_c=\sqrt{\frac{2\pi GI RC_{\text{gr}}}{H_0c^2}} \, ,
\end{equation}
which does not depend either on the frequency or on the ellipticity of the systems.
This fact reminds us Blandford's argument (first cited in \cite{[{Section 9.4.2 (b) of }] HawkingIsrael1987}, revisited in \cite{LIGO2007b} and \cite{KnispelAllen2008}).
According to this argument, the expected gravitational wave amplitude of the nearest system of a uniform galactic population of gravitars, is independent of the ellipticity and the spin frequency of the systems.
Some of the assumptions required to obtain Equation (\ref{eq:hcgl}) are different from those of Blandford's argument.
However, the result is similar: the expected gravitational wave characteristic amplitude of the background produced by a population of gravitars, uniformly distributed in the universe, is independent of the ellipticity and the spin frequency of the systems.

The assumptions needed for Blandford's argument to hold, as well as those needed for $h_c$ not to be a function of $\epsilon$ and $f$, are quite unrealistic.
Once one adopts more realistic assumptions on the galactic population (for example, that gravitars are not distributed on a two-dimensional disk), Blandford's argument vanishes \cite{KnispelAllen2008}.
Analogously, under more realistic assumptions on the ensemble of gravitars in the universe (for example, that they have a finite time to evolve, and a finite initial spin frequency), the characteristic amplitude in Equation (\ref{eq:hcgl}) depends on the ellipticity and on the frequency.

\bibliography{Rosado2012}

\begin{thebibliography}{85}%
\makeatletter
\providecommand \@ifxundefined [1]{%
 \@ifx{#1\undefined}
}%
\providecommand \@ifnum [1]{%
 \ifnum #1\expandafter \@firstoftwo
 \else \expandafter \@secondoftwo
 \fi
}%
\providecommand \@ifx [1]{%
 \ifx #1\expandafter \@firstoftwo
 \else \expandafter \@secondoftwo
 \fi
}%
\providecommand \natexlab [1]{#1}%
\providecommand \enquote  [1]{``#1''}%
\providecommand \bibnamefont  [1]{#1}%
\providecommand \bibfnamefont [1]{#1}%
\providecommand \citenamefont [1]{#1}%
\providecommand \href@noop [0]{\@secondoftwo}%
\providecommand \href [0]{\begingroup \@sanitize@url \@href}%
\providecommand \@href[1]{\@@startlink{#1}\@@href}%
\providecommand \@@href[1]{\endgroup#1\@@endlink}%
\providecommand \@sanitize@url [0]{\catcode `\\12\catcode `\$12\catcode
  `\&12\catcode `\#12\catcode `\^12\catcode `\_12\catcode `\%12\relax}%
\providecommand \@@startlink[1]{}%
\providecommand \@@endlink[0]{}%
\providecommand \url  [0]{\begingroup\@sanitize@url \@url }%
\providecommand \@url [1]{\endgroup\@href {#1}{\urlprefix }}%
\providecommand \urlprefix  [0]{URL }%
\providecommand \Eprint [0]{\href }%
\providecommand \doibase [0]{http://dx.doi.org/}%
\providecommand \selectlanguage [0]{\@gobble}%
\providecommand \bibinfo  [0]{\@secondoftwo}%
\providecommand \bibfield  [0]{\@secondoftwo}%
\providecommand \translation [1]{[#1]}%
\providecommand \BibitemOpen [0]{}%
\providecommand \bibitemStop [0]{}%
\providecommand \bibitemNoStop [0]{.\EOS\space}%
\providecommand \EOS [0]{\spacefactor3000\relax}%
\providecommand \BibitemShut  [1]{\csname bibitem#1\endcsname}%
\let\auto@bib@innerbib\@empty
\bibitem [{\citenamefont {Allen}\ and\ \citenamefont
  {Romano}(1999)}]{AllenRomano1999}%
  \BibitemOpen
  \bibfield  {author} {\bibinfo {author} {\bibfnamefont {B.}~\bibnamefont
  {Allen}}\ and\ \bibinfo {author} {\bibfnamefont {J.~D.}\ \bibnamefont
  {Romano}},\ }\href@noop {} {\bibfield  {journal} {\bibinfo  {journal} {Phys.
  Rev. D}\ }\textbf {\bibinfo {volume} {59}},\ \bibinfo {pages} {102001}
  (\bibinfo {year} {1999})}\BibitemShut {NoStop}%
\bibitem [{\citenamefont {Maggiore}(2000)}]{Maggiore2000}%
  \BibitemOpen
  \bibfield  {author} {\bibinfo {author} {\bibfnamefont {M.}~\bibnamefont
  {Maggiore}},\ }\href@noop {} {\bibfield  {journal} {\bibinfo  {journal}
  {Phys. Reports}\ }\textbf {\bibinfo {volume} {331}},\ \bibinfo {pages} {283}
  (\bibinfo {year} {2000})}\BibitemShut {NoStop}%
\bibitem [{\citenamefont {Lattimer}\ and\ \citenamefont
  {Prakash}(2004)}]{LattimerPrakash2004}%
  \BibitemOpen
  \bibfield  {author} {\bibinfo {author} {\bibfnamefont {J.~M.}\ \bibnamefont
  {Lattimer}}\ and\ \bibinfo {author} {\bibfnamefont {M.}~\bibnamefont
  {Prakash}},\ }\href@noop {} {\bibfield  {journal} {\bibinfo  {journal}
  {Science}\ }\textbf {\bibinfo {volume} {304}},\ \bibinfo {pages} {536}
  (\bibinfo {year} {2004})}\BibitemShut {NoStop}%
\bibitem [{\citenamefont {Deutsch}(1955)}]{Deutsch1955}%
  \BibitemOpen
  \bibfield  {author} {\bibinfo {author} {\bibfnamefont {A.~J.}\ \bibnamefont
  {Deutsch}},\ }\href@noop {} {\bibfield  {journal} {\bibinfo  {journal}
  {Annales d'Astrophysique}\ }\textbf {\bibinfo {volume} {18}},\ \bibinfo
  {pages} {1} (\bibinfo {year} {1955})}\BibitemShut {NoStop}%
\bibitem [{\citenamefont {Pacini}(1968)}]{Pacini1968}%
  \BibitemOpen
  \bibfield  {author} {\bibinfo {author} {\bibfnamefont {F.}~\bibnamefont
  {Pacini}},\ }\href@noop {} {\bibfield  {journal} {\bibinfo  {journal}
  {Nature}\ }\textbf {\bibinfo {volume} {219}},\ \bibinfo {pages} {145}
  (\bibinfo {year} {1968})}\BibitemShut {NoStop}%
\bibitem [{\citenamefont {Hawking}\ and\ \citenamefont
  {Israel}(1987)}]{HawkingIsrael1987}%
  \BibitemOpen
  \bibfield  {author} {\bibinfo {author} {\bibfnamefont {S.~W.}\ \bibnamefont
  {Hawking}}\ and\ \bibinfo {author} {\bibfnamefont {W.}~\bibnamefont
  {Israel}},\ }\href@noop {} {\emph {\bibinfo {title} {300 Years of
  Gravitation}}}\ (\bibinfo  {publisher} {Cambridge University Press},\
  \bibinfo {address} {Cambridge, UK},\ \bibinfo {year} {1987})\BibitemShut
  {NoStop}%
\bibitem [{\citenamefont {Prix}(2009)}]{Prix2009}%
  \BibitemOpen
  \bibfield  {author} {\bibinfo {author} {\bibfnamefont {R.}~\bibnamefont
  {Prix}},\ }in\ \href@noop {} {\emph {\bibinfo {booktitle} {Neutron Stars and
  Pulsars}}},\ \bibinfo {series} {Astrophysics and Space Science Library},
  Vol.\ \bibinfo {volume} {357},\ \bibinfo {editor} {edited by\ \bibinfo
  {editor} {\bibfnamefont {W.}~\bibnamefont {Becker}}}\ (\bibinfo  {publisher}
  {Springer Berlin Heidelberg},\ \bibinfo {year} {2009})\ pp.\ \bibinfo {pages}
  {651--685},\ \bibinfo {note} {10.1007/978-3-540-76965-1\_24}\BibitemShut
  {NoStop}%
\bibitem [{\citenamefont {Lorimer}\ and\ \citenamefont
  {Kramer}(2005)}]{LorimerKramer2005}%
  \BibitemOpen
  \bibfield  {author} {\bibinfo {author} {\bibfnamefont {D.}~\bibnamefont
  {Lorimer}}\ and\ \bibinfo {author} {\bibfnamefont {M.}~\bibnamefont
  {Kramer}},\ }\href@noop {} {\emph {\bibinfo {title} {Handbook of Pulsar
  Astronomy}}}\ (\bibinfo  {publisher} {Cambridge University Press},\ \bibinfo
  {address} {Cambridge, UK},\ \bibinfo {year} {2005})\BibitemShut {NoStop}%
\bibitem [{\citenamefont {Lorimer}(2008)}]{Lorimer2008}%
  \BibitemOpen
  \bibfield  {author} {\bibinfo {author} {\bibfnamefont {D.}~\bibnamefont
  {Lorimer}},\ }\href@noop {} {\bibfield  {journal} {\bibinfo  {journal}
  {Living Rev. Relativity}\ }\textbf {\bibinfo {volume} {11}},\ \bibinfo
  {pages} {8} (\bibinfo {year} {2008})}\BibitemShut {NoStop}%
\bibitem [{\citenamefont {Thompson}\ and\ \citenamefont
  {Duncan}(1993)}]{ThompsonDuncan1993}%
  \BibitemOpen
  \bibfield  {author} {\bibinfo {author} {\bibfnamefont {C.}~\bibnamefont
  {Thompson}}\ and\ \bibinfo {author} {\bibfnamefont {R.~C.}\ \bibnamefont
  {Duncan}},\ }\href@noop {} {\bibfield  {journal} {\bibinfo  {journal}
  {Astrophys. J.}\ }\textbf {\bibinfo {volume} {408}},\ \bibinfo {pages} {194}
  (\bibinfo {year} {1993})}\BibitemShut {NoStop}%
\bibitem [{\citenamefont {Thompson}\ and\ \citenamefont
  {Duncan}(1995)}]{ThompsonDuncan1995}%
  \BibitemOpen
  \bibfield  {author} {\bibinfo {author} {\bibfnamefont {C.}~\bibnamefont
  {Thompson}}\ and\ \bibinfo {author} {\bibfnamefont {R.~C.}\ \bibnamefont
  {Duncan}},\ }\href@noop {} {\bibfield  {journal} {\bibinfo  {journal} {Mon.
  Not. R. Astron. Soc.}\ }\textbf {\bibinfo {volume} {275}},\ \bibinfo {pages}
  {255} (\bibinfo {year} {1995})}\BibitemShut {NoStop}%
\bibitem [{\citenamefont {Thompson}\ \emph {et~al.}(2002)\citenamefont
  {Thompson}, \citenamefont {Lyutikov},\ and\ \citenamefont
  {Kulkarni}}]{ThompsonEtAl2002}%
  \BibitemOpen
  \bibfield  {author} {\bibinfo {author} {\bibfnamefont {C.}~\bibnamefont
  {Thompson}}, \bibinfo {author} {\bibfnamefont {M.}~\bibnamefont {Lyutikov}},
  \ and\ \bibinfo {author} {\bibfnamefont {S.~R.}\ \bibnamefont {Kulkarni}},\
  }\href@noop {} {\bibfield  {journal} {\bibinfo  {journal} {Astrophys. J}\
  }\textbf {\bibinfo {volume} {574}},\ \bibinfo {pages} {332} (\bibinfo {year}
  {2002})}\BibitemShut {NoStop}%
\bibitem [{\citenamefont {Harding}\ and\ \citenamefont
  {Lai}(2006)}]{HardingLai2006}%
  \BibitemOpen
  \bibfield  {author} {\bibinfo {author} {\bibfnamefont {A.~K.}\ \bibnamefont
  {Harding}}\ and\ \bibinfo {author} {\bibfnamefont {D.}~\bibnamefont {Lai}},\
  }\href@noop {} {\bibfield  {journal} {\bibinfo  {journal} {Rept.Prog.Phys.}\
  }\textbf {\bibinfo {volume} {69}},\ \bibinfo {pages} {2631} (\bibinfo {year}
  {2006})}\BibitemShut {NoStop}%
\bibitem [{\citenamefont {Cutler}(2002)}]{Cutler2002}%
  \BibitemOpen
  \bibfield  {author} {\bibinfo {author} {\bibfnamefont {C.}~\bibnamefont
  {Cutler}},\ }\href@noop {} {\bibfield  {journal} {\bibinfo  {journal} {Phys.
  Rev. D}\ }\textbf {\bibinfo {volume} {66}},\ \bibinfo {pages} {084025}
  (\bibinfo {year} {2002})}\BibitemShut {NoStop}%
\bibitem [{\citenamefont {Palomba}(2005)}]{Palomba2005}%
  \BibitemOpen
  \bibfield  {author} {\bibinfo {author} {\bibfnamefont {C.}~\bibnamefont
  {Palomba}},\ }\href@noop {} {\bibfield  {journal} {\bibinfo  {journal} {Mon.
  Not. R. Astron. Soc.}\ }\textbf {\bibinfo {volume} {359}},\ \bibinfo {pages}
  {1150} (\bibinfo {year} {2005})}\BibitemShut {NoStop}%
\bibitem [{\citenamefont {Knispel}\ and\ \citenamefont
  {Allen}(2008)}]{KnispelAllen2008}%
  \BibitemOpen
  \bibfield  {author} {\bibinfo {author} {\bibfnamefont {B.}~\bibnamefont
  {Knispel}}\ and\ \bibinfo {author} {\bibfnamefont {B.}~\bibnamefont
  {Allen}},\ }\href@noop {} {\bibfield  {journal} {\bibinfo  {journal} {Phys.
  Rev. D}\ }\textbf {\bibinfo {volume} {78}},\ \bibinfo {pages} {044031}
  (\bibinfo {year} {2008})}\BibitemShut {NoStop}%
\bibitem [{\citenamefont {Popov}\ \emph {et~al.}(2000)\citenamefont {Popov}
  \emph {et~al.}}]{PopovEtAl2000}%
  \BibitemOpen
  \bibfield  {author} {\bibinfo {author} {\bibfnamefont {S.~B.}\ \bibnamefont
  {Popov}} \emph {et~al.},\ }\href@noop {} {\bibfield  {journal} {\bibinfo
  {journal} {Astrophys. J}\ }\textbf {\bibinfo {volume} {530}},\ \bibinfo
  {pages} {896} (\bibinfo {year} {2000})}\BibitemShut {NoStop}%
\bibitem [{\citenamefont {Rosado}(2011)}]{Rosado2011}%
  \BibitemOpen
  \bibfield  {author} {\bibinfo {author} {\bibfnamefont {P.~A.}\ \bibnamefont
  {Rosado}},\ }\href@noop {} {\bibfield  {journal} {\bibinfo  {journal} {Phys.
  Rev. D}\ }\textbf {\bibinfo {volume} {84}},\ \bibinfo {pages} {084004}
  (\bibinfo {year} {2011})}\BibitemShut {NoStop}%
\bibitem [{\citenamefont {Owen}\ \emph {et~al.}(1998)\citenamefont {Owen} \emph
  {et~al.}}]{OwenEtAl1998}%
  \BibitemOpen
  \bibfield  {author} {\bibinfo {author} {\bibfnamefont {B.~J.}\ \bibnamefont
  {Owen}} \emph {et~al.},\ }\href@noop {} {\bibfield  {journal} {\bibinfo
  {journal} {Phys. Rev. D}\ }\textbf {\bibinfo {volume} {58}},\ \bibinfo
  {pages} {084020} (\bibinfo {year} {1998})}\BibitemShut {NoStop}%
\bibitem [{\citenamefont {Zhu}\ \emph {et~al.}(2011)\citenamefont {Zhu},
  \citenamefont {Fan},\ and\ \citenamefont {Zhu}}]{ZhuEtAl2011b}%
  \BibitemOpen
  \bibfield  {author} {\bibinfo {author} {\bibfnamefont {X.-J.}\ \bibnamefont
  {Zhu}}, \bibinfo {author} {\bibfnamefont {X.-L.}\ \bibnamefont {Fan}}, \ and\
  \bibinfo {author} {\bibfnamefont {Z.-H.}\ \bibnamefont {Zhu}},\ }\href@noop
  {} {\bibfield  {journal} {\bibinfo  {journal} {Astrophys. J}\ }\textbf
  {\bibinfo {volume} {729}},\ \bibinfo {pages} {59} (\bibinfo {year}
  {2011})}\BibitemShut {NoStop}%
\bibitem [{\citenamefont {Barack}\ and\ \citenamefont
  {Cutler}(2004)}]{BarackCutler2004}%
  \BibitemOpen
  \bibfield  {author} {\bibinfo {author} {\bibfnamefont {L.}~\bibnamefont
  {Barack}}\ and\ \bibinfo {author} {\bibfnamefont {C.}~\bibnamefont
  {Cutler}},\ }\href@noop {} {\bibfield  {journal} {\bibinfo  {journal} {Phys.
  Rev. D}\ }\textbf {\bibinfo {volume} {70}},\ \bibinfo {pages} {122002}
  (\bibinfo {year} {2004})}\BibitemShut {NoStop}%
\bibitem [{\citenamefont {Amaro-Seoane}\ \emph {et~al.}(2007)\citenamefont
  {Amaro-Seoane} \emph {et~al.}}]{AmaroEtAl2007}%
  \BibitemOpen
  \bibfield  {author} {\bibinfo {author} {\bibfnamefont {P.}~\bibnamefont
  {Amaro-Seoane}} \emph {et~al.},\ }\href@noop {} {\bibfield  {journal}
  {\bibinfo  {journal} {Clas. Quantum Grav.}\ }\textbf {\bibinfo {volume}
  {24}},\ \bibinfo {pages} {R113} (\bibinfo {year} {2007})}\BibitemShut
  {NoStop}%
\bibitem [{\citenamefont {Buonanno}\ \emph {et~al.}(2005)\citenamefont
  {Buonanno} \emph {et~al.}}]{BuonannoEtAl2005}%
  \BibitemOpen
  \bibfield  {author} {\bibinfo {author} {\bibfnamefont {A.}~\bibnamefont
  {Buonanno}} \emph {et~al.},\ }\href@noop {} {\bibfield  {journal} {\bibinfo
  {journal} {Phys. Rev. D}\ }\textbf {\bibinfo {volume} {72}},\ \bibinfo
  {pages} {084001} (\bibinfo {year} {2005})}\BibitemShut {NoStop}%
\bibitem [{\citenamefont {Marassi}\ \emph {et~al.}(2009)\citenamefont
  {Marassi}, \citenamefont {Schneider},\ and\ \citenamefont
  {Ferrari}}]{MarassiEtAl2009}%
  \BibitemOpen
  \bibfield  {author} {\bibinfo {author} {\bibfnamefont {S.}~\bibnamefont
  {Marassi}}, \bibinfo {author} {\bibfnamefont {R.}~\bibnamefont {Schneider}},
  \ and\ \bibinfo {author} {\bibfnamefont {V.}~\bibnamefont {Ferrari}},\
  }\href@noop {} {\bibfield  {journal} {\bibinfo  {journal} {Mon. Not. R.
  Astron. Soc.}\ }\textbf {\bibinfo {volume} {398}},\ \bibinfo {pages} {293}
  (\bibinfo {year} {2009})}\BibitemShut {NoStop}%
\bibitem [{\citenamefont {Kowalska}\ \emph {et~al.}(2012)\citenamefont
  {Kowalska}, \citenamefont {Bulik},\ and\ \citenamefont
  {Belczynski}}]{KowalskaEtAl2012}%
  \BibitemOpen
  \bibfield  {author} {\bibinfo {author} {\bibfnamefont {I.}~\bibnamefont
  {Kowalska}}, \bibinfo {author} {\bibfnamefont {T.}~\bibnamefont {Bulik}}, \
  and\ \bibinfo {author} {\bibfnamefont {K.}~\bibnamefont {Belczynski}},\
  }\href@noop {} {\bibfield  {journal} {\bibinfo  {journal}
  {Astron.Astrophys.}\ }\textbf {\bibinfo {volume} {541}},\ \bibinfo {pages}
  {A120} (\bibinfo {year} {2012})}\BibitemShut {NoStop}%
\bibitem [{\citenamefont {Allen}(1996)}]{Allen1996}%
  \BibitemOpen
  \bibfield  {author} {\bibinfo {author} {\bibfnamefont {B.}~\bibnamefont
  {Allen}},\ }\href@noop {} {\bibfield  {journal} {\bibinfo  {journal} {arXiv:
  gr-qc/9604033v3}\ } (\bibinfo {year} {1996})}\BibitemShut {NoStop}%
\bibitem [{\citenamefont {Buonanno}(2004)}]{Buonanno2004}%
  \BibitemOpen
  \bibfield  {author} {\bibinfo {author} {\bibfnamefont {A.}~\bibnamefont
  {Buonanno}},\ }\href@noop {} {\bibfield  {journal} {\bibinfo  {journal}
  {arXiv: gr-qc/0303085v2}\ } (\bibinfo {year} {2004})}\BibitemShut {NoStop}%
\bibitem [{\citenamefont {Cutler}\ and\ \citenamefont
  {Harms}(2006)}]{CutlerHarms2006}%
  \BibitemOpen
  \bibfield  {author} {\bibinfo {author} {\bibfnamefont {C.}~\bibnamefont
  {Cutler}}\ and\ \bibinfo {author} {\bibfnamefont {J.}~\bibnamefont {Harms}},\
  }\href@noop {} {\bibfield  {journal} {\bibinfo  {journal} {Phys. Rev. D}\
  }\textbf {\bibinfo {volume} {73}},\ \bibinfo {pages} {042001} (\bibinfo
  {year} {2006})}\BibitemShut {NoStop}%
\bibitem [{\citenamefont {Cutler}\ and\ \citenamefont
  {Holz}(2009)}]{CutlerHolz2009}%
  \BibitemOpen
  \bibfield  {author} {\bibinfo {author} {\bibfnamefont {C.}~\bibnamefont
  {Cutler}}\ and\ \bibinfo {author} {\bibfnamefont {D.~E.}\ \bibnamefont
  {Holz}},\ }\href@noop {} {\bibfield  {journal} {\bibinfo  {journal} {Phys.
  Rev. D}\ }\textbf {\bibinfo {volume} {80}},\ \bibinfo {pages} {104009}
  (\bibinfo {year} {2009})}\BibitemShut {NoStop}%
\bibitem [{\citenamefont {Yagi}\ and\ \citenamefont
  {Seto}(2011)}]{YagiSeto2011}%
  \BibitemOpen
  \bibfield  {author} {\bibinfo {author} {\bibfnamefont {K.}~\bibnamefont
  {Yagi}}\ and\ \bibinfo {author} {\bibfnamefont {N.}~\bibnamefont {Seto}},\
  }\href@noop {} {\bibfield  {journal} {\bibinfo  {journal} {Phys. Rev. D}\
  }\textbf {\bibinfo {volume} {83}},\ \bibinfo {pages} {044011} (\bibinfo
  {year} {2011})}\BibitemShut {NoStop}%
\bibitem [{\citenamefont {Regimbau}\ and\ \citenamefont
  {de~Freitas~Pacheco}(2001)}]{RegimbauFreitas2001}%
  \BibitemOpen
  \bibfield  {author} {\bibinfo {author} {\bibfnamefont {T.}~\bibnamefont
  {Regimbau}}\ and\ \bibinfo {author} {\bibfnamefont {J.~A.}\ \bibnamefont
  {de~Freitas~Pacheco}},\ }\href@noop {} {\bibfield  {journal} {\bibinfo
  {journal} {Astron. Astrophys.}\ }\textbf {\bibinfo {volume} {376}},\ \bibinfo
  {pages} {381} (\bibinfo {year} {2001})}\BibitemShut {NoStop}%
\bibitem [{\citenamefont {Regimbau}\ and\ \citenamefont
  {de~Freitas~Pacheco}(2006)}]{RegimbauFreitas2006}%
  \BibitemOpen
  \bibfield  {author} {\bibinfo {author} {\bibfnamefont {T.}~\bibnamefont
  {Regimbau}}\ and\ \bibinfo {author} {\bibfnamefont {J.~A.}\ \bibnamefont
  {de~Freitas~Pacheco}},\ }\href@noop {} {\bibfield  {journal} {\bibinfo
  {journal} {Astrophys. J}\ }\textbf {\bibinfo {volume} {642}},\ \bibinfo
  {pages} {455} (\bibinfo {year} {2006})}\BibitemShut {NoStop}%
\bibitem [{\citenamefont {Regimbau}\ and\ \citenamefont
  {Mandic}(2008)}]{RegimbauMandic2008}%
  \BibitemOpen
  \bibfield  {author} {\bibinfo {author} {\bibfnamefont {T.}~\bibnamefont
  {Regimbau}}\ and\ \bibinfo {author} {\bibfnamefont {V.}~\bibnamefont
  {Mandic}},\ }\href@noop {} {\bibfield  {journal} {\bibinfo  {journal} {Clas.
  Quantum Grav.}\ }\textbf {\bibinfo {volume} {25}},\ \bibinfo {pages} {184018}
  (\bibinfo {year} {2008})}\BibitemShut {NoStop}%
\bibitem [{\citenamefont {Marassi}\ \emph {et~al.}(2011)\citenamefont {Marassi}
  \emph {et~al.}}]{MarassiEtAl2011b}%
  \BibitemOpen
  \bibfield  {author} {\bibinfo {author} {\bibfnamefont {S.}~\bibnamefont
  {Marassi}} \emph {et~al.},\ }\href@noop {} {\bibfield  {journal} {\bibinfo
  {journal} {Mon. Not. R. Astron. Soc.}\ }\textbf {\bibinfo {volume} {411}},\
  \bibinfo {pages} {2549} (\bibinfo {year} {2011})}\BibitemShut {NoStop}%
\bibitem [{\citenamefont {Faucher-Giguere}\ and\ \citenamefont
  {Kaspi}(2006)}]{FaucherKaspi2006}%
  \BibitemOpen
  \bibfield  {author} {\bibinfo {author} {\bibfnamefont {C.-A.}\ \bibnamefont
  {Faucher-Giguere}}\ and\ \bibinfo {author} {\bibfnamefont {V.~M.}\
  \bibnamefont {Kaspi}},\ }\href@noop {} {\bibfield  {journal} {\bibinfo
  {journal} {Astrophys. J.}\ }\textbf {\bibinfo {volume} {643}},\ \bibinfo
  {pages} {332} (\bibinfo {year} {2006})}\BibitemShut {NoStop}%
\bibitem [{\citenamefont {Punturo}\ \emph {et~al.}(2010)\citenamefont {Punturo}
  \emph {et~al.}}]{PunturoEtAl2010}%
  \BibitemOpen
  \bibfield  {author} {\bibinfo {author} {\bibfnamefont {M.}~\bibnamefont
  {Punturo}} \emph {et~al.},\ }\href@noop {} {\bibfield  {journal} {\bibinfo
  {journal} {Clas. Quantum Grav.}\ }\textbf {\bibinfo {volume} {27}},\ \bibinfo
  {pages} {194002} (\bibinfo {year} {2010})}\BibitemShut {NoStop}%
\bibitem [{\citenamefont {Grote}(2010)}]{Grote2010}%
  \BibitemOpen
  \bibfield  {author} {\bibinfo {author} {\bibfnamefont {H.}~\bibnamefont
  {Grote}},\ }\href@noop {} {\bibfield  {journal} {\bibinfo  {journal} {Clas.
  Quantum Grav.}\ }\textbf {\bibinfo {volume} {27}},\ \bibinfo {pages} {084003}
  (\bibinfo {year} {2010})}\BibitemShut {NoStop}%
\bibitem [{\citenamefont {Accadia}\ and\ \citenamefont
  {Swinkels}(2010)}]{AccadiaSwinkels2010}%
  \BibitemOpen
  \bibfield  {author} {\bibinfo {author} {\bibfnamefont {T.}~\bibnamefont
  {Accadia}}\ and\ \bibinfo {author} {\bibfnamefont {B.~L.}\ \bibnamefont
  {Swinkels}},\ }\href@noop {} {\bibfield  {journal} {\bibinfo  {journal}
  {Clas. Quantum Grav.}\ }\textbf {\bibinfo {volume} {27}},\ \bibinfo {pages}
  {084002} (\bibinfo {year} {2010})}\BibitemShut {NoStop}%
\bibitem [{\citenamefont {{LIGO Scientific Collaboration}}(2009)}]{LIGO2009}%
  \BibitemOpen
  \bibfield  {author} {\bibinfo {author} {\bibnamefont {{LIGO Scientific
  Collaboration}}},\ }\href@noop {} {\bibfield  {journal} {\bibinfo  {journal}
  {RPP}\ }\textbf {\bibinfo {volume} {72}},\ \bibinfo {pages} {076901}
  (\bibinfo {year} {2009})}\BibitemShut {NoStop}%
\bibitem [{\citenamefont {Harry}(2010)}]{Harry2010}%
  \BibitemOpen
  \bibfield  {author} {\bibinfo {author} {\bibfnamefont {G.~M.}\ \bibnamefont
  {Harry}},\ }\href@noop {} {\bibfield  {journal} {\bibinfo  {journal} {Clas.
  Quantum Grav.}\ }\textbf {\bibinfo {volume} {27}},\ \bibinfo {pages} {084006}
  (\bibinfo {year} {2010})}\BibitemShut {NoStop}%
\bibitem [{\citenamefont {Kawamura}\ \emph {et~al.}(2011)\citenamefont
  {Kawamura} \emph {et~al.}}]{KawamuraEtAl2011}%
  \BibitemOpen
  \bibfield  {author} {\bibinfo {author} {\bibfnamefont {S.}~\bibnamefont
  {Kawamura}} \emph {et~al.},\ }\href@noop {} {\bibfield  {journal} {\bibinfo
  {journal} {Clas. Quantum Grav.}\ }\textbf {\bibinfo {volume} {28}},\ \bibinfo
  {pages} {094011} (\bibinfo {year} {2011})}\BibitemShut {NoStop}%
\bibitem [{\citenamefont {Riess}\ \emph {et~al.}(2009)\citenamefont {Riess}
  \emph {et~al.}}]{RiessEtAl2009}%
  \BibitemOpen
  \bibfield  {author} {\bibinfo {author} {\bibfnamefont {A.~G.}\ \bibnamefont
  {Riess}} \emph {et~al.},\ }\href@noop {} {\bibfield  {journal} {\bibinfo
  {journal} {Astrophys. J}\ }\textbf {\bibinfo {volume} {699}},\ \bibinfo
  {pages} {539} (\bibinfo {year} {2009})}\BibitemShut {NoStop}%
\bibitem [{\citenamefont {Riess}\ \emph {et~al.}(2011)\citenamefont {Riess}
  \emph {et~al.}}]{RiessEtAl2011}%
  \BibitemOpen
  \bibfield  {author} {\bibinfo {author} {\bibfnamefont {A.~G.}\ \bibnamefont
  {Riess}} \emph {et~al.},\ }\href@noop {} {\bibfield  {journal} {\bibinfo
  {journal} {Astrophys. J}\ }\textbf {\bibinfo {volume} {730}},\ \bibinfo
  {pages} {119} (\bibinfo {year} {2011})}\BibitemShut {NoStop}%
\bibitem [{\citenamefont {Jarosik}\ \emph {et~al.}(2011)\citenamefont {Jarosik}
  \emph {et~al.}}]{JarosikEtAl2011}%
  \BibitemOpen
  \bibfield  {author} {\bibinfo {author} {\bibfnamefont {N.}~\bibnamefont
  {Jarosik}} \emph {et~al.},\ }\href@noop {} {\bibfield  {journal} {\bibinfo
  {journal} {Astrophys. J. Suppl. Ser.}\ }\textbf {\bibinfo {volume} {192}},\
  \bibinfo {pages} {14} (\bibinfo {year} {2011})}\BibitemShut {NoStop}%
\bibitem [{\citenamefont {Phinney}(2001)}]{Phinney2001}%
  \BibitemOpen
  \bibfield  {author} {\bibinfo {author} {\bibfnamefont {E.~S.}\ \bibnamefont
  {Phinney}},\ }\href@noop {} {\bibfield  {journal} {\bibinfo  {journal}
  {arXiv: astro-ph/0108028v1}\ } (\bibinfo {year} {2001})}\BibitemShut
  {NoStop}%
\bibitem [{Note1()}]{Note1}%
  \BibitemOpen
  \bibinfo {note} {A more thorough definition of the resolvability, that takes
  into account the difference in amplitude of the signals, can be the subject
  of a future work.}\BibitemShut {Stop}%
\bibitem [{\citenamefont {Regimbau}\ \emph {et~al.}(2012)\citenamefont
  {Regimbau} \emph {et~al.}}]{RegimbauEtAl2012}%
  \BibitemOpen
  \bibfield  {author} {\bibinfo {author} {\bibfnamefont {T.}~\bibnamefont
  {Regimbau}} \emph {et~al.},\ }\href@noop {} {\bibfield  {journal} {\bibinfo
  {journal} {arXiv: gr-qc/1201.3563v1}\ } (\bibinfo {year} {2012})}\BibitemShut
  {NoStop}%
\bibitem [{\citenamefont {Maggiore}(2008)}]{Maggiore2008}%
  \BibitemOpen
  \bibfield  {author} {\bibinfo {author} {\bibfnamefont {M.}~\bibnamefont
  {Maggiore}},\ }\href@noop {} {\emph {\bibinfo {title} {Gravitational Waves
  Volume 1: Theory and Experiments}}}\ (\bibinfo  {publisher} {Oxford
  University Press},\ \bibinfo {address} {New York, USA},\ \bibinfo {year}
  {2008})\BibitemShut {NoStop}%
\bibitem [{\citenamefont {Schutz}(2011)}]{Schutz2011}%
  \BibitemOpen
  \bibfield  {author} {\bibinfo {author} {\bibfnamefont {B.~F.}\ \bibnamefont
  {Schutz}},\ }\href {http://arxiv.org/abs/1102.5421} {\bibfield  {journal}
  {\bibinfo  {journal} {Clas. Quantum Grav.}\ }\textbf {\bibinfo {volume}
  {28}},\ \bibinfo {pages} {125023} (\bibinfo {year} {2011})}\BibitemShut
  {NoStop}%
\bibitem [{\citenamefont {Finn}\ \emph {et~al.}(2009)\citenamefont {Finn},
  \citenamefont {Larson},\ and\ \citenamefont {Romano}}]{FinnEtAl2009}%
  \BibitemOpen
  \bibfield  {author} {\bibinfo {author} {\bibfnamefont {L.~S.}\ \bibnamefont
  {Finn}}, \bibinfo {author} {\bibfnamefont {S.~L.}\ \bibnamefont {Larson}}, \
  and\ \bibinfo {author} {\bibfnamefont {J.~D.}\ \bibnamefont {Romano}},\
  }\href@noop {} {\bibfield  {journal} {\bibinfo  {journal} {Phys. Rev. D}\
  }\textbf {\bibinfo {volume} {79}},\ \bibinfo {pages} {062003} (\bibinfo
  {year} {2009})}\BibitemShut {NoStop}%
\bibitem [{\citenamefont {{LIGO Scientific Collaboration}}(2010)}]{LIGO2010e}%
  \BibitemOpen
  \bibfield  {author} {\bibinfo {author} {\bibnamefont {{LIGO Scientific
  Collaboration}}},\ }\href@noop {} {\emph {\bibinfo {title} {Advanced LIGO
  anticipated sensitivity curves}}},\ \bibinfo {type} {Tech. Rep.}\ \bibinfo
  {number} {LIGO-T1000414-v13}\ (\bibinfo  {institution} {LIGO Scientific
  Collaboration},\ \bibinfo {year} {2010})\BibitemShut {NoStop}%
\bibitem [{Note2()}]{Note2}%
  \BibitemOpen
  \bibinfo {note} {The spectral strain sensitivities of ETB and ETD were kindly
  provided by Tania Regimbau in a private communication.}\BibitemShut {Stop}%
\bibitem [{\citenamefont {Nishizawa}\ \emph {et~al.}(2012)\citenamefont
  {Nishizawa} \emph {et~al.}}]{NishizawaEtAl2012}%
  \BibitemOpen
  \bibfield  {author} {\bibinfo {author} {\bibfnamefont {A.}~\bibnamefont
  {Nishizawa}} \emph {et~al.},\ }\href@noop {} {\bibfield  {journal} {\bibinfo
  {journal} {Phys. Rev. D}\ }\textbf {\bibinfo {volume} {85}},\ \bibinfo
  {pages} {044047} (\bibinfo {year} {2012})}\BibitemShut {NoStop}%
\bibitem [{\citenamefont {Peebles}(1993)}]{Peebles1993}%
  \BibitemOpen
  \bibfield  {author} {\bibinfo {author} {\bibfnamefont {P.~J.~E.}\
  \bibnamefont {Peebles}},\ }\href@noop {} {\emph {\bibinfo {title} {Principles
  of Physical Cosmology}}},\ Princeton series in physics\ (\bibinfo
  {publisher} {Princeton University Press},\ \bibinfo {address} {Princeton,
  USA},\ \bibinfo {year} {1993})\BibitemShut {NoStop}%
\bibitem [{Note3()}]{Note3}%
  \BibitemOpen
  \bibinfo {note} {By imposing $\xi =0$, the function $\protect \mathcal
  {T}(z)$ gives the age of the universe at the instant when the waves of
  redshift $z$ were emitted, as in Equation (13.20) of \cite
  {Peebles1993}.}\BibitemShut {Stop}%
\bibitem [{\citenamefont {Knispel}(2011)}]{Knispel2011}%
  \BibitemOpen
  \bibfield  {author} {\bibinfo {author} {\bibfnamefont {B.}~\bibnamefont
  {Knispel}},\ }\emph {\bibinfo {title} {Pulsar Discoveries by Volunteer
  Distributed Computing}},\ \href@noop {} {Ph.D. thesis},\ \bibinfo  {school}
  {Leibniz Universit{\"a}t Hannover} (\bibinfo {year} {2011})\BibitemShut
  {NoStop}%
\bibitem [{Note4()}]{Note4}%
  \BibitemOpen
  \bibinfo {note} {One can prove it by partially differentiating Equation (\ref
  {eq:finidef}) with respect to $t_e$.}\BibitemShut {Stop}%
\bibitem [{\citenamefont {Regimbau}(2011)}]{Regimbau2011}%
  \BibitemOpen
  \bibfield  {author} {\bibinfo {author} {\bibfnamefont {T.}~\bibnamefont
  {Regimbau}},\ }\href@noop {} {\bibfield  {journal} {\bibinfo  {journal}
  {Research in Astron. Astrophys.}\ }\textbf {\bibinfo {volume} {11}},\
  \bibinfo {pages} {369} (\bibinfo {year} {2011})}\BibitemShut {NoStop}%
\bibitem [{Note5()}]{Note5}%
  \BibitemOpen
  \bibinfo {note} {In this regard, one should read Section \ref
  {sec:minmaxfreq}; the plots in this section, for instance the ones in Figure
  (\ref {fig:zfplotmd}), can be qualitatively compared with Figures (2) and (4)
  of \cite {Rosado2011}.}\BibitemShut {Stop}%
\bibitem [{\citenamefont {Lattimer}(2010)}]{Lattimer2010}%
  \BibitemOpen
  \bibfield  {author} {\bibinfo {author} {\bibfnamefont {J.~M.}\ \bibnamefont
  {Lattimer}},\ }\href@noop {} {\bibfield  {journal} {\bibinfo  {journal} {New
  Astronomy Reviews}\ }\textbf {\bibinfo {volume} {54}},\ \bibinfo {pages}
  {101} (\bibinfo {year} {2010})}\BibitemShut {NoStop}%
\bibitem [{\citenamefont {Tr{\"u}mper}(2011)}]{Truemper2011}%
  \BibitemOpen
  \bibfield  {author} {\bibinfo {author} {\bibfnamefont {J.~E.}\ \bibnamefont
  {Tr{\"u}mper}},\ }\href@noop {} {\bibfield  {journal} {\bibinfo  {journal}
  {Progress in Particle and Nuclear Physics}\ }\textbf {\bibinfo {volume}
  {66}},\ \bibinfo {pages} {674} (\bibinfo {year} {2011})}\BibitemShut
  {NoStop}%
\bibitem [{\citenamefont {Brown}(2000)}]{Brown2000}%
  \BibitemOpen
  \bibfield  {author} {\bibinfo {author} {\bibfnamefont {J.~D.}\ \bibnamefont
  {Brown}},\ }\href@noop {} {\bibfield  {journal} {\bibinfo  {journal}
  {Phys.Rev.D}\ }\textbf {\bibinfo {volume} {62}},\ \bibinfo {pages} {084024}
  (\bibinfo {year} {2000})}\BibitemShut {NoStop}%
\bibitem [{\citenamefont {Lorimer}(2011)}]{Lorimer2011}%
  \BibitemOpen
  \bibfield  {author} {\bibinfo {author} {\bibfnamefont {D.~R.}\ \bibnamefont
  {Lorimer}},\ }\href {http://arxiv.org/abs/1008.1928} {\emph {\bibinfo {title}
  {High-Energy Emission from Pulsars and their Systems}}},\ edited by\ \bibinfo
  {editor} {\bibfnamefont {D.}~\bibnamefont {Torres}}\ and\ \bibinfo {editor}
  {\bibfnamefont {N.}~\bibnamefont {Rea}},\ Astrophysics and Space Science
  Proceedings\ (\bibinfo  {publisher} {Springer Verlag},\ \bibinfo {address}
  {Berlin Heidelberg},\ \bibinfo {year} {2011})\ pp.\ \bibinfo {pages}
  {21--36}\BibitemShut {NoStop}%
\bibitem [{\citenamefont {Manchester}\ \emph {et~al.}(2005)\citenamefont
  {Manchester} \emph {et~al.}}]{ManchesterEtAl2005}%
  \BibitemOpen
  \bibfield  {author} {\bibinfo {author} {\bibfnamefont {R.~N.}\ \bibnamefont
  {Manchester}} \emph {et~al.},\ }\href@noop {} {\bibfield  {journal} {\bibinfo
   {journal} {Astron.J.}\ }\textbf {\bibinfo {volume} {129}},\ \bibinfo {pages}
  {1993} (\bibinfo {year} {2005})}\BibitemShut {NoStop}%
\bibitem [{\citenamefont {Backer}\ \emph {et~al.}(1982)\citenamefont {Backer}
  \emph {et~al.}}]{BackerEtAl1982}%
  \BibitemOpen
  \bibfield  {author} {\bibinfo {author} {\bibfnamefont {D.~C.}\ \bibnamefont
  {Backer}} \emph {et~al.},\ }\href@noop {} {\bibfield  {journal} {\bibinfo
  {journal} {Nature}\ }\textbf {\bibinfo {volume} {300}},\ \bibinfo {pages}
  {615} (\bibinfo {year} {1982})}\BibitemShut {NoStop}%
\bibitem [{\citenamefont {Shapiro}\ and\ \citenamefont
  {Teukolsky}(1983)}]{ShapiroTeukolsky1983}%
  \BibitemOpen
  \bibfield  {author} {\bibinfo {author} {\bibfnamefont {S.~L.}\ \bibnamefont
  {Shapiro}}\ and\ \bibinfo {author} {\bibfnamefont {S.~A.}\ \bibnamefont
  {Teukolsky}},\ }\href@noop {} {\emph {\bibinfo {title} {Black Holes, White
  Dwarfs, and Neutron Stars: The Physics of Compact Objects}}},\ Physics
  textbook\ (\bibinfo  {publisher} {Wiley},\ \bibinfo {address} {New York,
  USA},\ \bibinfo {year} {1983})\BibitemShut {NoStop}%
\bibitem [{\citenamefont {Madau}\ \emph {et~al.}(1998)\citenamefont {Madau},
  \citenamefont {Valle},\ and\ \citenamefont {Panagia}}]{MadauEtAl1998}%
  \BibitemOpen
  \bibfield  {author} {\bibinfo {author} {\bibfnamefont {P.}~\bibnamefont
  {Madau}}, \bibinfo {author} {\bibfnamefont {M.~D.}\ \bibnamefont {Valle}}, \
  and\ \bibinfo {author} {\bibfnamefont {N.}~\bibnamefont {Panagia}},\
  }\href@noop {} {\bibfield  {journal} {\bibinfo  {journal} {Mon. Not. R.
  Astron. Soc.}\ }\textbf {\bibinfo {volume} {297}},\ \bibinfo {pages} {L17}
  (\bibinfo {year} {1998})}\BibitemShut {NoStop}%
\bibitem [{\citenamefont {Porciani}\ and\ \citenamefont
  {Madau}(2001)}]{PorcianiMadau2001}%
  \BibitemOpen
  \bibfield  {author} {\bibinfo {author} {\bibfnamefont {C.}~\bibnamefont
  {Porciani}}\ and\ \bibinfo {author} {\bibfnamefont {P.}~\bibnamefont
  {Madau}},\ }\href@noop {} {\bibfield  {journal} {\bibinfo  {journal}
  {Astrophys. J}\ }\textbf {\bibinfo {volume} {548}},\ \bibinfo {pages} {522}
  (\bibinfo {year} {2001})}\BibitemShut {NoStop}%
\bibitem [{\citenamefont {2dFGRS Team}(2001)}]{2dFGRS2001}%
  \BibitemOpen
  \bibfield  {author} {\bibinfo {author} {\bibnamefont {2dFGRS Team}},\
  }\href@noop {} {\bibfield  {journal} {\bibinfo  {journal} {Mon. Not. R.
  Astron. Soc.}\ }\textbf {\bibinfo {volume} {326}},\ \bibinfo {pages} {255}
  (\bibinfo {year} {2001})}\BibitemShut {NoStop}%
\bibitem [{\citenamefont {Strolger}\ \emph {et~al.}(2004)\citenamefont
  {Strolger} \emph {et~al.}}]{StrolgerEtAl2004}%
  \BibitemOpen
  \bibfield  {author} {\bibinfo {author} {\bibfnamefont {L.~G.}\ \bibnamefont
  {Strolger}} \emph {et~al.},\ }\href@noop {} {\bibfield  {journal} {\bibinfo
  {journal} {Astrophys. J}\ }\textbf {\bibinfo {volume} {613}},\ \bibinfo
  {pages} {200} (\bibinfo {year} {2004})}\BibitemShut {NoStop}%
\bibitem [{\citenamefont {Hopkins}\ and\ \citenamefont
  {Beacom}(2006)}]{HopkinsBeacom2006}%
  \BibitemOpen
  \bibfield  {author} {\bibinfo {author} {\bibfnamefont {A.~M.}\ \bibnamefont
  {Hopkins}}\ and\ \bibinfo {author} {\bibfnamefont {J.~F.}\ \bibnamefont
  {Beacom}},\ }\href@noop {} {\bibfield  {journal} {\bibinfo  {journal}
  {Astrophys. J}\ }\textbf {\bibinfo {volume} {651}},\ \bibinfo {pages} {142}
  (\bibinfo {year} {2006})}\BibitemShut {NoStop}%
\bibitem [{\citenamefont {Nagamine}\ \emph {et~al.}(2006)\citenamefont
  {Nagamine} \emph {et~al.}}]{NagamineEtAl2006}%
  \BibitemOpen
  \bibfield  {author} {\bibinfo {author} {\bibfnamefont {K.}~\bibnamefont
  {Nagamine}} \emph {et~al.},\ }\href@noop {} {\bibfield  {journal} {\bibinfo
  {journal} {Astrophys. J}\ }\textbf {\bibinfo {volume} {653}},\ \bibinfo
  {pages} {881} (\bibinfo {year} {2006})}\BibitemShut {NoStop}%
\bibitem [{\citenamefont {Fardal}\ \emph {et~al.}(2007)\citenamefont {Fardal}
  \emph {et~al.}}]{FardalEtAl2007}%
  \BibitemOpen
  \bibfield  {author} {\bibinfo {author} {\bibfnamefont {M.~A.}\ \bibnamefont
  {Fardal}} \emph {et~al.},\ }\href@noop {} {\bibfield  {journal} {\bibinfo
  {journal} {Mon. Not. R. Astron. Soc.}\ }\textbf {\bibinfo {volume} {379}},\
  \bibinfo {pages} {985} (\bibinfo {year} {2007})}\BibitemShut {NoStop}%
\bibitem [{\citenamefont {Wilkins}\ \emph {et~al.}(2008)\citenamefont
  {Wilkins}, \citenamefont {Trentham},\ and\ \citenamefont
  {Hopkins}}]{WilkinsEtAl2008}%
  \BibitemOpen
  \bibfield  {author} {\bibinfo {author} {\bibfnamefont {S.~M.}\ \bibnamefont
  {Wilkins}}, \bibinfo {author} {\bibfnamefont {N.}~\bibnamefont {Trentham}}, \
  and\ \bibinfo {author} {\bibfnamefont {A.~M.}\ \bibnamefont {Hopkins}},\
  }\href@noop {} {\bibfield  {journal} {\bibinfo  {journal} {Mon. Not. R.
  Astron. Soc.}\ }\textbf {\bibinfo {volume} {385}},\ \bibinfo {pages} {687}
  (\bibinfo {year} {2008})}\BibitemShut {NoStop}%
\bibitem [{\citenamefont {Salpeter}(1955)}]{Salpeter1955}%
  \BibitemOpen
  \bibfield  {author} {\bibinfo {author} {\bibfnamefont {E.~E.}\ \bibnamefont
  {Salpeter}},\ }\href@noop {} {\bibfield  {journal} {\bibinfo  {journal}
  {Astrophys. J}\ }\textbf {\bibinfo {volume} {121}},\ \bibinfo {pages} {161}
  (\bibinfo {year} {1955})}\BibitemShut {NoStop}%
\bibitem [{\citenamefont {Johnston}\ and\ \citenamefont
  {Galloway}(1999)}]{JohnstonGalloway1999}%
  \BibitemOpen
  \bibfield  {author} {\bibinfo {author} {\bibfnamefont {S.}~\bibnamefont
  {Johnston}}\ and\ \bibinfo {author} {\bibfnamefont {D.}~\bibnamefont
  {Galloway}},\ }\href@noop {} {\bibfield  {journal} {\bibinfo  {journal} {Mon.
  Not. R. Astron. Soc.}\ }\textbf {\bibinfo {volume} {306}},\ \bibinfo {pages}
  {L50} (\bibinfo {year} {1999})}\BibitemShut {NoStop}%
\bibitem [{\citenamefont {Arzoumanian}\ \emph {et~al.}(2002)\citenamefont
  {Arzoumanian}, \citenamefont {Chernoff},\ and\ \citenamefont
  {Cordes}}]{ArzoumanianEtAl2002}%
  \BibitemOpen
  \bibfield  {author} {\bibinfo {author} {\bibfnamefont {Z.}~\bibnamefont
  {Arzoumanian}}, \bibinfo {author} {\bibfnamefont {D.~F.}\ \bibnamefont
  {Chernoff}}, \ and\ \bibinfo {author} {\bibfnamefont {J.~M.}\ \bibnamefont
  {Cordes}},\ }\href@noop {} {\bibfield  {journal} {\bibinfo  {journal}
  {Astrophys. J.}\ }\textbf {\bibinfo {volume} {568}},\ \bibinfo {pages} {289}
  (\bibinfo {year} {2002})}\BibitemShut {NoStop}%
\bibitem [{\citenamefont {Regimbau}\ and\ \citenamefont
  {de~Freitas~Pacheco}(2000)}]{RegimbauFreitas2000}%
  \BibitemOpen
  \bibfield  {author} {\bibinfo {author} {\bibfnamefont {T.}~\bibnamefont
  {Regimbau}}\ and\ \bibinfo {author} {\bibfnamefont {J.~A.}\ \bibnamefont
  {de~Freitas~Pacheco}},\ }\href {http://arxiv.org/abs/astro-ph/0005043}
  {\bibfield  {journal} {\bibinfo  {journal} {Astron. Astrophys.}\ }\textbf
  {\bibinfo {volume} {359}},\ \bibinfo {pages} {242} (\bibinfo {year}
  {2000})}\BibitemShut {NoStop}%
\bibitem [{\citenamefont {Popov}\ \emph {et~al.}(2010)\citenamefont {Popov}
  \emph {et~al.}}]{PopovEtAl2010}%
  \BibitemOpen
  \bibfield  {author} {\bibinfo {author} {\bibfnamefont {S.~B.}\ \bibnamefont
  {Popov}} \emph {et~al.},\ }\href@noop {} {\bibfield  {journal} {\bibinfo
  {journal} {Mon. Not. R. Astron. Soc.}\ }\textbf {\bibinfo {volume} {401}},\
  \bibinfo {pages} {2675} (\bibinfo {year} {2010})}\BibitemShut {NoStop}%
\bibitem [{\citenamefont {Perna}\ \emph {et~al.}(2008)\citenamefont {Perna}
  \emph {et~al.}}]{PernaEtAl2008}%
  \BibitemOpen
  \bibfield  {author} {\bibinfo {author} {\bibfnamefont {R.}~\bibnamefont
  {Perna}} \emph {et~al.},\ }\href@noop {} {\bibfield  {journal} {\bibinfo
  {journal} {Mon. Not. R. Astron. Soc.}\ }\textbf {\bibinfo {volume} {384}},\
  \bibinfo {pages} {1638} (\bibinfo {year} {2008})}\BibitemShut {NoStop}%
\bibitem [{\citenamefont {Gonthier}\ \emph {et~al.}(2011)\citenamefont
  {Gonthier} \emph {et~al.}}]{GonthierEtAl2011}%
  \BibitemOpen
  \bibfield  {author} {\bibinfo {author} {\bibfnamefont {P.~L.}\ \bibnamefont
  {Gonthier}} \emph {et~al.},\ }\href@noop {} {\bibfield  {journal} {\bibinfo
  {journal} {AIP Conf. Proc.}\ }\textbf {\bibinfo {volume} {1357}},\ \bibinfo
  {pages} {245} (\bibinfo {year} {2011})}\BibitemShut {NoStop}%
\bibitem [{Note6()}]{Note6}%
  \BibitemOpen
  \bibinfo {note} {Some studies predict that magnetars are formed with fast
  initial spins \cite {ThompsonDuncan1993}. Since we are interested in
  obtaining an upper limit, we assume that all magnetars start emitting at
  $f_{\protect \text {max}}$.}\BibitemShut {Stop}%
\bibitem [{\citenamefont {Nishizawa}\ \emph {et~al.}(2010)\citenamefont
  {Nishizawa}, \citenamefont {Taruya},\ and\ \citenamefont
  {Kawamura}}]{NishizawaEtAl2010}%
  \BibitemOpen
  \bibfield  {author} {\bibinfo {author} {\bibfnamefont {A.}~\bibnamefont
  {Nishizawa}}, \bibinfo {author} {\bibfnamefont {A.}~\bibnamefont {Taruya}}, \
  and\ \bibinfo {author} {\bibfnamefont {S.}~\bibnamefont {Kawamura}},\
  }\href@noop {} {\bibfield  {journal} {\bibinfo  {journal} {Phys. Rev. D}\
  }\textbf {\bibinfo {volume} {81}},\ \bibinfo {pages} {104043} (\bibinfo
  {year} {2010})}\BibitemShut {NoStop}%
\bibitem [{\citenamefont {Kotera}(2011)}]{Kotera2011}%
  \BibitemOpen
  \bibfield  {author} {\bibinfo {author} {\bibfnamefont {K.}~\bibnamefont
  {Kotera}},\ }\href@noop {} {\bibfield  {journal} {\bibinfo  {journal} {Phys.
  Rev. D}\ }\textbf {\bibinfo {volume} {84}},\ \bibinfo {pages} {023002}
  (\bibinfo {year} {2011})}\BibitemShut {NoStop}%
\bibitem [{\citenamefont {{LIGO Scientific Collaboration}}(2007)}]{LIGO2007b}%
  \BibitemOpen
  \bibfield  {author} {\bibinfo {author} {\bibnamefont {{LIGO Scientific
  Collaboration}}},\ }\href@noop {} {\bibfield  {journal} {\bibinfo  {journal}
  {Phys. Rev. D}\ }\textbf {\bibinfo {volume} {76}},\ \bibinfo {pages} {082001}
  (\bibinfo {year} {2007})}\BibitemShut {NoStop}%
\end{thebibliography}%

\end{document}